\documentclass[preprint,12pt]{elsarticle}

\usepackage{amsmath,amsfonts,amsthm,amssymb,paralist,subfigure,graphicx,amsbsy,float,epsfig,color, bm}
\usepackage{cuted,mathtools,lipsum}

\usepackage{algorithm,algpseudocode}

\usepackage{stmaryrd,url}

\usepackage{amsthm}

\def\mb{\mathbf}

\def\mc{\mathcal}

\journal{Annual Reviews in Control}

\begin{document}
	
	\begin{frontmatter}
		
		\title{Distributed Algorithms for Filtering, Estimation, and Fault Detection over Cyber-Physical-Systems: A Tutorial and Survey}

		\author[Sem]{Mohammadreza Doostmohammadian}
		\affiliation[Sem]{Faculty of Mechanical Engineering, Semnan University, Semnan, Iran, doost@semnan.ac.ir.}
		\author[HRR1]{Mahdi Shamsi}
		\affiliation[HRR1]{Department of Electrical Engineering, Sharif University of Technology, Tehran, Iran, Mahdi.Shamsi@alum.sharif.edu.}
		\author[Qom]{Hadi Zayyani}
		\affiliation[Qom]{Faculty of Electrical and Computer Engineering, Qom University of Technology, Qom, Iran, zayyani@qut.ac.ir.}
		\author[NM]{ Nader Meskin}
		\affiliation[NM]{Electrical Engineering Department, Qatar University, Doha, Qatar, nader.meskin@qu.edu.qa.}
		\author[HRR]{ Hamid R. Rabiee}
		\affiliation[HRR]{Department of Computer Engineering, Sharif University of Technology, Tehran, Iran, rabiee@sharif.edu.}
		\author[SP]{ Sergio Pequito}
		\affiliation[SP]{Department of Electrical and Computer Engineering and Institute for Systems and Robotics, Instituto Superior Tecnico, University of Lisbon, Portugal, sergio.pequito@tecnico.ulisboa.pt.}
		\author[UK]{ Usman A. Khan}
		\affiliation[UK]{Computer Science Department, Boston College, Boston, USA, {usman.khan@bc.edu}}

		\begin{abstract}
	This survey provides a comprehensive overview of distributed estimation, filtering, and fault detection techniques in the context of cyber-physical systems (CPS). Distributed algorithms are crucial for large-scale system monitoring as they enable parallel data processing, local fault identification, and real-time analysis across multiple nodes.
	
	To establish a strong foundation, we first define essential aspects of linear dynamical systems and related graph theoretic concepts, emphasizing observability conditions that are key for local state estimation. We introduce the mathematical framework necessary to understand both the theoretical underpinnings and practical implementations of distributed algorithms in CPS environments.
	
	After discussing consensus algorithms, this survey highlights single-time and double-time-scale consensus-based estimation and filtering approaches. We provide a detailed comparative analysis of these methodologies, examining their computational requirements, communication overhead, and performance characteristics in resource-constrained environments. We further explore different diffusion-based estimation techniques and observationally redundant designs to enhance resilience and robustness against failures and adversarial attacks.
	
	In addition, we investigate distributed fault detection methods that enable local isolation of faults over large-scale CPS. We present both stateless and stateful detection mechanisms, along with threshold-based techniques that balance detection accuracy and false alarm rates. These approaches are essential for maintaining system integrity and preventing cascading failures in critical infrastructure.
	
	This survey concludes with an exploration of diverse real-world applications. We examine implementation challenges and algorithm adaptations in smart grid and power networks, social systems, target tracking and localization, and intelligent transportation systems. Each application domain demonstrates how theoretical advances translate into practical solutions for complex monitoring problems.
	
	By bridging theoretical insights with practical applications, this survey and tutorial provides valuable understanding and research directions in the field of distributed algorithm design for CPS, offering both newcomers and experienced researchers a comprehensive resource for addressing current challenges and future opportunities.
\end{abstract}

\begin{graphicalabstract}
	\includegraphics{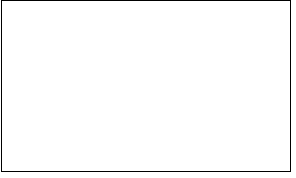}
\end{graphicalabstract}

\begin{highlights}
	\item Delivering a comprehensive theoretical foundation on linear dynamical systems, algebraic graph theory, observability analysis, and consensus algorithms with concrete mathematical frameworks essential for understanding modern distributed estimation and filtering approaches.
	\item Systematically analyzing and categorizing consensus-based filtering algorithms through a novel classification framework that distinguishes between single-time-scale and double-time-scale approaches, including a quantitative comparison of their computational and communication requirements.
	\item Advancing the understanding of diffusion-based filtering algorithms with detailed analysis of their convergence properties, robustness characteristics, and superior performance in dynamic network topologies with varying connectivity.
	\item Unveiling methodologies for distributed fault detection with threshold-based techniques that enable real-time fault identification and isolation, addressing a critical security vulnerability in large-scale CPS implementations.
	\item Bridging theory and practice through in-depth analysis of high-impact applications in smart grids, social systems, target tracking, and intelligent transportation systems, with implementable algorithms that demonstrate significant performance improvements over traditional centralized approaches.
\end{highlights}

\end{frontmatter}

\section{Introduction} \label{sec_intro}
In recent years, the development of cyber-physical systems (CPS) has revolutionized various sectors, including manufacturing, transportation, healthcare, and smart cities \cite{9695482}. CPS integrate computational algorithms, communication networks, and physical processes, enabling complex interactions between hardware and software components. As these systems become increasingly prevalent, ensuring their reliability, safety, and efficiency has emerged as a critical challenge.

Distributed algorithms play a pivotal role in addressing these challenges by enabling robust filtering, accurate estimation, and localized fault detection across decentralized architectures. The dynamic and often unpredictable nature of CPS environments necessitates innovative approaches to data processing and decision-making. Traditional centralized methods can be vulnerable to single points of failure, latency issues, and scalability constraints. In contrast, distributed algorithms facilitate localized processing, where data is aggregated and analyzed across multiple nodes, enhancing system resilience against faults while supporting real-time processing essential for applications ranging from autonomous vehicles \cite{safi2022resilient,xie2022distributed} to industrial automation \cite{lesi2021security,chen2010distributed} and even social networks \cite{isj_cyber,9723301}.

\begin{figure} 
\centering
\includegraphics[width=4in]{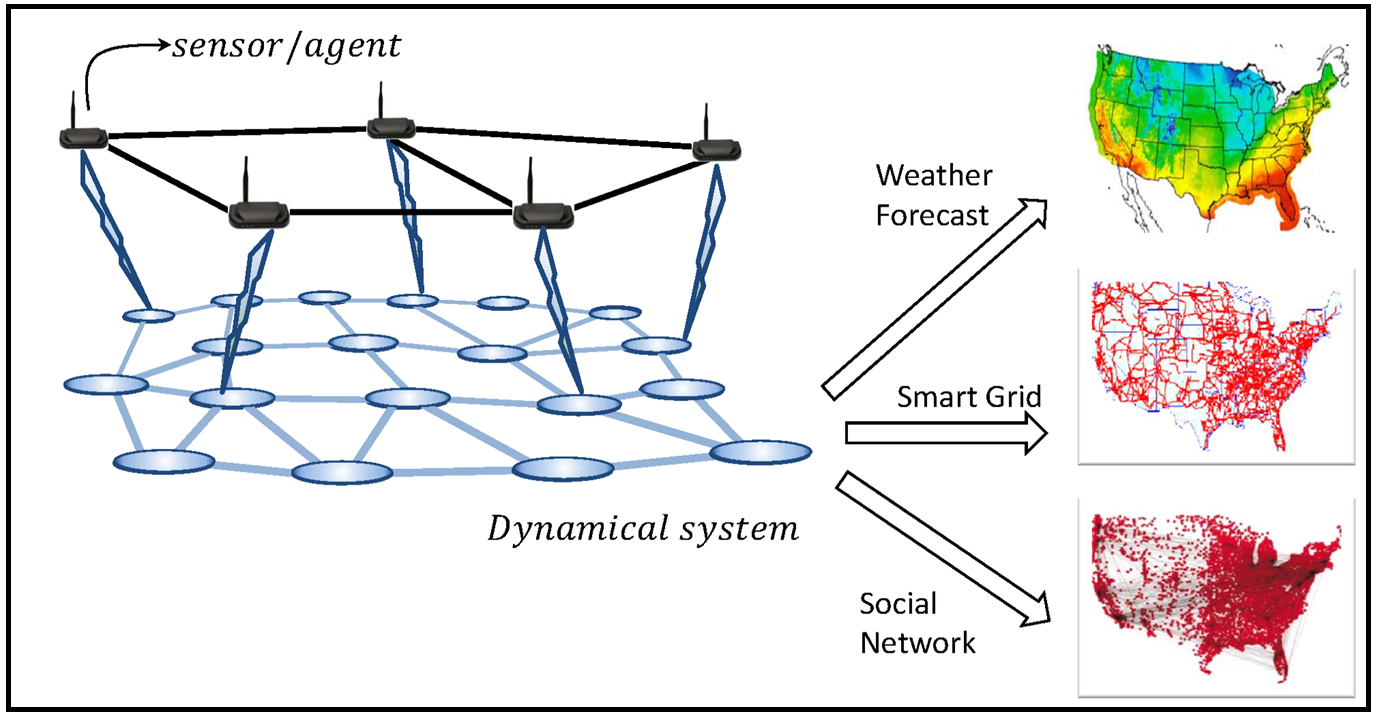}
\caption{An illustrative distributed CPS setup with the cyber-layer consisting of a network of agents/sensors (multi-agent network or sensor network) monitoring the physical-layer modelled as a large-scale dynamical system (e.g., social network, smart grid, or weather system).
} \label{fig_cps}
\end{figure}

The capability to enhance system reliability and scalability in complex interconnected setups, along with localized processing and synchronization of information, has increased interest in distributed filtering, estimation, and fault detection. Advanced algorithms such as Kalman filters and particle filters are commonly adapted to distributed settings \cite{kumari2021use}, allowing for practical state estimation while accommodating constraints like bandwidth limitations \cite{5438273} and varying communication conditions (switching networks) \cite{taoufik2020distributed}.

Furthermore, as CPS becomes more integrated and interconnected, distributed fault detection algorithms are gaining more interest as they allow local identification and isolation of anomalies, thereby minimizing the impact of faults on overall system performance \cite{taoufik2022distributed}. The advantages of adopting distributed algorithms over CPS include the following:

\begin{itemize}
\item \textbf{Scalability:} Distributed algorithms are inherently scalable, making them suitable for large networks of nodes typical in CPS. As the number of sensors, agents, or system components increases, these algorithms can adaptively analyze additional data without significant performance degradation;
\item \textbf{Local decision-making and redundancy:} Individual processing nodes can operate in parallel, making local decisions based on localized data. This reduces the risk of a single point of failure affecting the overall system and enhances the resilience of CPS;
\item \textbf{Real-time processing:} Distributed algorithms enable real-time data processing by allowing agents/nodes to process data locally without relying on a centralized processor. This is crucial for applications requiring immediate responses, such as autonomous navigation or intelligent transportation networks;
\item \textbf{Resource efficiency:} By processing data closer to its source and only transmitting essential information, distributed algorithms significantly reduce bandwidth usage. This is particularly beneficial in resource-constrained setups where communication costs are of primary concern;
\item \textbf{Collaborative information sharing:} Distributed algorithms can leverage information from different nodes to refine estimates. For instance, in multi-agent systems, agents can share their local observations to converge toward a more accurate global estimate.
\end{itemize}

These advantages make distributed algorithms valuable in many technical scenarios, including resource allocation and scheduling \cite{doostmohammadian2025survey,jiang2015survey}, optimization \cite{nedic2018distributed,yang2019survey,molzahn2017survey,Zheng2022review,jfi},  data mining \cite{zeng2012distributed,gan2017data}, and machine learning \cite{peteiro2013survey,verbraeken2020survey}. For filtering applications, distributed algorithms efficiently aggregate and process data from various sensors to improve decision-making quality. Consensus filters \cite{demetriou2010design,olfati2005consensus} can operate across networked nodes, enabling them to reach a common estimate despite local data inconsistencies. Moreover, distributed filtering techniques can robustly handle noise, disturbances, and data uncertainties prevalent in sensor measurements \cite{hedayati2020robust}. 

In addition, distributed fault detection and isolation (FDI) algorithms can monitor system performance by handling data from multiple nodes via parallel processing \cite{6848128,Bakhtiaridoust}. These algorithms facilitate the isolation of faults to specific nodes or components, enabling targeted interventions and preventing further damage to the multi-agent system.

In summary, this survey and tutorial present a comprehensive examination of \textit{distributed} estimation, filtering, and fault detection techniques within the context of CPS.
While existing surveys comprehensively cover distributed algorithms for optimization tasks--including multi-agent coordination \cite{yang2019survey}, power grid control \cite{molzahn2017survey}, resource allocation \cite{doostmohammadian2025survey}, task scheduling \cite{jiang2015survey}, and machine learning \cite{verbraeken2020survey,peteiro2013survey}--they do not address the fundamentally different challenges that arise in distributed estimation and detection problems. Unlike optimization, which seeks to minimize a global cost function, estimation and fault detection require agents to reconstruct unobservable states and distinguish genuine system dynamics from malicious attacks, introducing unique challenges in observability analysis, consensus under faulty data, and robustness guarantees. This paper fills this gap by systematically addressing the modelling, observability, and robustness challenges specific to state/parameter estimation and fault/attack detection in decentralized sensor networks.

Specifically, our survey and tutorial include the following:

\begin{itemize}
\item  \textbf{Comprehensive Framework:} We provide a thorough overview of the current state of research by integrating foundational concepts of linear dynamical systems with advanced graph theoretic principles, creating a framework for understanding observability conditions critical to distributed system monitoring;
\item \textbf{Consensus Algorithm Analysis:} We explore consensus algorithms that serve as the backbone for distributed estimation, filtering, and fault detection techniques, clearly differentiating between single-time-scale and double-time-scale approaches and their applicability in resource-constrained environments;
\item \textbf{Resilient Design Strategies:} We review observationally redundant designs and diffusion-based filtering algorithms, highlighting their effectiveness in enhancing system resilience against failures and robustness against disturbances;
\item \textbf{Complexity Assessment:} We analyze the communication and computation complexity of existing methods, providing insights that illuminate future research directions in the field;
\item \textbf{Fault Detection Methodologies:} We examine distributed fault detection methods that address the challenges of monitoring in the presence of faults, attacks, or anomalies, with a particular focus on observer-based detection and isolation of faulty nodes to prevent cascading failures;
\item \textbf{Practical Applications:} We investigate diverse applications including smart grid and power networks, social systems, target tracking and localization, and intelligent transportation systems, demonstrating the practical implications of these theoretical approaches.
\end{itemize}

\begin{figure} 
	\centering
	\includegraphics[width=4in]{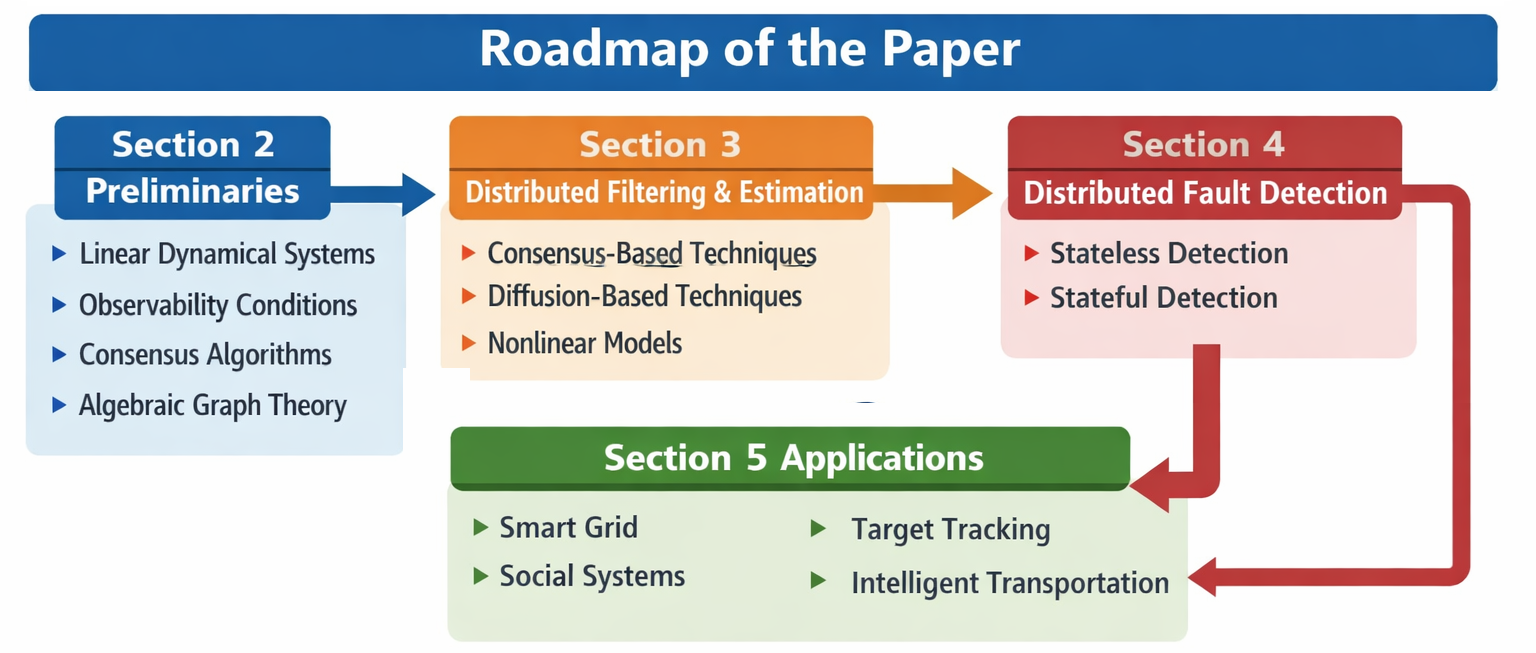}
	\caption{This figure presents the roadmap of our survey.
	} \label{fig_roadmap}
\end{figure}

Through this comprehensive review, we bridge the gap between theoretical advancements and practical applications, motivating further research on distributed algorithm design for CPS. Fig.~\ref{fig_roadmap} provides the roadmap of the paper showing the section-wise structure and thematic progression of this survey.

\section{Preliminary Concepts and Background}
This section establishes the foundational framework necessary for understanding distributed algorithms in cyber-physical systems. In Subsection \ref{sec_lindyn}, we examine linear dynamical systems and their graph-theoretic representations, which model the physical processes being monitored. Subsection \ref{sec_obsrv} explores observability conditions -- both algebraic and structural -- that determine when a system's state can be inferred from available measurements. In Subsection \ref{sec_consensus}, we delve into consensus algorithms, the backbone of many distributed techniques, explaining how multiple agents can reach agreement through local interactions. Finally, Subsection \ref{sec_alggraph} investigates key concepts from algebraic graph theory, including Laplacian matrices and algebraic connectivity, which provide critical insights into the convergence behavior and resilience of distributed systems under various network topologies.

\subsection{Linear Dynamical Systems} \label{sec_lindyn}
Linear dynamical systems are the main mathematical model considered for the distributed setups, describing the evolution of the underlying (physical) system over time. These systems are widely used in various applications, including control theory \cite{Kalman63,robinson2012introduction}, signal processing \cite{gajic2003linear}, and communication systems \cite{you2015}.
Linear systems are defined by their state-space representation, transfer function representation, or both \cite{ogata2009modern}. This section presents the state-space representation, which is particularly useful in distributed algorithms for filtering, estimation, and fault detection.

\paragraph{State-Space Representation} A linear dynamical system can be described using a state-space model represented by the following equations \cite{ogata2009modern}:
\begin{align}  \label{eq_A}
\mathbf{x}(t+1) &= A \mathbf{x}(t) + B \mathbf{u}(t) + \nu(t),
\\ \label{eq_C}
\mathbf{y}(t) &= C \mathbf{x}(t) + \mu(t),
\end{align}
where $x(t) \in \mathbb{R}^n$ is the state vector at time $t$ representing the physical parameters involved in the dynamical system, $u(t) \in \mathbb{R}^m$ is the input vector (control input) at time $t$, $y(t) \in \mathbb{R}^N$ is the output vector at time $t$, $A \in \mathbb{R}^{n\times n}$ is the state transition matrix, $B \in \mathbb{R}^{n\times m}$ is the input matrix, and $C \in \mathbb{R}^{N\times n}$ is the output matrix. The state transition matrix $A$ governs the dynamic behavior of the system, while the matrices $B$ and $C$ define how the inputs affect the state and how the state contributes to the output, respectively.

\paragraph{Noise Modeling}
In practical applications, system models and measurements are  subject to noise. To capture this realistic condition:

\begin{itemize}
\item $\boldsymbol{\nu}(t) \in \mathbb{R}^n$ is the process noise, typically assumed to be zero-mean Gaussian noise
\item $\boldsymbol{\mu}(t) \in \mathbb{R}^N$ is the measurement noise, also assumed to be zero-mean Gaussian noise
\end{itemize}

This formulation is the typical model for filtering techniques, such as the Kalman filter, and fault detection over CPS. Understanding these linear systems is foundational for designing and implementing distributed filtering and estimation techniques that can efficiently track the (physical) state of the system in noise-corrupted setups.

\paragraph{Structured Systems Theory and Graph Representation}

A representation of linear systems and the study of their properties can be effectively approached through structured systems theory \cite{jcn,RAMOS2022110229,DION20031125,ijss,Obsrv_IoT}. In structured systems theory, the behavior and dynamics of a linear system can be visually represented using a directed graph, known as the system digraph. This representation highlights the interconnections among state variables and outputs, and models the structural information contained in the system matrix $A$ and output matrix $C$ through their zero/nonzero patterns  \cite{BOUKHOBZA2006629}.

A system digraph is a directed graph that represents the relationships between the state variables and the outputs of a system, as shown in Fig.~\ref{fig_A} for an illustrative example. The construction process follows these principles:

\begin{figure} 
\centering
\includegraphics[width=1.5in]{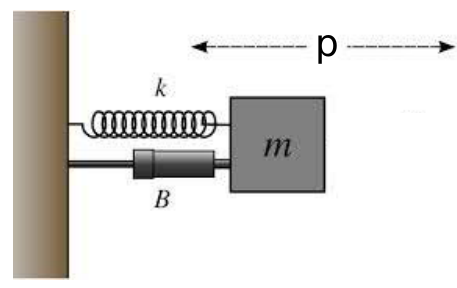}
\includegraphics[width=1.5in]{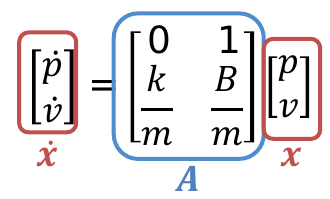}
\includegraphics[width=1.5in]{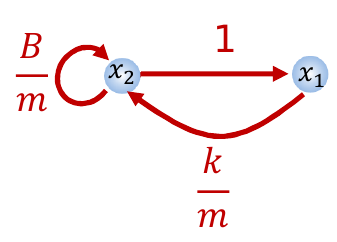}
\caption{The left figure shows a simple mass-spring-damper as a linear dynamical system. The figure in the middle represents its state-space dynamical system representation following Eq.~\eqref{eq_A}, where the state variables are position $p$ and velocity $v$. The right figure shows its system digraph representation with two state nodes denoting the position and velocity state variables and three links as the nonzero entries of the system matrix $A$.
} \label{fig_A}
\end{figure}

\begin{itemize}
\item \textbf{Nodes:} The nodes in the digraph are associated with the state variables, representing the individual components of the system state vector;

\item \textbf{Directed links:} The directed links denote the influence or control relationships among these variables as defined by the entries of the system matrix; and

\item \textbf{Construction rules:} The system digraph is constructed by creating nodes for each state variable and drawing directed links according to the nonzero entries in the matrix $\mc{A}$.
\end{itemize}

\paragraph{Structural Influence Relationships}

The entries of the system matrix $A$ dictate how the state variables interact with each other, i.e.,

\begin{itemize}
\item If an entry $a_{ij}$ is nonzero, there is a directed link from state node $j$ to state node $i$ in the digraph, indicating that the state $x_j$ influences the state $x_i$;

\item Conversely, if $a_{ij} = 0$, there is no direct influence from state $x_j$ to state $x_i$ in the system.
\end{itemize}

Similar to the system matrix, the output matrix $C$ can be represented by its zero/nonzero pattern as follows:

\begin{itemize}
\item A nonzero entry $c_{ij}$ indicates a directed link from state node $j$ to output node $i$ in the digraph, representing that the state $x_j$ directly influences the output $y_i$; and

\item This establishes how each state variable contributes to the observable outputs, which is crucial for tasks such as estimation and fault detection.
\end{itemize}

This graph-theoretic representation provides valuable insights into the structural properties of the system, enabling the analysis of observability, controllability, and other fundamental system characteristics that are essential for distributed algorithm design.

\subsection{Observability Conditions} \label{sec_obsrv}
Observability is a fundamental concept in control theory and estimation, referring to the ability to determine the complete internal state of a dynamical system by observing its outputs over time \cite{bay}. More formally, a system is said to be observable if, for every possible sequence of states, the current state can be determined in a finite number of steps from the output measurements.
For a linear system described by the state-space equations given by Eq.~\eqref{eq_A}-\eqref{eq_C}, the system is said to be observable \textit{if and only if} the \textit{observability matrix} (or \textit{Grammian matrix}) $\mathcal{O}$ has full rank which  is defined as follows \cite{bay}:
\begin{align}  \label{eq_O}
\mathcal{O} =
\begin{bmatrix}
	C \\
	CA \\
	CA^2 \\
	\vdots \\
	CA^{n-1}
\end{bmatrix},
\end{align}
where $n$ is the dimension of the state vector $\mathbf{x}$ (or the size of the system).  If $\mbox{rank}(\mathcal{O}) < n$, the system is unobservable, meaning that some states cannot be inferred/estimated from the output measurements.

\subsubsection{Structural Observability}
\paragraph{Graph-Based Representation}
Structural observability offers a broader perspective that goes beyond numerical input-output relationships by examining the zero-nonzero pattern of the matrices involved in the system \cite{liu2018partial,sundaram2012structural}. In a structural sense, we analyze the system based on the connectivity and sparsity of its representation, which can be beneficial in cases where the exact numerical values of the matrices are uncertain or when the system is subject to changes \cite{RAMOS2022110229,DION20031125}.

Similar to the traditional observability matrix defined by (3), the structural observability matrix is derived from the zero-nonzero pattern of the system matrices. A matrix entry is marked as nonzero if it is possible for that entry to contain a nonzero value for some realization of the system parameters. The structural observability matrix can be formed similarly, using the same structure as $O$ in (3).

The system is structurally observable if the corresponding structural observability matrix, derived from the patterns of nonzero entries of $C$ and $A$, has full structural rank. Structural rank (or generic rank) refers to the maximum rank that a matrix achieves when its nonzero entries are allowed to take arbitrary values. This concept is fundamental in structured systems theory, as it characterizes system properties that are determined by the pattern of interconnections rather than their specific numerical values.

Structural observability is particularly useful in systems where components or parameters are uncertain \cite{sobsrv}, allowing for a qualitative assessment of the system's ability to infer states based on output measurements. It is more convenient to analyze the structural observability using graph theory, providing a visual and analytical approach to assess the observability of the dynamical system. By representing the system as a directed graph, we can derive conditions that ensure structural observability based on the connectivity of system states and outputs \cite{liu2013observability,tnse18}.

Formally, let $G_A = (V_A, E_A)$ be the directed graph, where $V_A$ comprises both state nodes $x$ and output nodes $y$. For structural observability, two key conditions must hold on $G_A$ \cite{liu2013observability,tnse18}:

\begin{enumerate}
\item For every state node $i \in V_A$, there must be a path $P_{i\rightarrow j}$ such that $j$ is an output node, indicating that the dynamics of state $i$ affect at least one of the system output. This condition ensures that changes in the state can influence the output, which is crucial for observability.

\item There must exist a family of disjoint cycles and/or output-connected paths that govern all the state nodes present in the system. This means that for each state node, there can be either a cycle involving that node or a connected path to an output node that includes the state node. Specifically, let $C$ be the set of cycles in the graph and $P$ be the set of paths. Each state node $i$ should either belong to some cycle $c \in C$ or have a path $P_{i\rightarrow j}$ to an output node $j$ such that the output can be influenced by the state node through the cycles and paths.
\end{enumerate}

To check for structural observability in a directed graph $G_A$ of a dynamical system $A$ with outputs $C$, these two conditions need to be verified. The existence of output-connected paths can be checked through depth-first search (DFS) or breadth-first search (BFS) algorithms \cite{algorithm}, starting from each state node to check for paths to output nodes. On the other hand, the algorithms to check the structural rank of $A$ verify the existence of family cycles over $G_A$, for example, see \cite{harary}. It should be noted that many existing results are stated for the dual problem of structural controllability \cite{lin,isj_minimal,7112630,10589428}, which can be simply extended to structural analysis for observability, for example, by reversing the conditions for the direction of paths/links.

Certain properties of the system can be understood from this linear model, particularly in the generic sense. The graph representation modeling the zero-nonzero pattern of the system allows efficient checking of generic rank and structural properties. One main condition for estimation and filtering is to verify the observability of the system pair $(A, C)$, which can be effectively achieved via structured systems theory as described above.

\subsection{Consensus Algorithms} \label{sec_consensus}
In distributed systems, consensus algorithms are fundamental for achieving agreement among multiple agents or nodes by sharing states or making collective decisions despite the presence of uncertainties or failures. These algorithms play a critical role in coordination tasks across various domains, including sensor networks, robotics, and multi-agent systems, where they serve as the backbone for decentralized filtering, estimation, and fault detection mechanisms.

The consensus problem involves a group of agents (or processing nodes) that need to converge to a common value (reach agreement) \cite{cons,ren2005consensus,nonlincons}, which may represent an estimated state, a measurement, or the output of a decentralized decision-making process. The agents operate on their local information, communicate with each other, and rely on a set of rules to update their states based on the networked interactions. The primary challenges arise from the distributed nature of the network and communication network of agents (modelled by a graph topology) that might be subject to delays, asynchronicity, or potential packet drops and link failure.

Mathematically, the consensus problem can be formulated as follows:
\begin{enumerate}[1.]
\item \textbf{Agent state initialization}: Each agent $i$ in a network of $n$ agents is initialized with a state $x_i(0)$ at time $t = 0$. This state could be a scalar or vector representing the information each agent possesses.
\item \textbf{Communication topology}: Agents communicate according to a directed or undirected graph represented by $\mc{G}(\mc{V}, \mc{E})$, where $\mc{V}$ is the set of nodes (agents) and $\mc{E}$ is the set of links (communications). Each link $(i, j) \in \mc{E}$ indicates that agent $i$ can exchange information directly with agent $j$.
\item  \textbf{Update rule}: Every agent $i$ updates its states iteratively based on information received from its neighbours, denoted by $\mc{N}_i$. The most common linear consensus update rule is as follows \cite{olfati_rev,SensNets:Olfati04,ren2005consensus}:
\begin{align}
	z_i(t+1) = z_i(t) + \sum_{j \in N_i} w_{ij} (z_j(t) - z_i(t)),
\end{align}
where  $w_{ij}$ is the weight assigned to the information received from agent $j$ with $W=[w_{ij}]$ as the weight matrix and $z_i(t)$ is the state at time $t$.
One can reformulate the solution in Laplacian form by defining the Laplacian matrix $L=[l_{ij}]$ as:
\begin{align}
	L = \mc{D} - \mc{A},
\end{align}
where $\mc{D}$ is the diagonal degree matrix, where each diagonal entry $d_{ii}$ is the degree of node $i$, and
$\mc{A}$ is the adjacency matrix, where each entry $a_{ij}$ is $1$ if there is a link between nodes $i$ and $j$ and $0$ otherwise. Then, the consensus dynamics is described by:
\begin{align}
	\mathbf{z}(t+1) = \mathbf{z}(t) - \epsilon L \mathbf{z}(t) = (I -\epsilon L) \mathbf{z}(t),
\end{align}
where $\epsilon$ is a small positive constant (step size) that controls the convergence rate and the column vector $\mathbf{z} = [z_1,\dots,z_N]^\top$ as the state variable. This equation shows that each agent updates its value based on the differences between its own state and the states of its neighbours, influenced by the structure of the communication graph captured by $L$.

\item \textbf{Convergence criteria}: The goal is for all agents to converge to a common consensus state $z^*$ such that:
\begin{align}
	\lim_{t \to \infty} z_i(t) = z^*(t), \quad \forall i \in \mc{V}.
\end{align}
\end{enumerate}

In the context of consensus algorithms, weight design plays a crucial role in determining how agents combine received information from their neighbours during state updates.  Stochastic weight design is particularly key in consensus, i.e., the weight matrix $W=[w_{ij}]$ satisfies row/column/bi-stochasticity depending on the network structure (directed or undirected). A row-stochastic consensus matrix satisfies the following:
\begin{align} \label{eq_stochastic}
\sum_{j \in N_i} w_{ij} = 1.
\end{align}
Similarly, column-stochasticity is over the columns of $W$. Bi-stochasticity implies both row and column stochastic weights. There are different algorithms in the literature to design stochastic weights, namely Metropolis-Hastings algorithm \cite{schwarz2014convergence}, Wasserstein average consensus \cite{xin2022distributed}, or simply set $w_{ij} = \frac{1}{N_i}$ \cite{SensNets:Olfati04}.
Note that under certain conditions these weights $w_{ij}$ can vary over time and the network topology might be also switching.

Consensus algorithms can be categorized based on their specific characteristics:
\begin{enumerate}[1.]
\item \textbf{Asymptotic vs. Finite-time Convergence}: Finite/fixed-time algorithms guarantee convergence to a unique value in a finite time horizon (or finite number of iterations), while asymptotic algorithms converge asymptotically over time. The existing linear algorithms mostly converge assymptotically \cite{olfati_rev,SensNets:Olfati04,ren2005consensus}, while other nonlinear finite-time \cite{Scientia2011,taes,li2011finite,wang2010finite,rikos2022distributed}, fixed-time \cite{ning2022fixed,li2020fixed,liu2022overview,ni2022fixed}, and prescribed-time \cite{ning2022fixed,ren2021prescribed,chen2020prescribed,gong2020distributed} algorithms are proposed in the literature.

\item \textbf{Synchronous vs. Asynchronous}: In synchronous algorithms \cite{olfati_rev,SensNets:Olfati04,ren2005consensus}, all agents update their states simultaneously based on the latest available data, whereas asynchronous algorithms \cite{fang2005asynchronous,carron2014asynchronous,li2020distributed,zhu2020asynchronous,zhao2021leader} allow agents to update at different times, accommodating delays and improving resilience.

\item \textbf{Linear vs. Nonlinear}: Linear consensus algorithms \cite{olfati_rev,SensNets:Olfati04,ren2005consensus} use linear combinations of neighbor states (and initial states), while nonlinear variants are designed to converge to a nonlinear function of initial states \cite{bauso2006non,nonlincons,NOSRATI20122262}.
\end{enumerate}

Furthermore, the consensus convergence is defined under several conditions, including the properties of the communication graph (e.g., switching connectivity \cite{ZHOU20091455,7270266}, packet loss \cite{Fagnani060676866,6426252}, or potential delays \cite{6571230,MUNZ20101252}).

\subsection{Algebraic Graph Theory} \label{sec_alggraph}

In distributed systems, multi-agent networks (or the network of computing nodes) can be modelled by graphs, denoted by $\mc{G}(\mc{V}, \mc{E})$,  where the set of agents $\mc{V}$ correspond to nodes and communication links between them correspond to links $\mc{E}$. This graph representation facilitates the analysis of consensus algorithms, as it allows us to use concepts from graph theory to understand how agents interact and converge to the consensus value. This is discussed in the previous subsection.

The behaviour of a multi-agent network can be described using the Laplacian matrix $L$ (or the weight matrix $W$), which encodes the structure of the underlying graph \cite{hornjohnson}. The consensus process is typically represented through linear iterative updates driven by the topology represented by $L$ (or $W$) \cite{olfati_rev,SensNets:Olfati04}.
The Laplacian matrix $L$ has several important properties that are critical for analyzing consensus algorithms:

\begin{enumerate} [1.]
\item \textbf{Symmetry and Positive Semi-Definiteness}:
The Laplacian matrix for undirected graphs is symmetric, which follows from the symmetric nature of the adjacency matrix and the diagonal degree matrix. Additionally, $L$ is positive semi-definite \cite{olfati_rev,SensNets:Olfati04}, meaning for any vector $\mathbf{z} \in \mathbb{R}^{n \times n}$ such that:
\begin{align}
	\mathbf{z}^\top L \mathbf{z} \geq 0.
\end{align}
This property ensures that the quadratic form derived from the Laplacian does not take negative values, which is essential for stability in many algorithms.

\item \textbf{Zero Eigenvalue}:
The matrix $L$ always has at least one eigenvalue equal to zero. The multiplicity of this eigenvalue corresponds to the number of connected components in the graph. A connected graph will have exactly one zero eigenvalue \cite{godsil}.

\item \textbf{Eigenvalues and Convergence}:
The remaining eigenvalues of $L$ are positive, and their magnitudes give information about the convergence rates of the consensus process. The eigenvalue $\lambda_2$ denoting the second smallest eigenvalue (also known as the algebraic connectivity) plays a key role in convergence; higher values of $\lambda_2$ indicate faster convergence to consensus \cite{olfati_rev,SensNets:Olfati04} and optimization \cite{doostmohammadian2024clustering}.

\end{enumerate}

Algebraic connectivity $\lambda_2$ is an important property as it is crucial in understanding the resilience and convergence behaviour of multi-agent networks:

\begin{enumerate} [1.]
\item \textbf{Algebraic Connectivity}: Algebraic connectivity, $\lambda_2$, measures the connectivity of the graph. If $\lambda_2 > 0$, the graph is connected, meaning there is a path between any two nodes. Conversely, if $\lambda_2 = 0$, the graph is disconnected, indicating the presence of multiple connected components \cite{godsil}.

\item \textbf{Relevance to Consensus}: For consensus algorithms to function optimally, the underlying graph must be connected. A connected graph ensures that information can flow between all agents, thus allowing them to eventually reach a common consensus/agreement value. The magnitude of $\lambda_2$ also provides insight into the speed of convergence -- the larger the value of $\lambda_2$, the faster the agents will converge to consensus \cite{SensNets:Olfati04}.

\item \textbf{Graph Connectivity}: Graph connectivity refers to the minimum number of link deletions required to make a graph disconnected. A graph is said to be $k$-connected if at least $k$ links must be removed to disconnect the graph \cite{augment_book}. The algebraic connectivity $\lambda_2$ provides a spectral characterization of this property. Higher algebraic connectivity suggests greater resilience against failures and more robust consensus dynamics.
\end{enumerate}

Based on these, the Laplacian matrix serves as a fundamental tool in algebraic graph theory for modelling multi-agent networks. The properties of the Laplacian matrix, along with the concepts of algebraic connectivity and graph connectivity, provide significant insights into the convergence rate and stability of distributed algorithms. By studying these properties, one can better understand the dynamics of agreement processes in distributed systems.

\section{Distributed Filtering and Estimation}
This section explores how multiple nodes in a network can collaboratively estimate system states through data sharing. In Section~\ref{sec_cons_est}, the paper examines consensus-based techniques, which are further divided into single time-scale algorithms, where nodes perform one communication step between consecutive system dynamics updates, and double time-scale algorithms, where nodes perform multiple consensus iterations between system updates. In Section \ref{sec_obsrv_red}, the paper discusses observationally redundant design, which enhances system reliability by incorporating multiple equivalent sensors to maintain functionality despite failures. Section \ref{subsec_dlms} extensively discusses diffusion-based techniques as an alternative approach. Finally, Section \ref{sec_nonlin}, addresses nonlinear models, covering extensions of distributed filtering to nonlinear systems through methods like Consensus + Innovation Filtering, Extended Kalman Filters, Unscented Kalman Filters, and Distributed Particle Filters. Throughout these subsections, the paper analyzes the trade-offs between communication efficiency, observability requirements, and estimation accuracy.

It should be noted that distributed observer design constitutes a relevant line of research in distributed estimation and filtering. In contrast to stochastic Kalman-like formulations, distributed observers are typically developed within a deterministic framework and aim to asymptotically reconstruct the system state using local measurements and inter-agent communication. Many distributed estimation and filtering algorithms can be interpreted as stochastic extensions of such observer-based schemes, sharing similar information exchange structures, consensus mechanisms, and convergence objectives. For this reason, the literature on distributed observers is closely tied with distributed estimation and filtering literature. In this paper, we adopt this integrated perspective and discuss distributed observer and estimation-based methods jointly, as they form a coherent methodological framework for the studied distributed filtering approach.

\subsection{Consensus-based Techniques} \label{sec_cons_est}
Consensus-based techniques form the backbone of many distributed estimation systems, enabling multiple agents to collaboratively determine the state of a monitored system. In this subsection, we explore the fundamental principles of consensus estimation, examine the key differences between single and double time-scale approaches, analyze their computational and communication requirements, and discuss resilient design strategies for handling network failures.

Consensus-based techniques are collaborative algorithms used in distributed systems to achieve agreement among a group of agents or sensors on local estimates of states or parameters of interest, given that system observability holds. These methods leverage iterative communication protocols where agents exchange information to refine their estimates based on both local observations and the estimates from neighboring nodes.

Depending on the number of consensus iterations used for filtering the data, two main scenarios are adopted in the literature: single time-scale and double time-scale algorithms \cite{camsap11}. The primary difference relates to how the estimation and consensus processes are synchronized and how sensors handle updates over time.

Single time-scale methods operate under a one-time update schedule where all sensors update their estimates simultaneously once between two consecutive samples of system dynamics \cite{asilomar11}. This approach minimizes communication overhead but may require stronger observability conditions.

\begin{figure} 
\centering
\includegraphics[width=2in]{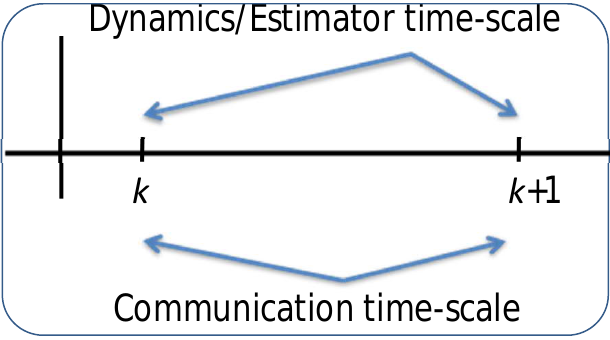}	
\includegraphics[width=2in]{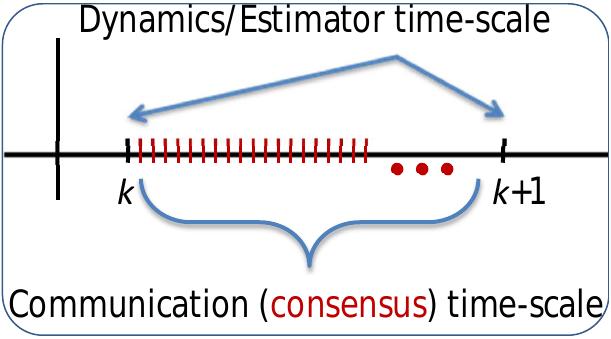}
\caption{Two main scenarios for consensus-based distributed estimation are compared in this figure. The left figure illustrates the single time-scale scenario with only one iteration of communication/consensus between two consecutive sampling times of the system dynamics. The right figure illustrates the double time-scale scenario with many iterations of consensus and communications (inner consensus loop) between two consecutive sampling times of the system dynamics.
} \label{fig_scale}
\end{figure}

In contrast, the double time-scale scenario allows sensors to update their local consensus estimates at a faster rate than the system dynamics, creating what is referred to as an ``inner consensus loop" \cite{4434303,olfati2005consensus}. This approach typically improves estimation accuracy but requires more communication resources. The distinction between these approaches is illustrated in Fig.~\ref{fig_scale}.

One primary advantage of consensus-based approaches is their robustness to communication challenges commonly encountered in CPS environments, including network delays \cite{lcss22,consdelay} and unreliable connections \cite{he2019secure,he2020secure}. This inherent resilience makes them particularly valuable for real-world deployments.

Additionally, these techniques demonstrate excellent scalability, allowing for the integration of numerous data-processing nodes without significant performance degradation \cite{tang2014information}. This property is crucial for large-scale monitoring applications such as smart grids and environmental sensing networks.

Consensus-based methods also support adaptive filtering capabilities, making them well-suited for time-varying scenarios where system dynamics evolve over time \cite{priel2023distributed}. As sensors iterate toward consensus on estimates, they naturally converge to the accurate global estimate, enhancing resilience against network perturbations \cite{rezaei2022event,ugrinovski,8419768}. Furthermore, many consensus algorithms offer inherent protection against denial-of-service (DoS) attacks, an increasingly important consideration in security-critical applications \cite{ali2024fully,asilomar14}. Interval-based (set-membership) distributed estimation is another interesting line of research. In this setup, each node maintains confidence sets (intervals or convex sets) that contain the true state given bounded noise and model uncertainty, and then, updates local feasible sets with information received from neighbors to shrink the intervals \cite{mumtaz2025peer}. Similarly, in a stochastic filtering/security framework, the work \cite{basit2024event} uses interval/robust reasoning implicitly when constructing bounds and resilient thresholds for attack detection and guaranteed error sets under bounded attack/noise assumptions.
Some distributed estimation methods involve estimation of certain unknown underlying parameters or inputs. In \cite{asad2024distributed}, joint state-and-parameter estimation is framed with $H_\infty$ objectives to provide worst-case (robust) guarantees under energy/bandwidth constraints while reducing the effect of uncertainty on the $H_\infty$ error bound. The paper \cite{basit2022adaptive} proposes an event-triggered distributed state estimator that handles unknown parameters by using adaptive observer gains.

Event-triggered distributed estimation are further of interest in the literature. The work \cite{li2016event} proposes a communication-efficient protocol where agents transmit quantized state updates only when local event conditions trigger, combining asynchronous event rules with finite-bit quantization to meet bandwidth constraints.  Via local computation, time-varying events to transmit measurements/estimates is considered in \cite{basit2023dynamic}, reducing communication while preserving estimation performance. 
Local event-triggered observer mechanisms to detect/mitigate DoS effects and to maintain consensus on estimates despite packet losses is considered in \cite{basit2022distributed}.
Zeno-free event-triggered protocol under hybrid attacks via $H_\infty$ observers is considered in \cite{han2022local}.

In this paper mainly discrete-time algorithms are considered
for three main reasons: (i) most practical implementations of estimation and filtering in CPS run on digital hardware that naturally operate in discrete time; (ii) discrete-time models simplify handling sampled measurements, packet-based communications, and network-induced effects, which are essential to distributed estimation problems; (iii) many continuous-time results can be discretized. However, some papers in the literature consider continuous-time dynamics as follows. Distributed extended Kalman filter via inter-nodal transformation theory is considered in  \cite{rostami2017distributed} to estimate the dynamic states of power systems.  A distributed filtering scheme is proposed in \cite{duan2022distributed} that fuses neighbor information by explicit handling of correlated measurement noise between sensors. Distributed state estimation for jointly observable multi-agent systems over periodic communication networks is considered in
\cite{WANG2024111564}. Continuous-time distributed observers
with discrete communication is discussed in \cite{dorfler2013continuous}. Distributed asynchronous Kalman filter for continuous-time stochastic processes is developed in \cite{kowalczuk2013asynchronous}.

In the following subsections, we delve deeper into the specific implementations of these consensus approaches. Section~\ref{sec_single} examines single time-scale algorithms, which prioritize communication efficiency while maintaining estimation accuracy. Section~\ref{sec_double} explores double time-scale algorithms that leverage multiple consensus iterations to achieve enhanced performance. Finally, Section~\ref{sec_obsrv_red} discusses observationally redundant designs that further improve robustness by incorporating strategic sensor redundancy to maintain functionality despite potential failures.

\subsubsection{Single Time-Scale Algorithms} \label{sec_single}
Single time-scale algorithms balance estimation accuracy with communication efficiency by performing a single consensus iteration between consecutive system measurements. This approach is particularly valuable in bandwidth-constrained environments while still maintaining robust state estimation capabilities.

Given a linear dynamical system in the form~\eqref{eq_A} with sensor measurements as described in~\eqref{eq_C}, distributed estimation aims to locally infer the global state of the system using only partial state measurements available at each node. Each agent must combine its local observations with information received from neighboring nodes to construct an accurate global state estimate.

Below, we present a representative example of a single time-scale distributed estimation technique that illustrates the core principles of this approach. This distributed estimator operates through a structured two-step process that alternates between consensus on predictions and measurement updates:
\begin{enumerate} [1.]
\item \textbf{Consensus on a-priori estimates:}
Agents (i.e., an entity comprised of a sensor and communication capabilities) share a-priori estimates (or predictions \cite{jstsp,acc13}) over the network $\mc{G}$ as follows:
\begin{eqnarray}\label{eq_p}
	\widehat{\mb{x}}_i(t|t-1) = \sum_{j\in\mathcal{N}_i} w_{ij}A\widehat{\mb{x}}_j(t-1|t-1),
\end{eqnarray}
where $\widehat{\mb{x}}_i(t|t-1)$ represents the priori estimate of state~$\mb{x}$ at time $t$, using all the mesurements of node~$i$, and its neighbors $\mathcal{N}_i$ at time~$t-1$. The consensus matrix $W=[w_{ij}]$ satisfies the stochasticity conditions described in Section~\ref{sec_consensus}. As it is clear from Eq.~\eqref{eq_p}, only one step of consensus is performed between two steps of system dynamics $t-1$ and $t$ (two consecutive sampling times), which confirms the single time-scale setup as illustrated in Fig.~\ref{fig_scale}.

\item \textbf{Measurement update:} Agents/sensors share their measurements over the network $\mc{G}$ and update their priori estimates as follows:
\begin{eqnarray}\label{eq_m}
	\widehat{\mb{x}}_i(t|t) =\widehat{\mb{x}}_i(t|t-1) + K_i \sum_{j\in \mc{N}_i}C_j^\top \left(\mb{y}_j(t)-C_j\widehat{\mb{x}}_i(t|t-1)\right),
\end{eqnarray}
with $\mb{y}_j(t)$ as the measurement of node $j$ at time-step $t$ and $K_i$ as the \textit{local gain} matrix at node $i$.  This step is also called \textit{innovation-update} in some literature \cite{das2016consensus,kar6934985,mohammadi2014distributed,Sign,kar2013consensus}, just to mention a few.
\end{enumerate}

The dynamics of the distributed estimation error evolves as
\begin{align}\label{eq_err1}
\mb{e}(t) = (W\otimes A - KD_C(W\otimes A))\mb{e}(t-1) +
\mb{\zeta}(t),
\end{align}
where the error vector $\mb{e}(t)$ (at all nodes) is defined as,
\begin{align}\nonumber
\mb{e}(t) = \left( \begin{array}{c}
	\mb{e}_1(t)\\
	\vdots \\
	\mb{e}_N(t)
\end{array}\right),
\end{align}
${\zeta}(t)$ collects the noise terms as
\begin{align} \label{eq_zeta}
{\zeta}(t) = \nu(t-1)-\sum_{j\in \mathcal{N}_i}C_j^\top C_j \nu(t-1) - \sum_{j\in \mathcal{N}_i}C_j^\top \mu_i(t),
\end{align}	
$D_C$ is defined as
\begin{align} \label{eq_D_C}
D_C := \left(
\begin{array}{cccc}
	\sum_{j\in \mathcal{N}_1} C_j^\top C_j\\
	&\ddots\\
	& &\sum_{j\in \mathcal{N}_N} C_j^\top C_j\
\end{array}
\right)
\end{align}
and the block-diagonal gain matrix in the form
\begin{align}
K := \left(
\begin{array}{cccc}
	K_1\\
	&\ddots\\
	& &K_N\
\end{array}
\right).
\end{align}


\paragraph{Distributed Gain Matrix Requirements:} 
The block-diagonal design of the gain matrix $K$ is essential for ensuring the estimation setup remains truly distributed. However, this matrix often cannot be computed locally using standard procedures employed in traditional Kalman-type estimation. To address this constraint (the block-diagonal structure of $K$), researchers have developed specialized techniques based on iterative cone-complementarity optimization algorithms using Linear Matrix Inequality (LMI) approaches \cite{usman_cdc:11,rami:97,5717159}.

\paragraph{Stability and Distributed Observability:}
Based on Kalman filtering theory \cite{kalman:61}, equation \eqref{eq_err1} is steady-state Schur stabilizable \emph{if and only if} the pair $(W \otimes A, D_C)$ is observable. This property, known as distributed observability \cite{globalsip14}, can be analyzed using the graph-theoretic results presented in Section~\ref{sec_obsrv} and structural composite network design principles \cite{kronecker_TSIPN,cartesian_TSIPN}.

\paragraph{Varying Observability Requirements in Literature:} The assumption of distributed observability differs significantly across the existing literature:

\begin{itemize}
\item In semi-centralized estimation scenarios, when the system is observable, transmitting all measurement data makes the system globally observable to all processing nodes; for example see \cite{commault-recovery,sauter:09};

\item However, in single-time scale estimators, each agent has access to only local measurements at each step. This limited local information may not contain the necessary data to guarantee observability; for example see \cite{acc13,globalsip14}.
\end{itemize}

\paragraph{Network Design for Observability Recovery:}
The key challenge becomes designing the structure of the communication network $G$ according to the underlying fusion rules to recover distributed observability at every node \cite{boukhobza-recovery,spl17}. Several approaches have been proposed:

\begin{enumerate}
\item \textbf{Data-Sharing Approach:} A straightforward solution is to share all necessary measurements for observability at every step of system dynamics, as implemented in semi-centralized estimation approaches \cite{commault-recovery,sauter:09};

\item \textbf{Hybrid Communication Strategy:} A more communication-efficient approach involves transmitting both measurements $y_i(t)$ and predictions $\hat{x}_i(t|t)$ among nodes;

\item \textbf{Rank-Based Analysis:} The structural rank of the system matrix $A$ plays a crucial role in determining communication requirements. For structurally rank-deficient dynamical systems, more extensive communication and data-sharing are needed to satisfy distributed observability conditions \cite{icassp13}.
\end{enumerate}

\paragraph{Bounded Error and Local Observability:}
Related work in \cite{usman_cdc:10} demonstrates that a particular distributed estimator maintains bounded error if the two-norm of the system matrix is less than the Network Tracking Capacity (NTC) -- a quantity determined by the communication network and system measurement model. Along similar lines, many approaches, including \mbox{consensus + innovation} techniques \cite{das2016consensus,kar6934985,mohammadi2014distributed,Sign,kar2013consensus}, make the simplifying assumption that the underlying system is locally observable within the neighborhood of every sensor node $\mc{N}_i$.

\subsubsection*{Literature Review on Single Time-Scale Approaches}

The existing CPS literature on single time-scale distributed estimation and filtering encompasses diverse approaches addressing specific challenges inherent in distributed systems. Below, we categorize and review the key contributions in this domain:

\paragraph{Performance-Focused Innovations:}
Finite-time data fusion techniques \cite{SILM2020104707,usman_acc:11,10178290} enable multiple sensors to achieve accurate state estimation within a specified time frame, eliminating the asymptotic convergence limitations of traditional approaches. Similarly, resilient $H_\infty$ filtering methods \cite{8742899} maintain robust performance in the presence of disturbances and uncertainties, with recent extensions addressing attack mitigation over sensor networks \cite{SU2024111370,deghat2019detection}. These filtering approaches minimize worst-case estimation error, providing robustness against both internal uncertainties and external adversarial actions.

\paragraph{Resource-Efficient Communication Strategies:}
Event-triggered approaches \cite{9599480,li2021distributed,liu2020event} significantly reduce communication overhead by enabling nodes to transmit data only when predefined events occur—particularly valuable in scenarios with infrequent data changes or limited bandwidth. Complementing these, delay-tolerant methods \cite{abdelmawgoud2020distributed,jenabzadeh2020distributed,liu2017distributed,lcss22,9292963,scl25} are essential in wireless sensor networks with variable latency, employing techniques such as time-stamping, buffer management, and predictive algorithms to compensate for communication delays while maintaining accurate state estimation.

\paragraph{Resilience to Network Failures:}
Network reliability challenges are addressed by solutions designed to withstand link failures and unreliable communications \cite{yu2020distributed,4663899,kar2012distributed,ejc} or node failures \cite{spl17,doostmohammadian2025design}. These approaches typically involve filters that dynamically adapt to evolving network topologies. Adaptive consensus algorithms \cite{9633017,8113573} enable networks to reconfigure themselves in response to failures, ensuring continuous data fusion and state estimation despite partial system degradation.

\paragraph{Heterogeneity and Optimization Considerations:}
In practical CPS deployments, sensor nodes often have different capabilities (sensing ranges, processing power, communication bandwidth), requiring distributed filtering techniques that account for this heterogeneity to ensure reliable data fusion \cite{8003378}. Cost optimization strategies balance performance with resource expenditure by minimizing communication costs and energy consumption while achieving desired estimation accuracy. Techniques including sensor selection and optimal placement algorithms \cite{kruzick2017structurally,spl18,tnse19} ensure that the most informative sensors from an observability perspective are strategically utilized.

These diverse approaches collectively advance single time-scale distributed estimation, each addressing specific operational challenges while maintaining the fundamental single time-scale communication paradigm.

\subsubsection{Double Time-Scale Algorithms} \label{sec_double}
Double time-scale algorithms represent a fundamentally different approach to distributed estimation, characterized by their intensified communication pattern between system dynamics updates. In these scenarios, agents perform multiple iterations of consensus and communication between consecutive time steps of system dynamics, as illustrated in Fig.~\ref{fig_scale}. This communication-intensive phase is commonly referred to as the "consensus loop" in distributed filtering literature.

This approach presents a clear trade-off in distributed estimation design: while it imposes significantly higher communication load and network traffic on the multi-agent system, it offers two substantial advantages. First, it effectively relaxes the strict observability requirements that constrain single time-scale methods. Second, it demonstrates superior error performance, achieving more accurate state estimates compared to single time-scale approaches under equivalent system conditions. The key is that agents run multiple consensus rounds--more than the network diameter--in the interval between two consecutive system measurements. This multi-step consensus has opposing effects: it imposes higher communication/computation burden on agents, but simultaneously ensures that information originating from any agent reaches all other agents, effectively making the entire network's measurements globally observable. This global information sharing compensates for limited local observability. 

Below, we present a canonical implementation of a double time-scale distributed filter that exemplifies the core principles of this approach, as detailed in seminal works by Olfati-Saber and He \emph{et al.} \cite{olfati:cdc09,he2020secure}. This distributed estimator operates through two distinct phases that separate local prediction from network-wide consensus:	
\begin{enumerate} [1.]
\item \textbf{Local priori estimate:} This step includes no consensus iteration and is performed locally at every node $i$ as follows:
\begin{align} \label{eq_fil}
	\widehat{\mb{x}}_i(t)= A \widehat{\mb{x}}_i(t-1)+{C}_{i}^\top(\mb{y}_i(t)-C_{i}A\widehat{\mb{x}}_i(t-1)),
\end{align}
where  $\widehat{\mb{x}}_i(t)$ represents the priori estimated state of node $i$ and $\mb{y}_i(t)$ is its measurement at time-step $t$.

\item \textbf{Consensus loop update:}
$\mc{L}$ iterations of consensus update are then performed to average the estimate values as follows:

\begin{align} \label{eq_con}
	\widehat{\mb{x}}_{i,l}(t)=\widehat{\mb{x}}_{i,l-1}(t)-\epsilon\sum_{j \in \mc{N}_i}w_{ij}\left(\widehat{\mb{x}}_{i,l-1}(t)-\widehat{\mb{x}}_{j,l-1}(t)\right),
\end{align}
where $\widehat{\mb{x}}_{i,l}(t)$ represents the state estimate of node $i$ after $1\leq l\leq \mc{L}$ communication and consensus iterations at time-step $t$, while ${\epsilon}$ is a small positive constant that controls the rate of convergence. This consensus loop is over the neighbouring set ${\mc{N}_i}$ of node $i$, and the last consensus term  $\epsilon\sum_{j \in \mc{N}_i}w_{ij}\left(\widehat{\mb{x}}_{i,l-1}(t)-\widehat{\mb{x}}_{j,l-1}(t)\right)$ ensures that the estimate of node $i$ moves toward the average of its own and neighbours' local estimates after $\mc{L}$ iterations. For this, the consensus matrix needs to satisfy the stochastic condition in Section~\ref{sec_consensus}.
As it is clear from Eq.~\eqref{eq_con}, $\mc{L}$ steps of consensus are performed between two steps of system dynamics $t-1$ and $t$ (two consecutive sampling times) and then the algorithm moves to the next sampling time-step $t+1$. This represents the double time-scale setup as illustrated in Fig.~\ref{fig_scale}.
\end{enumerate}

\subsubsection*{Observability Advantages of Double Time-Scale Approaches}

A critical parameter in double time-scale algorithms is the number of consensus iterations $L$ performed between consecutive system dynamics updates. In the literature, it is typically assumed that $L \geq d_G$, where $d_G$ denotes the diameter of the sensor network $G$ (defined as the longest shortest path between any two nodes in the network). This requirement ensures complete information propagation across the network -- every node's data eventually reaches every other node during a single system time step.

This comprehensive information sharing fundamentally addresses the observability challenges that plague single time-scale approaches. By executing multiple consensus iterations, the double time-scale approach effectively transforms a partially observable system into a fully observable one from each node's perspective. The consensus loop essentially functions as an information diffusion mechanism, redistributing measurements throughout the network and ensuring that every agent has access to sufficient information to reconstruct the global state.

The observability benefits of this approach yield two significant advantages:

\begin{enumerate}
\item \textbf{Relaxed Network Connectivity Requirements:} Strong connectivity of the network $G$ becomes sufficient to ensure both observability and error stability. Unlike single time-scale approaches where specific topological structures must be carefully designed to guarantee observability \cite{usman_cdc:10,usman_cdc:11}, double time-scale methods are more tolerant of arbitrary network configurations.

\item \textbf{Enhanced Estimation Stability:} The $L$ iterations of communication and consensus make the information of every sensor accessible to every other sensor between consecutive system dynamics updates. This comprehensive data sharing ensures the system becomes observable to every sensor, eliminating the observability concerns that persist in single time-scale estimation.
\end{enumerate}

Table~\ref{tab_compare} provides a comparative analysis of single time-scale and double time-scale protocols, highlighting the fundamental differences in computation rate, network connectivity requirements, and communication overhead per sample. This comparison illustrates the explicit trade-off between communication efficiency and observability guarantees that system designers must consider when selecting an appropriate distributed estimation approach.

\begin{table} [hbpt!]
\centering
\caption{Comparison of single-time-scale and double-time-scale distributed filters in terms of computation complexity and network-connectivity $\times$ communication-rate per sample. }
\label{tab_compare}
\begin{tabular}{|c|c|c|c|}
	\hline
	Reference & time-scale & computation & links $\times$ rate    \\
	\hline
	\cite{sauter:09} & single & 1 &  $n(n-1)\times 1$   \\
	\hline
	\cite{das2016consensus,kar6934985,mohammadi2014distributed,Sign,kar2013consensus}
	& single & 1 &  $3n\times 1$ \\\hline
	\cite{isj_cyber,jstsp,acc13,globalsip14,icassp13}	 & single  & 1 &  $n \times 1$ \\ \hline
	\cite{olfati:cdc09,he2020secure,zou2020moving,dong2022consensus,farina-mhe,BATTISTELLI201875,21M1405083} &	double & $\mc{L}$ &  $n\times \mc{L}$ with $\mc{L} \geq d_G$  \\
	\hline
	\hline
\end{tabular}
\end{table}

\subsubsection*{Literature Survey on Double Time-Scale Techniques}

The literature on double time-scale estimation protocols encompasses diverse approaches addressing various challenges in distributed filtering. Below, we categorize and analyze key contributions across several technical domains:

\paragraph{Advanced Estimation Frameworks:}
Distributed Moving Horizon Estimation (MHE) \cite{zou2020moving,dong2022consensus,farina-mhe} represents a sophisticated optimization-based approach that estimates system states over a sliding time window. This technique allows nodes to leverage recent measurements and control inputs while considering system constraints, and optimizing state estimates within defined horizons. In distributed implementations, each node computes local estimates and engages in consensus exchanges with neighbors, iteratively refining results through collaborative optimization. This approach is particularly valuable for systems with complex dynamics or constraints that traditional filtering methods struggle to accommodate.

\paragraph{Communication-Efficient Adaptations:}
Despite the inherently communication-intensive nature of double time-scale methods, several approaches aim to reduce unnecessary data transmission. Event-triggered strategies \cite{8675461,BATTISTELLI201875,21M1405083} define specific thresholds that govern when nodes transmit updates, activating communication only when significant changes occur or estimation errors exceed predefined bounds. These techniques are particularly advantageous in bandwidth-constrained or energy-limited environments, offering substantial communication savings while preserving estimation performance. By intelligently managing the communication-accuracy trade-off, these methods extend the practical applicability of double time-scale approaches to resource-constrained settings.

\paragraph{Adaptability to Heterogeneous Networks:}
Real-world sensor networks often exhibit significant heterogeneity in terms of sensing capabilities, processing power, and communication bandwidth. Advanced filtering techniques \cite{ZHENG2024111839} explicitly exploit this diversity, employing adaptive algorithms that adjust filtering parameters based on individual sensor reliability and performance characteristics. These approaches dynamically weight sensor contributions according to their demonstrated accuracy, effectively leveraging the strengths of different nodes while minimizing the impact of less reliable measurements.

\paragraph{Robustness Against Network Imperfections:}
Network reliability presents significant challenges in practical distributed estimation. Several robust filtering approaches have been developed to address:

\begin{itemize}
\item \textbf{Link Failures and Unreliable Channels:} Distributed filters with inherent fault tolerance \cite{battilotti2023consensus,alonso2016adaptive,patterson2010convergence,8374821,9395240} maintain functionality despite communication link failures or packet losses. These approaches often incorporate redundancy mechanisms and adaptive consensus strategies that reconfigure information flow paths when network disruptions occur. Additionally, methods such as gossip algorithms \cite{Boyd-GossipInfTheory,Riccati-weakcons,dimakiskarmourarabbatscaglione-11} can be employed, where nodes iteratively share information to converge on a common estimate, enhancing robustness against network unreliability.

\item \textbf{Denial-of-Service Mitigation:} Specialized filtering algorithms \cite{battistelli2023stability,sun2020event,liu2021resilient} operate effectively under partial information conditions resulting from DoS attacks. These methods employ resilient filtering techniques and adaptive consensus mechanisms that prioritize reliable information sources while identifying and isolating compromised nodes.

\item \textbf{Adversarial Input Protection:} Robust filtering approaches \cite{forti2020joint,he2019secure,he2020secure,huang2023security} specifically address the challenge of adversarial inputs that might compromise estimation accuracy. These techniques implement statistical validation procedures and outlier rejection mechanisms to preserve estimation integrity despite malicious data injection.

\item \textbf{Communication Delay Handling:} Delay-tolerant consensus filters \cite{zou2019moving,mast2023unified} accommodate network latency through sophisticated time-stamping, buffering, and prediction mechanisms. These approaches ensure consistent estimation performance despite variable communication delays, making them particularly suitable for wireless and satellite-based sensing applications.
\end{itemize}

Other existing literature includes distributed consensus filtering for monitoring time-varying systems \cite{chen2019distributed}, often employing adaptive algorithms that can update their parameters based on observed data. Further, in many practical applications, noise statistics may not be fully known, complicating the estimation process. Distributed filtering techniques in such scenarios often rely on robust filtering methods that can operate under uncertainty. Techniques such as robust Kalman filters \cite{4547458} and consensus-based algorithms \cite{dong2021adaptive} that incorporate uncertainty quantification can be employed. 

Another concern in practical scenarios is that sensors may only be able to transmit quantized information due to communication constraints. Distributed filtering algorithms must be designed to operate practically under these conditions. Techniques such as quantization-aware filtering \cite{zhu2022adaptive,HE2020108842} incorporate quantization into the estimation process and maintain accuracy while adhering to the constraints of digital communication systems.
Further, cost-optimal algorithms are designed in \cite{10750405} to balance the trade-off between estimation accuracy and communication efficiency.

\subsubsection{Observationally Redundant Design} \label{sec_obsrv_red}
The observationally redundant design represents a sophisticated approach to enhancing reliability and fault tolerance in distributed estimation systems. This methodology systematically incorporates strategic sensor redundancy based on observational equivalence principles, creating inherently resilient monitoring networks capable of maintaining estimation performance despite sensor failures or attacks.

The fundamental concept behind observationally redundant design is the strategic deployment of sensors that observe structurally or functionally equivalent aspects of the system state. When multiple sensors can provide observationally equivalent information about critical system states, the network maintains complete observability even if individual sensors malfunction or are compromised. This architectural redundancy fundamentally differs from simple replication, as it is based on mathematical equivalence relationships that preserve the system's structural properties.

The notion of equivalence relation,~`$\sim$', in set theory and abstract algebra is defined as having three properties: reflexivity, symmetry, and transitivity~\cite{abstractalgebra}. Towards observational equivalence in state estimation, reflexivity implies that every state is equivalent to itself, i.e.~$x_i \sim x_i$; symmetry implies that if~$x_i \sim x_j$ then~$x_j \sim x_i$; and transitivity implies that if~$x_i \sim x_j$ and~$x_j \sim x_m$, then~$x_i \sim x_m$. With these notations, the observational equivalence of two state nodes and the associated measurements are defined as follows. Let~$C_i$ denote a row vector of size~$n$ with only non-zero at~$i$th entry denoting measurement of state $x_i$. Observational equivalence among two states,~$x_i\sim x_j$, is defined as
\begin{align}
\mbox{rank}~\mc{O}(A, C_i) = \mbox{rank}~\mc{O}(A, C_j) = \mbox{rank}~\mc{O}\left(A, \left(
\begin{array}{c}
	C_i\\
	C_j
\end{array}
\right)
\right).
\end{align}
It can be easily verified that the above definition follows three properties of transitivity, reflexivity, and symmetry.

\paragraph{Graph-Theoretic Classification for Observational Equivalence}

To establish a formal foundation for observational equivalency, we draw upon key structural system properties from graph theory. The rank deficiency of system matrix $A$ and the strong-connectivity characteristics of its associated system digraph $G_A$ give rise to specific structural observability properties that enable systematic sensor classification.

Following structured systems theory and generic analysis frameworks \cite{DION20031125,liu-pnas,icassp2016,isj_minimal}, we can develop a precise classification of sensors/agents based solely on the structural zero-nonzero pattern of the system matrix $A$ and its corresponding digraph representation $G_A$. This classification does not depend on specific parameter values, making it robust against system uncertainties and modeling inaccuracies.

Within this framework, we classify agents (and their associated observations) into three fundamental categories—Type-$\alpha$, Type-$\beta$, and Type-$\gamma$ -- based on their position within the system's structural components:

\begin{itemize}
\item In $G_A$, we define a strongly-connected-component (SCC) as a maximal subgraph in which a directed path exists between every pair of nodes. This concept captures regions of the system with complete internal information flow.

\item An SCC is designated as a parent component, denoted by $S^p_l$, if it has no outgoing links to other SCCs, representing terminal information aggregation points in the system.

\item We define a contraction $C_l$ as a component for which $|N_{G_A}(C_l)| < |C_l|$, where $N_{G_A}(C_l) = \{b | a \rightarrow b, a \in C_l\}$ represents the set of nodes that receive direct connections from nodes within $C_l$. Contractions correspond to regions where information flow experiences dimensional reduction.
\end{itemize}

Based on these precisely defined graph components, we establish the following agent classification scheme:

\begin{itemize}
\item \textbf{Type-$\alpha$}: an agent with observation of a state node in a contraction $\mc{C}_l$.
\item \textbf{Type-$\beta$}: an agent with observation of a state node in a parent SCC $\mc{S}^p_l$.
\item \textbf{Type-$\gamma$}: any agent which is neither $\alpha$ nor $\beta$.
\end{itemize}

\paragraph{Algebraic Interpretation and Redundant Observability Design}

\begin{figure} 
\centering
\includegraphics[width=4in]{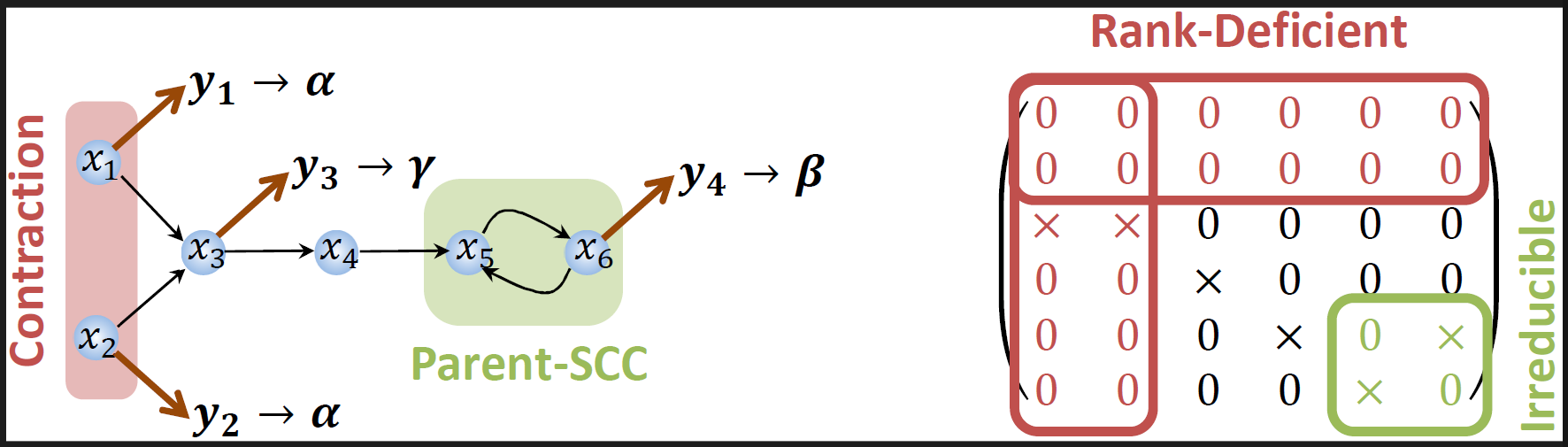}
\caption{The left figure shows a simple system digraph representation and its agent classification based on the observationally equivalent nodes in contractions and parent SCCs. The right figure shows the system matrix of the same digraph and its irreducible and rank-deficient parts respectively associated with the parent SCC and contraction on the digraph.
} \label{fig_class}
\end{figure}

From an algebraic graph-theoretic perspective, the system's structural properties have precise mathematical interpretations that inform redundant design strategies. The contractions in $G_A$ directly correspond to rank-deficiency regions in the system matrix $A$, while the SCC decomposition reflects the irreducibility properties of $A$. Fig.~\ref{fig_class} provides a visual illustration of these correspondences, demonstrating how graph structures map to algebraic properties.

A fundamental result in structural observability theory establishes that for a given system digraph $G_A$, outputs from a strategically minimal set of nodes are sufficient to ensure full system observability:

\begin{itemize}
\item One state node from every parent SCC $S^p_l$;
\item One state node from every contraction $C_l$.
\end{itemize}

This minimal output set guarantees $(A,C)$-observability. More significantly, states within the same contraction $C_l$ or within the same parent SCC $S^p_l$ exhibit observational equivalence -- they provide structurally identical information about the system's behavior.

\paragraph{Designing $q$-Redundant Observable Systems}

Leveraging this observational equivalence principle, we can systematically design sensor networks with guaranteed fault tolerance. Specifically, including $q+1$ different state outputs from each contraction and each parent SCC creates a $q$-redundant observable system. By distributing these outputs across $q+1$ distinct sensors/agents, we establish observational redundancy that preserves system observability even under sensor failures.

This redundancy ensures that after the removal or failure of any $q$ sensors (or outputs), the remaining sensor set still contains at least one output from every $S^p_l$ and one from every $C_l$, thereby maintaining the necessary conditions for $(A,C)$-observability. The property is particularly valuable in critical monitoring applications where sensor failures could otherwise compromise system visibility.

\paragraph{Network Connectivity for Distributed Implementation}

For practical implementation in distributed estimation, the communication network topology $G$ must support this observational redundancy. Specifically, to achieve \mbox{$q$-redundant} $(W \otimes A, D_C)$-observability in a distributed observer design, the multi-agent network $G$ should be designed with:

\begin{itemize}
\item $q$-node-connectivity: The network remains connected after the removal of any $q-1$ nodes;
\item $q$-link-connectivity: The network remains connected after the removal of any $q-1$ links.
\end{itemize}

This connectivity specification ensures that the network topology supports robust information flow even under multiple node or link failures, a property commonly referred to as ``survivable network design" \cite{jabal2021approximation}. 

Numerous computationally efficient algorithms exist for such network augmentation and topology design, including methods for incremental connectivity enhancement \cite{augment_book}, minimal-cost connectivity augmentation \cite{frederickson1981approximation}, structured augmentation approaches \cite{vegh2010connectivity}, and distributed construction techniques \cite{wu2008construction}.

\subsection{Diffusion-based Techniques}
\label{subsec_dlms}
Diffusion-based techniques represent a fundamentally different paradigm for distributed estimation compared to consensus-based approaches. These methods derive from adaptive filtering theory and are particularly well-suited for parameter estimation in dynamic environments. In this section, we explore the theoretical foundations, implementation variants, and performance characteristics of diffusion strategies for distributed estimation over networks.

\subsubsection*{From Centralized to Distributed Estimation}

Classical Least-Mean-Square (LMS) formulations traditionally operate under centralized processing assumptions, where either
\begin{itemize}
	\item All sensor observations are collected and processed at a central node, or
	\item Local LMS estimators operate independently with their outputs subsequently fused at a central location.
\end{itemize}

This centralized paradigm, while conceptually straightforward, faces significant scalability and reliability limitations in modern cyber-physical systems. Contemporary applications increasingly demand truly distributed processing architectures, where
\begin{itemize}
	\item Data is inherently distributed across a network of agents or sensors;
	\item Each node has access only to its local measurements and information from immediate neighbors;
	\item No central fusion center exists or is desirable due to robustness concerns; and
	\item Global parameter estimation must emerge from purely local computations and limited inter-node communication.
\end{itemize}

\subsubsection*{Distributed Optimization Framework}

Following the seminal diffusion adaptation framework developed in \cite{chen2012diffusion, sayed2014diffusion, sayed2014adaptation}, we consider a connected network of $N$ agents indexed by $k \in \{1, 2, \ldots, N\}$. Each agent $k$ maintains the following:
\begin{itemize}
	\item A twice-differentiable local cost function $J_k(\bm\omega) \in \mathbb{R}$, often referred to as its utility function;
	\item Access to local measurements that inform this cost function; and
	\item Communication links with neighboring agents that enable collaborative estimation.
\end{itemize}

\begin{figure} 
	\centering
	\includegraphics[width=.4\textwidth]{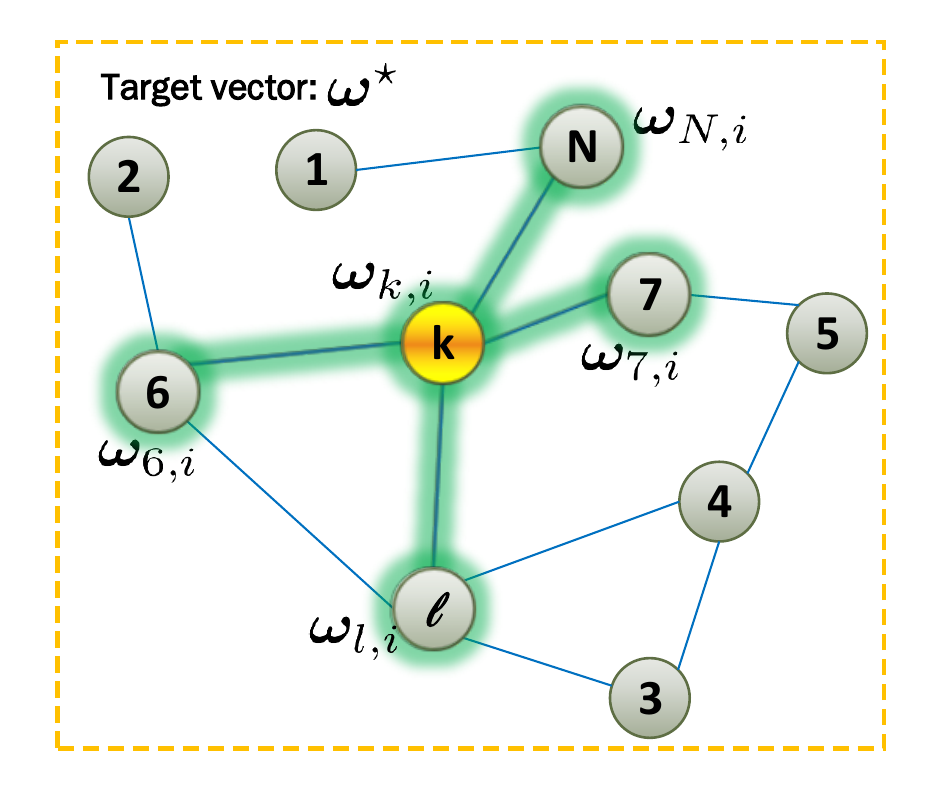}
	\caption{A  network of $N=10$ estimators is shown, with each numbered node representing an individual estimator. Blue lines indicate communication links between connected nodes. The highlighted node $k$ is connected to its neighboring nodes $6$, $7$, $\ell$, and $N$, forming the neighborhood $\mathcal{N}_k = \{k, 6, 7, \ell, N\}$. The shaded region emphasizes these neighbouring nodes.}
	\label{fig_network}
\end{figure}

Fig.~\ref{fig_network} illustrates a typical network topology for such systems, with bi-directional communication links between neighboring agents represented by single lines. This network structure fundamentally shapes how information diffuses through the system during the estimation process.

In this distributed setting, the collective goal is to estimate a global parameter vector that optimizes the aggregate cost function across all agents given by
\begin{equation}
	J^{\text{glob}}(\bm{\omega}) \triangleq \sum_{k=1}^{N} J_k(\bm{\omega}),
	\label{eq:global_cost}
\end{equation}
where the goal is to identify its unique minimizer, denoted by $ \bm{\omega}^\star $.

To enable fully distributed estimation, we introduce a set of non-negative combination weights $\{c_{k\ell} \geq 0\}$ that govern information exchange among neighboring nodes. These weights must satisfy
\begin{equation} 	\label{eq:combination_weights}
	\sum_{\ell=1}^{N} c_{k\ell} = 1, \quad c_{k\ell} = 0 \text{ if } \ell \notin \mathcal{N}_k,
\end{equation}
for each node $k = 1,2,\ldots,N$, where $\mathcal{N}_k$ denotes the neighborhood of node $k$ (including $k$ itself). This constraint ensures that the resulting combination matrix $C$ is right-stochastic, a property essential for the convergence of diffusion strategies.

Using these weights, we define for each node $\ell$ a localized cost function that aggregates weighted costs from its neighborhood, i.e., 
\begin{equation} 	\label{eq:combination_weights}
	J_{\ell}^{\text{loc}}(\bm\omega) \triangleq \sum_{k \in \mathcal{N}_{\ell}} c_{k\ell}J_k(\bm\omega).
\end{equation}

This formulation allows us to re-express the global cost function in terms of these localized costs as follows:
\begin{align} \nonumber
	J_{\text{glob}}(\bm\omega) &= \sum_{k=1}^{N} J_k(\bm\omega) = \sum_{\ell=1}^{N}\sum_{k=1}^{N} c_{k\ell}J_k(\bm\omega)\\
	&= \sum_{\ell=1}^{N} J_{\ell}^{\text{loc}}(\bm\omega).
	\label{eq:global_cost_rewritten}
\end{align}

\paragraph{Gradient-Based Distributed Optimization}

For practical implementation, we can further refine this formulation to explicitly incorporate the global minimizer $\bm\omega^*$ into a modified cost function:
\begin{equation} \label{eq:global_cost_modified2}
	J_{k}^{\text{glob}'}(\bm\omega) = \sum_{\ell \in \mathcal{N}_k} c_{\ell k}J_{\ell}(\bm\omega) + \sum_{\ell \in \mathcal{N}_k \setminus \{k\}} b_{\ell k}\|\bm\omega - \bm\omega^*\|^2.
\end{equation}

Although this expression contains the unknown variable $\bm\omega^*$, all other terms depend solely on information available to node $k$ and its neighborhood, making it amenable to distributed implementation.

Each node $k$ can then apply a steepest-descent iteration to minimize its local approximation of the global cost. Let $\bm\omega_{k,i}$ denote the estimate for $\bm\omega^*$ at time $i$ computed by node $k$. Starting from an initial condition $\omega_{k,-1}$, the update proceeds iteratively:

\begin{align}
	\bm\omega_{k,i} =& \bm\omega_{k,i-1} - \mu_k \nabla_{\omega}J_{k}^{\text{glob}'}(\bm\omega_{k,i-1}), \quad i \geq 0,	\nonumber\\
	=\;& \displaystyle \bm\omega_{k,i-1}
		- \mu_k \sum_{l \in \mathcal{N}_k} c_{l,k} \nabla_\omega J_l(\bm\omega_{k,i-1})
		-
		\displaystyle\mu_k \sum_{l \in \mathcal{N}_k\backslash \{k\}}
		2b_{l,k}(\bm\omega_{k,i-1}-\bm\omega^*),	
		\label{eq:diff_adapt_gd}			
\end{align}	
where $\mu_k$ is a small positive step-size parameter, and $\nabla_{\omega}J(\bm\omega)$ denotes the gradient vector of $J(\bm\omega)$ with respect to $\bm\omega$.

	Eq.~\eqref{eq:diff_adapt_gd} updates the estimate 
	$\bm\omega_{k,i-1}$ by adding two distinct correction terms to obtain $\bm\omega_{k,i}$.
	These corrections can be applied sequentially by decomposing the update into the following two steps:
	\begin{align}
		\label{eq:diff_adapt_int}
		\bm\psi_{k,i}	&=	 \displaystyle \bm\omega_{k,i-1} - \mu_k \sum_{l \in \mathcal{N}_k} c_{l,k} \nabla_\omega J_l(\bm\omega_{k,i-1}),	\\
		\label{eq:diff_adapt_cint}
		\bm\omega_{k,i}	&=	 \displaystyle \bm\psi_{k,i}
		- \mu_k \sum_{l \in \mathcal{N}_k\backslash\{k\}} 2b_{l,k} (\bm\omega_{k,i-1}-\bm\omega^*).
	\end{align}
	Step~\eqref{eq:diff_adapt_int} performs a \emph{local adaptation} by updating $\bm\omega_{k,i-1}$ to an intermediate estimate $\bm\psi_{k,i}$ using a weighted aggregation of gradient vectors associated with the cost functions of the neighboring nodes.
	Step~\eqref{eq:diff_adapt_cint} subsequently applies a \emph{coupling correction} that enforces similarity among neighboring estimates through a quadratic regularization term centered at the global minimizer $\bm\omega^*$.
	
	The update in~\eqref{eq:diff_adapt_cint} is not directly implementable since the optimal parameter vector $\bm\omega^*$ is unknown.
	To obtain a realizable recursion, as explained in \cite{chen2012diffusion}, the following substitutions are introduced:
	
	\begin{enumerate}[i)]
		\item
		The unknown vector $\bm\omega^*$ is replaced by the neighboring intermediate estimates
		$\{\bm\psi_{l,i}\}_{l \in \mathcal{N}_k}$.
		Since each intermediate estimate $\bm\psi_{l,i}$ results from a local gradient adaptation step, it constitutes a locally available approximation of $\bm\omega^*$.
		This substitution enables the diffusion of information across the network by allowing each node to exploit the most recent estimates produced by its neighbors.
		
		\item
		The previous estimate $\bm\omega_{k,i-1}$ is replaced by the intermediate estimate $\bm\psi_{k,i}$.
		Because $\bm\psi_{k,i}$ already incorporates the current gradient information, it provides a more accurate and up-to-date approximation of $\bm\omega^*$ than $\bm\omega_{k,i-1}$.
		This replacement reduces higher-order approximation errors and improves the stability and convergence performance of the resulting algorithm.
	\end{enumerate}
	Applying the substitutions described in items (i) and (ii) to \eqref{eq:diff_adapt_cint} yields the implementable diffusion update:
	\begin{align}
		\bm\omega_{k,i}	&=	 \displaystyle \bm\psi_{k,i}
		- \mu_k \sum_{l \in \mathcal{N}_k\backslash\{k\}} 2b_{l,k} (\bm\psi_{k,i}-\bm\psi_{l,i}).
	\end{align}
	
	The combination coefficients can be considered as
	\begin{align}
		\label{eq:comb_coeff}
		a_{l,k} \triangleq 2 \mu_k b_{l,k}	\quad (l \! \neq \! k),
		\quad
		a_{k,k} \triangleq 1 \!-\!  \mu_k \!\! \sum_{l \in \mathcal{N}_k\backslash\{k\}} 2b_{l,k}.
	\end{align}
	By construction, the coefficients $\{a_{l,k}\}$ are nonnegative for $l \neq k$. 
	Furthermore, for sufficiently small step-sizes $\mu_k$, the self-weight $a_{k,k}$ is also nonnegative.
	The resulting set of coefficients $\{a_{l,k}\}$ satisfies the following properties:
	\begin{align}
		\label{Equ:AdaptiveDiffusion:Condition_a}
		\displaystyle
		\sum_{l=1}^N a_{l,k} = 1, \quad a_{l,k} = 0~\mathrm{if}~l \notin \mathcal{N}_{k},	
	\end{align}
	which implies that the matrix $A = [a_{l,k}]$ is left-stochastic and conforms to the network topology.

This gradient-based distributed optimization procedure forms the foundation for the specific diffusion adaptation strategies we examine in subsequent sections, including \emph{adapt-then-combine} (ATC)
and \emph{combine-then-adapt} (CTA) variants
that offer different performance characteristics in practical implementations.

\subsubsection{Diffusion LMS}
Diffusion Least Mean Squares (DLMS) algorithms represent a practical implementation of the distributed optimization framework introduced previously. These methods enable networks of agents to collaboratively estimate a global parameter vector through a combination of local adaptation and strategic information exchange with neighboring nodes. By extending the classical LMS algorithm to decentralized network settings, diffusion strategies achieve robust estimation performance while maintaining the computational simplicity that makes LMS attractive for real-time applications.

\paragraph{Algorithmic Variants and Information Flow Patterns}

The defining characteristic of diffusion strategies is the introduction of a specific combination step that facilitates structured information flow across the network. This combination operation allows estimates to diffuse throughout the network, enabling all nodes to benefit from measurements collected across the entire system. Two principal variants of diffusion LMS have emerged in the literature, each with distinct information processing sequences:

\begin{itemize}
	\item \textbf{Adapt-then-combine (ATC):} In this variant, each node first performs a local adaptation step using its own measurement data to update its intermediate estimate. Following this adaptation, the node combines this locally-improved estimate with those received from neighboring nodes to produce its final estimate for the current iteration. This sequence prioritizes the integration of fresh local information before network-wide fusion:
	\begin{align}
			\boxed{
				\label{eq:ATC_box}
				\begin{array}{l}
					\bm\psi_{k,i}	=	 \displaystyle
					\bm\omega_{k,i-1} - \mu_k \sum_{l \in \mathcal{N}_k} c_{l,k}
					\nabla_\omega J_l(\bm\omega_{k,i-1})	\\
					\bm\omega_{k,i}	=	\displaystyle \sum_{l \in \mathcal{N}_k} a_{l,k} \bm\psi_{l,i}
				\end{array}
			}.
	\end{align}
	\item \textbf{Combine-then-adapt (CTA):} The CTA approach reverses this sequence. Each node first aggregates prior estimates from its neighborhood (including its own previous estimate) through a combination step. This fused intermediate result is then refined using the node's local observations in an adaptation step. This sequence emphasizes the incorporation of network-wide information before local refinement:
	\begin{align}
			\boxed{
				\label{eq:CTA_box}
				\begin{array}{l}
					\bm\psi_{k,i\!-\!1}	=	\displaystyle \sum_{l \in \mathcal{N}_k} a_{l,k} \bm\omega_{l,i-1}		 \\
					\bm\omega_{k,i}	=	 \displaystyle\bm\psi_{k,i-1}
					- \mu_k \sum_{l \in \mathcal{N}_k} c_{l,k} \nabla_\omega J_l(\bm\psi_{k,i-1})
				\end{array}
			}.
	\end{align}
\end{itemize}

These two information flow patterns lead to algorithms with different convergence characteristics and robustness properties, providing system designers with options that can be selected based on specific application requirements.

	It is worth noting that both the ATC and the CTA diffusion strategies yield \emph{unbiased} estimates of the optimal parameter vector under standard assumptions.
	However, in the mean-square-error (MSE) sense, the ATC strategy consistently outperforms CTA.
	This performance advantage arises because, in ATC, local error damping through adaptation precedes spatial mixing, which results in a smaller spectral radius of the error recursion and reduced steady-state noise amplification.
	In contrast, CTA performs spatial averaging prior to adaptation, which leads to	less effective suppression of gradient noise.
	
	Despite this performance gap, CTA remains of practical interest since it trades	optimal MSE performance for architectural simplicity, reduced communication	requirements, improved privacy characteristics, and closer compatibility with consensus-based distributed optimization frameworks.
	Formal performance comparisons and rigorous proofs of these claims can be found	in Chapters~9 and~11 of~\cite{sayed2014adaptation}.

To illustrate these differences, assume that the data at node~$l$ satisfy the linear regression model:
	\begin{align}
		\label{eq:reg_model}
		d_l(i) =   \bm{u}_{l,i}^\top\bm\omega^* + \bm{z}_l(i),
	\end{align}
	where the regression vectors $\{\bm{u}_{l,i}\}$ are zero-mean and temporally independent with covariance matrix	$R_{u,l} = \mathbb{E}\{\bm{u}_{l,i} \bm{u}_{l,i}^\top\}$ and cross-correlation vector $\bm{r}_{du,l} = \mathbb{E}\{d_l(i)\bm{u}_{l,i}\}$.
	The noise sequence $\{\bm{z}_l(i)\}$ is assumed to be zero-mean, white, with variance $\sigma_{z,l}^2$, and independent of the regressors $\{\bm{u}_{l,i}\}$ for all nodes $l$ and all time indices.
	
	Consider a network in which the loss function is given by the quadratic form:
	\begin{align}
		Q_l(\bm\omega,\{\bm{u}_{l,i},d_l(i)\})=|d_l(i)-\bm{u}_{l,i}^\top\bm\omega|^2,
		\label{eq:loss_f}
	\end{align}
	where $\{d_l(i)\}$ are scalar measurements and $\{\bm{u}_{l,i}\}$ are $M \times 1$ regression vectors.
	The associated cost function is, therefore,
	\begin{align}
		\label{eq:cost_f}
		J_l(\bm\omega)  &=  \mathbb{E} \{|d_l(i)-\bm{u}_{l,i}^\top\bm\omega|^2\}.
	\end{align}

\paragraph{ATC Algorithm Formulation}

Let $\alpha_{\ell k}$ denote the combination weights that govern how node $k$ incorporates estimates from its neighbors (including itself). The ATC diffusion strategy follows the recursion for $i \geq 0$:
\begin{align}
	\bm{\psi}_{k,i} &= \bm{\omega}_{k,i-1} + \mu_k \sum_{\ell \in \mathcal{N}_k} c_{\ell k} (\bm{r}_{du,\ell} - R_{u,\ell} \bm{\omega}_{k,i-1}),
	\label{eq:atc_psi_update} \\
	\bm{\omega}_{k,i} &= \sum_{\ell \in \mathcal{N}_k} \alpha_{\ell k} \bm{\psi}_{\ell,i},
	\label{eq:atc_w_update}
\end{align}
where $\{c_{\ell k}, \alpha_{\ell k}\}$ are non-negative coefficients satisfying the conditions:
\begin{equation}
	C \mathbf{1} = \mathbf{1}, \quad A^\top \mathbf{1} = \mathbf{1},
	\label{eq:CA_conditions}
\end{equation}
with $\mathbf{1}$ denoting the all-ones vector.
At each iteration $ i $, the ATC strategy \eqref{eq:atc_psi_update}-\eqref{eq:atc_w_update} involves two key steps:
\begin{itemize}
	\item \textbf{Information Exchange:} Each node $ k $ receives statistical moments $\{R_{u,\ell}, \bm{r}_{du,\ell}\}$ from its neighbours, aggregates them, and updates its estimate $\bm{\omega}_{k,i-1}$ to an intermediate value $\bm{\psi}_{k,i}$. This step is performed concurrently by all nodes.
	\item \textbf{Combination:} Each node $ k $ combines the intermediate estimates $\{\bm{\psi}_{\ell,i}\}$ received from its neighbors to obtain the updated estimate $\bm{\omega}_{k,i}$.
\end{itemize}
In the special case where the combination matrix $ C = I $, no information exchange occurs in the adaptation step. The update simplifies to:
\begin{align}
	\bm{\psi}_{k,i} &= \bm{\omega}_{k,i-1} + \mu_k (\bm{r}_{du,k} - R_{u,k} \bm{\omega}_{k,i-1}), \label{eq:atc_no_info_psi} \\
	\bm{\omega}_{k,i} &= \sum_{\ell \in \mathcal{N}_k} \alpha_{\ell k} \bm{\psi}_{\ell,i}, \label{eq:atc_no_info_w}
\end{align}
relying solely on local statistics.
Similarly, by adding the second correction term first, we arrive at the CTA strategy, as summarized in Table \ref{tab:atc_cta_four_versions}.
\begin{table}[h]
	\centering
	\caption{Steps of ATC and CTA Diffusion Strategies}
	\renewcommand{\arraystretch}{1.5}
	\begin{tabular}{|c|l|}
		\hline
		\textbf{Strategy} & \textbf{Steps} \\
		\hline
		\textbf{ATC} &
		\begin{tabular}[t]{@{}l@{}}
			\quad $\bm{\psi}_{k,i} = \bm{\omega}_{k,i-1} + \mu_k \sum\limits_{\ell \in \mathcal{N}_k} c_{\ell k} (\bm{r}_{du,\ell} - R_{u,\ell} \bm{\omega}_{k,i-1})$ \\[6pt]
			\quad $\bm{\omega}_{k,i} = \sum\limits_{\ell \in \mathcal{N}_k} \alpha_{\ell k} \bm{\psi}_{\ell,i}$
		\end{tabular}
		\\
		\hline
		\textbf{General ATC} &
		\begin{tabular}[t]{@{}l@{}}
			\quad $\bm{\psi}_{k,i} = \bm{\omega}_{k,i-1} - \mu_k \sum\limits_{\ell \in \mathcal{N}_k} c_{\ell k} \nabla_\omega J_\ell(\bm{\omega}_{k,i-1})$ \\[6pt]
			\quad $\bm{\omega}_{k,i} = \sum\limits_{\ell \in \mathcal{N}_k} \alpha_{\ell k} \bm{\psi}_{\ell,i}$
		\end{tabular}
		\\
		\hline
		\textbf{CTA} &
		\begin{tabular}[t]{@{}l@{}}
			\quad $\bm{\psi}_{k,i-1} = \sum\limits_{\ell \in \mathcal{N}_k} \alpha_{\ell k} \bm{\omega}_{\ell,i-1}$ \\[6pt]
			\quad $\bm{\omega}_{k,i} = \bm{\psi}_{k,i-1} + \mu_k \sum\limits_{\ell \in \mathcal{N}_k} c_{\ell k} (\bm{r}_{du,\ell} - R_{u,\ell} \bm{\psi}_{k,i-1})$
		\end{tabular}
		\\
		\hline
		\textbf{General CTA} &
		\begin{tabular}[t]{@{}l@{}}
			\quad $\bm{\psi}_{k,i-1} = \sum\limits_{\ell \in \mathcal{N}_k} \alpha_{\ell k} \bm{\omega}_{\ell,i-1}$ \\[6pt]
			\quad $\bm{\omega}_{k,i} = \bm{\psi}_{k,i-1} - \mu_k \sum\limits_{\ell \in \mathcal{N}_k} c_{\ell k} \nabla_\omega J_\ell(\bm{\psi}_{k,i-1})$
		\end{tabular}
		\\
		\hline
	\end{tabular}
	\label{tab:atc_cta_four_versions}
\end{table}

\subsubsection{Adaptive Diffusion Strategies}
While the diffusion strategies discussed previously provide a solid theoretical foundation, their practical implementation faces a significant challenge: the need for statistical moments $\{R_{u,k}, \bm r_{du,k}\}$ to evaluate gradient vectors. In real-world applications, these moments are rarely available a priori and must be estimated from available measurements. This section examines how stochastic approximation techniques transform theoretical diffusion algorithms into practical adaptive implementations suitable for deployment in uncertain environments.

\paragraph{From Exact to Stochastic Gradient Approximations}

The distributed ATC and CTA steepest-descent strategies summarized in Table~\ref{tab:atc_cta_four_versions} represent idealized implementations that assume perfect knowledge of statistical moments. To develop practical algorithms that operate with real-time measurements, by considering the quadratic loss function in Eq.~\eqref{eq:loss_f}, we replace exact gradients with instantaneous approximations based on stochastic observations available at each node.

Table~\ref{tab:adaptive_atc_cta} presents the resulting adaptive diffusion strategies, where $\nabla_{\omega}\hat{J}_{\ell}(\cdot)$ denotes an instantaneous gradient estimate constructed from local stochastic observations. This substitution transforms the deterministic optimization procedure into a stochastic approximation algorithm that converges to the optimal solution through repeated measurements and iterative refinement.

These adaptive implementations typically initialize with $\omega_{\ell,-1} = 0$ for all nodes $\ell$, though other suitable initialization values may be used depending on available prior information about the parameter being estimated. The convergence behavior and steady-state performance of these algorithms depend on both the network topology and the chosen step-size parameters, with smaller step sizes generally providing better steady-state accuracy at the cost of slower convergence.

\paragraph{Simplified Implementation Variants}

In many practical scenarios, it may be beneficial to reduce communication overhead by limiting information exchange during certain phases of the algorithm. In particular, if we eliminate information exchange during the adaptation step (implementing a ``combination only" approach), the adaptive ATC and CTA strategies reduce to simplified variants that maintain core functionality while requiring less inter-node communication:
\begin{itemize}
	\item[(a)] Adaptive ATC without Information Exchange, see Fig. \ref{fig_dlms_atc}:
	\begin{align}
		\text{Adaptation: }&\bm{\psi}_{k,i} = \bm{\omega}_{k,i-1} + \mu_k \bm{u}_{k,i} \left[d_k(i) - \bm{u}_{k,i}^\top \bm{\omega}_{k,i-1}\right], \notag \\
		\text{Combination: }&\bm{\omega}_{k,i} = \sum_{\ell \in \mathcal{N}_k} \alpha_{\ell k} \bm{\psi}_{\ell,i}.
	\end{align}
	\begin{figure} 
		\centering
		\includegraphics[width=.8\textwidth]{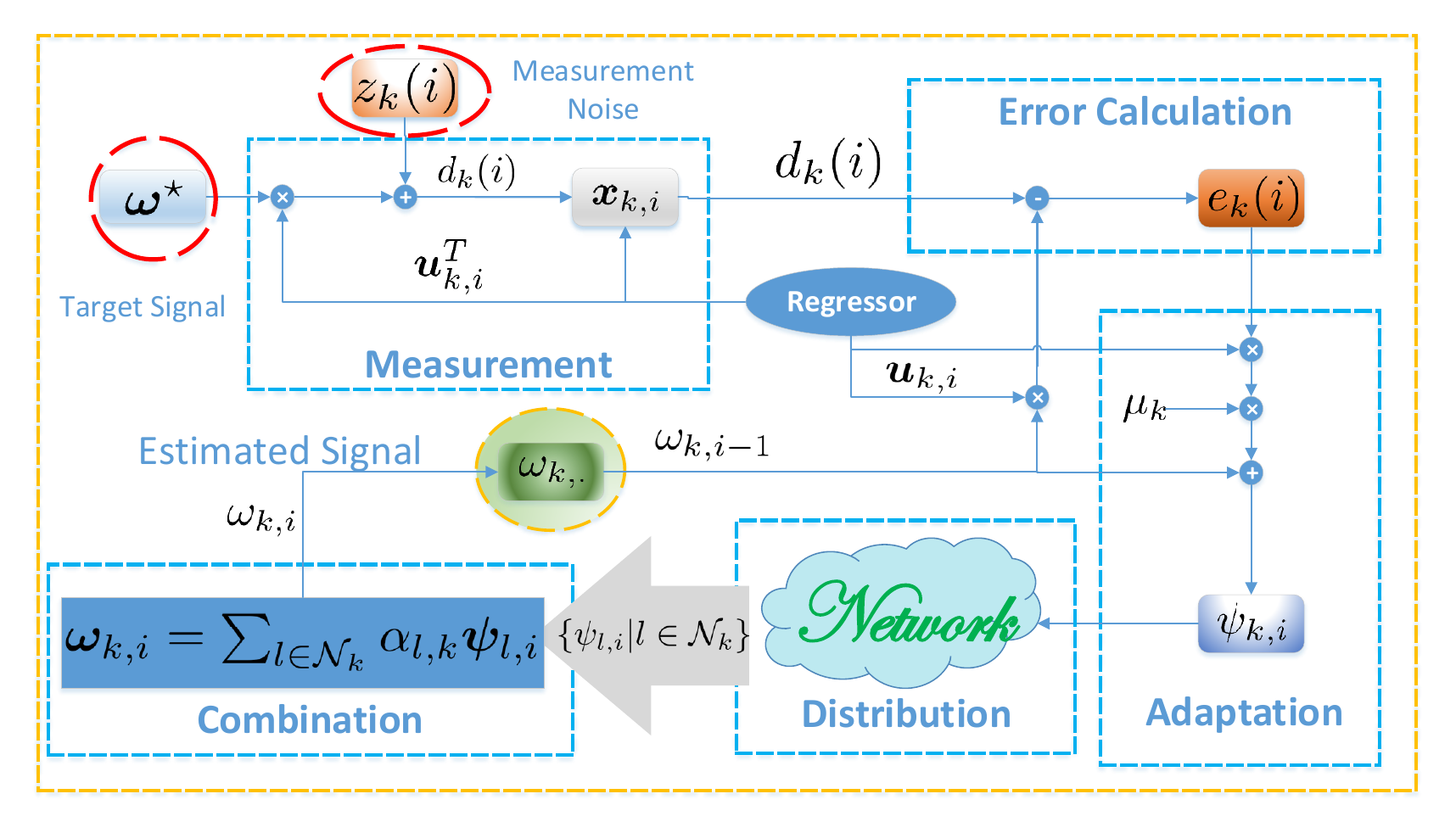}
		\caption{ATC strategy in diffusion LMS: Each node adapts its local estimate and then combines information from its neighbours.}
		\label{fig_dlms_atc}
	\end{figure}
	\item[(b)] Adaptive CTA without Information Exchange, see Fig. \ref{fig_dlms_cta}:
	\begin{align}
		\text{Combination: }&\bm{\psi}_{k,i-1} = \sum_{\ell \in \mathcal{N}_k} \alpha_{\ell k} \bm{\omega}_{\ell,i-1}, \notag \\
		\text{Adaptation: }&\bm{\omega}_{k,i} = \bm{\psi}_{k,i-1} + \mu_k \bm{u}_{k,i} \left[d_k(i) - \bm{u}_{k,i}^\top \bm{\psi}_{k,i-1}\right].
	\end{align}
	\begin{figure} 
		\centering
		\includegraphics[width=.8\textwidth]{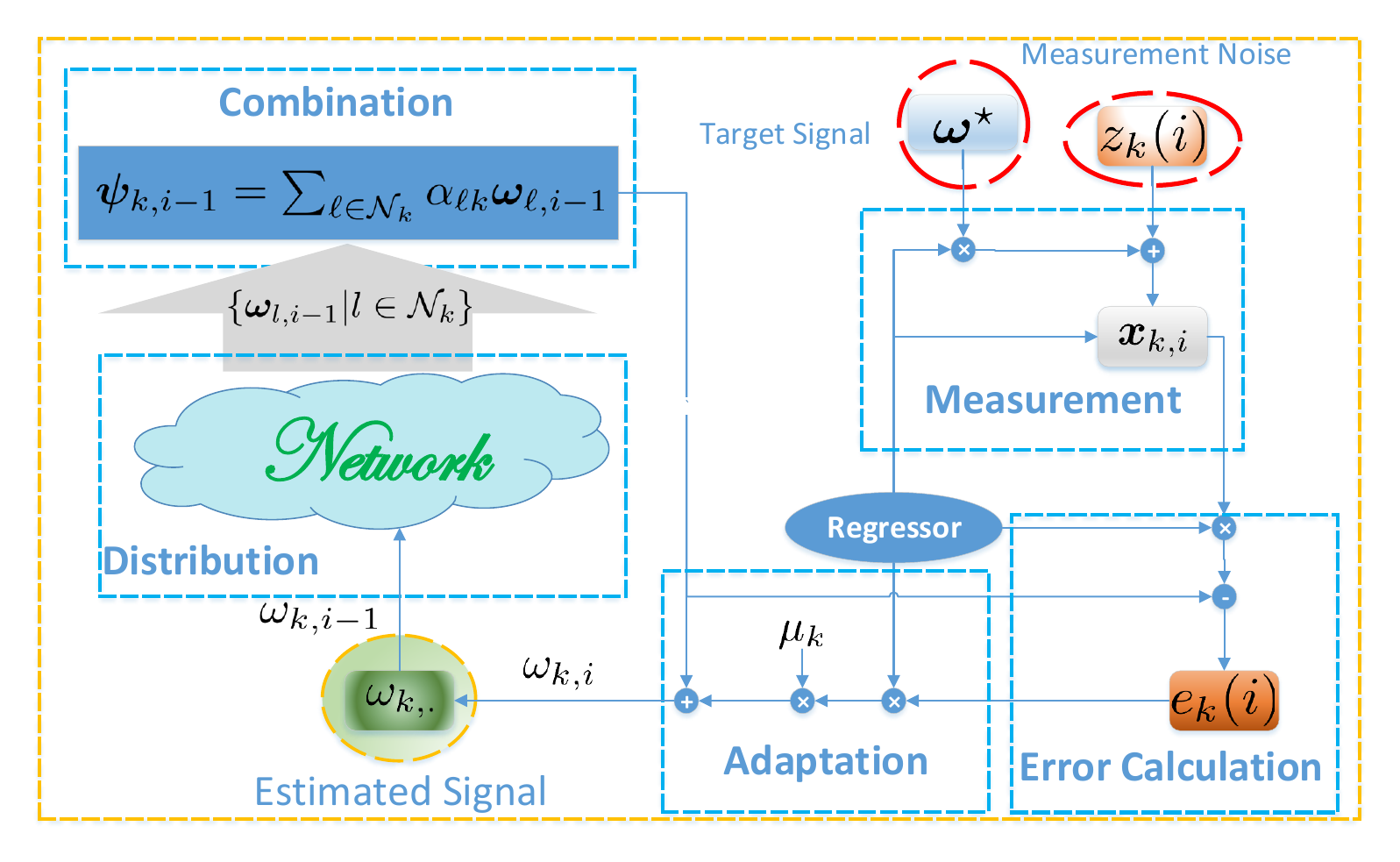}
		\caption{CTA strategy in diffusion LMS: Each node shares information with its neighbours for a subsequent adaptation step.}
		\label{fig_dlms_cta}
	\end{figure}
\end{itemize}
\begin{table}[h]
	\centering
	\caption{Steps of Adaptive ATC and CTA Diffusion Strategies}
	\renewcommand{\arraystretch}{1.5}
	\begin{tabular}{|c|l|}
		\hline
		\textbf{Strategy} & \textbf{Steps} \\
		\hline
		\textbf{Adaptive ATC} &
		\begin{tabular}[t]{@{}l@{}}
			\quad $\bm{\psi}_{k,i} = \bm{\omega}_{k,i-1} + \mu_k \sum\limits_{\ell \in \mathcal{N}_k} c_{\ell k} \bm{u}_{\ell,i} \left[ d_\ell(i) - \bm{u}_{\ell,i} \bm{\omega}_{k,i-1} \right]$ \\[6pt]
			\quad $\bm{\omega}_{k,i} = \sum\limits_{\ell \in \mathcal{N}_k} \alpha_{\ell k} \bm{\psi}_{\ell,i}$
		\end{tabular}
		\\
		\hline
		\begin{tabular}[t]{c@{}l@{}}
			\textbf{General} \\
			\textbf{Adaptive ATC}
		\end{tabular} &
		\begin{tabular}[t]{@{}l@{}}
			\quad $\bm{\psi}_{k,i} = \bm{\omega}_{k,i-1} - \mu_k \sum\limits_{\ell \in \mathcal{N}_k} c_{\ell k} \nabla_\omega \widehat{J}_\ell(\bm{\omega}_{k,i-1})$ \\[6pt]
			\quad $\bm{\omega}_{k,i} = \sum\limits_{\ell \in \mathcal{N}_k} \alpha_{\ell k} \bm{\psi}_{\ell,i}$
		\end{tabular}
		\\
		\hline
		\textbf{Adaptive CTA} &
		\begin{tabular}[t]{@{}l@{}}
			\quad $\bm{\psi}_{k,i-1} = \sum\limits_{\ell \in \mathcal{N}_k} \alpha_{\ell k} \bm{\omega}_{\ell,i-1}$ \\[6pt]
			\quad $\bm{\omega}_{k,i} = \bm{\psi}_{k,i-1} + \mu_k \sum\limits_{\ell \in \mathcal{N}_k} c_{\ell k} \bm{u}_{\ell,i} \left[ d_\ell(i) - \bm{u}_{\ell,i} \bm{\psi}_{k,i-1} \right]$
		\end{tabular}
		\\
		\hline
		\begin{tabular}[t]{c@{}l@{}}
			\textbf{General} \\
			\textbf{Adaptive CTA}
		\end{tabular} &
		\begin{tabular}[t]{@{}l@{}}
			\quad $\bm{\psi}_{k,i-1} = \sum\limits_{\ell \in \mathcal{N}_k} \alpha_{\ell k} \bm{\omega}_{\ell,i-1}$ \\[6pt]
			\quad $\bm{\omega}_{k,i} = \bm{\psi}_{k,i-1} - \mu_k \sum\limits_{\ell \in \mathcal{N}_k} c_{\ell k} \nabla_\omega \widehat{J}_\ell(\bm{\psi}_{k,i-1})$
		\end{tabular}
		\\
		\hline
	\end{tabular}
	\label{tab:adaptive_atc_cta}
\end{table}
These simplified implementations offer attractive trade-offs between estimation performance and communication efficiency, making them particularly suitable for resource-constrained applications where bandwidth limitations or energy considerations restrict the feasible communication volume.
A comprehensive convergence analysis of the DLMS algorithm can be found in \cite{chen2012diffusion}, which the reader is encouraged to consult.

\subsubsection{Multitask Diffusion Problem}

The diffusion strategies discussed thus far assume that all network nodes collaborate to estimate a single global parameter vector. However, many practical applications involve scenarios where different nodes need to estimate distinct yet related parameter vectors -- a paradigm known as multitask learning. This section examines how diffusion strategies can be extended to support collaborative estimation in multitask environments while exploiting inter-task relationships to enhance overall system performance.

\paragraph{Multitask Learning Framework}

In multitask networks, nodes work toward potentially different optimization objectives while benefiting from knowledge transfer across related tasks. This framework enables more sophisticated modeling of complex systems, where

\begin{itemize}
	\item Different regions of the network may monitor distinct but correlated phenomena;
	\item Nodes may have varying objectives but share underlying structural similarities; and
	\item Parameter vectors across the network exhibit partial correlation rather than identity.
\end{itemize}

Diffusion strategies support this collaborative learning paradigm by exploiting inter-node task similarity through carefully designed information exchange mechanisms \cite{chen2014multitask,plata2015distributed,gogineni2021performance,marano2021decision}. By incorporating regularization terms that promote appropriate similarity between neighboring tasks, these approaches balance individual task accuracy with beneficial knowledge transfer.

\paragraph{Mathematical Formulation}

In the multitask setting, the data at node $k$ follows a linear model:
\begin{equation} \label{eq:datamodel}
	d_k(n) = \bm{u}_{k,i}^\top \bm{\omega}_k^\star + z_k(n),
\end{equation}
where $\omega_k^*$ represents the target parameter vector specific to node $k$, and $z_k(n)$ denotes observation noise. Unlike the single-task scenario, each node now pursues its own optimal parameter vector, though these vectors may exhibit varying degrees of similarity across the network.

\paragraph{Taxonomy of Distributed Learning Architectures}

The estimation objectives and resulting algorithms depend fundamentally on the network's collaborative structure, as illustrated in Fig.~\ref{fig_MT_DLMS}. This figure presents three principal architectural paradigms as follows:

\begin{figure} 
	\centering
	\includegraphics[trim={0 0 0 0}, scale=0.24]{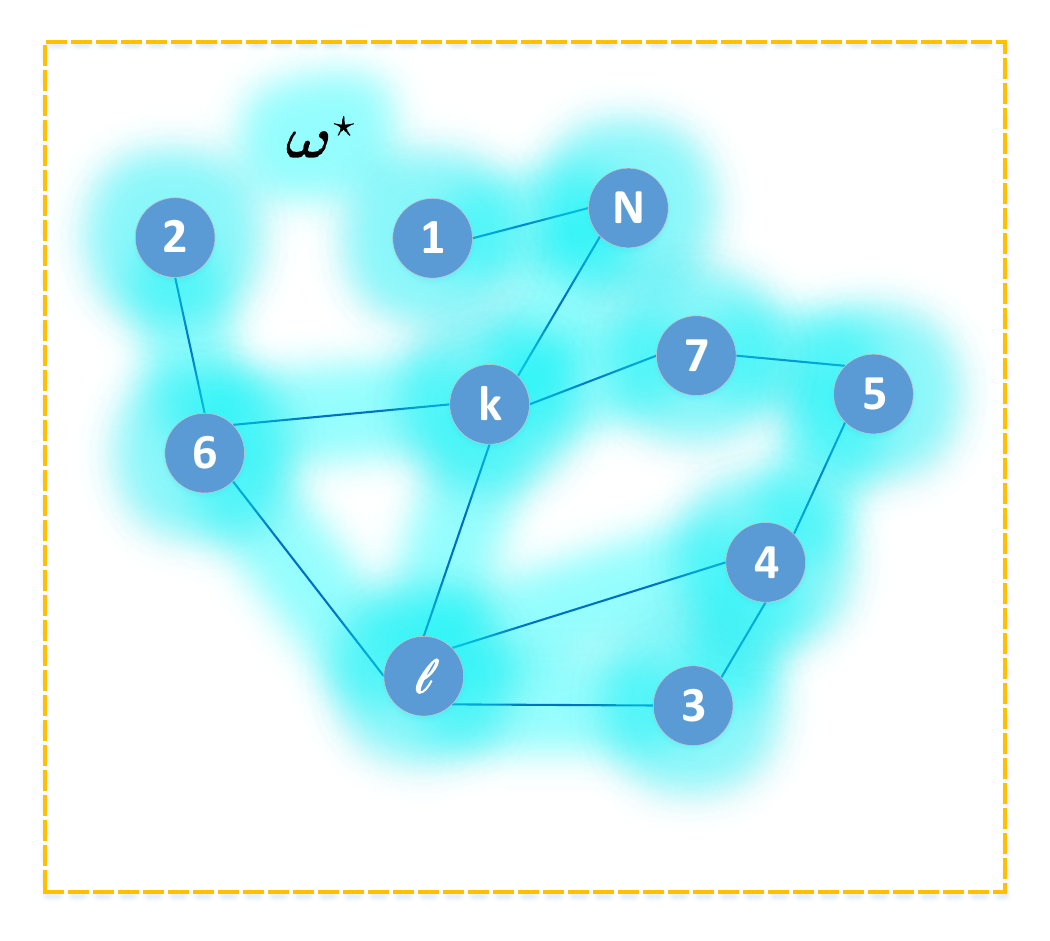}
	\includegraphics[trim={0 0 0 0}, scale=0.24]{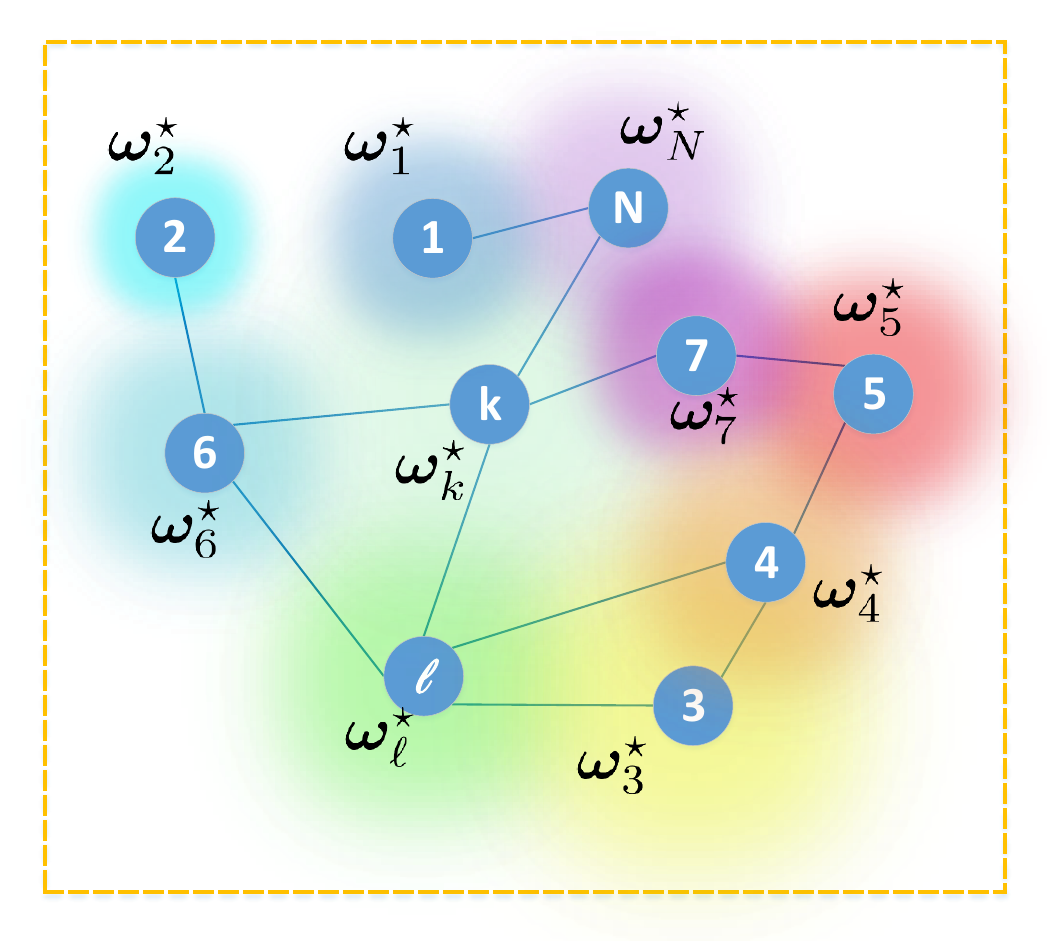}
	\includegraphics[trim={0 0 0 0}, scale=0.24]{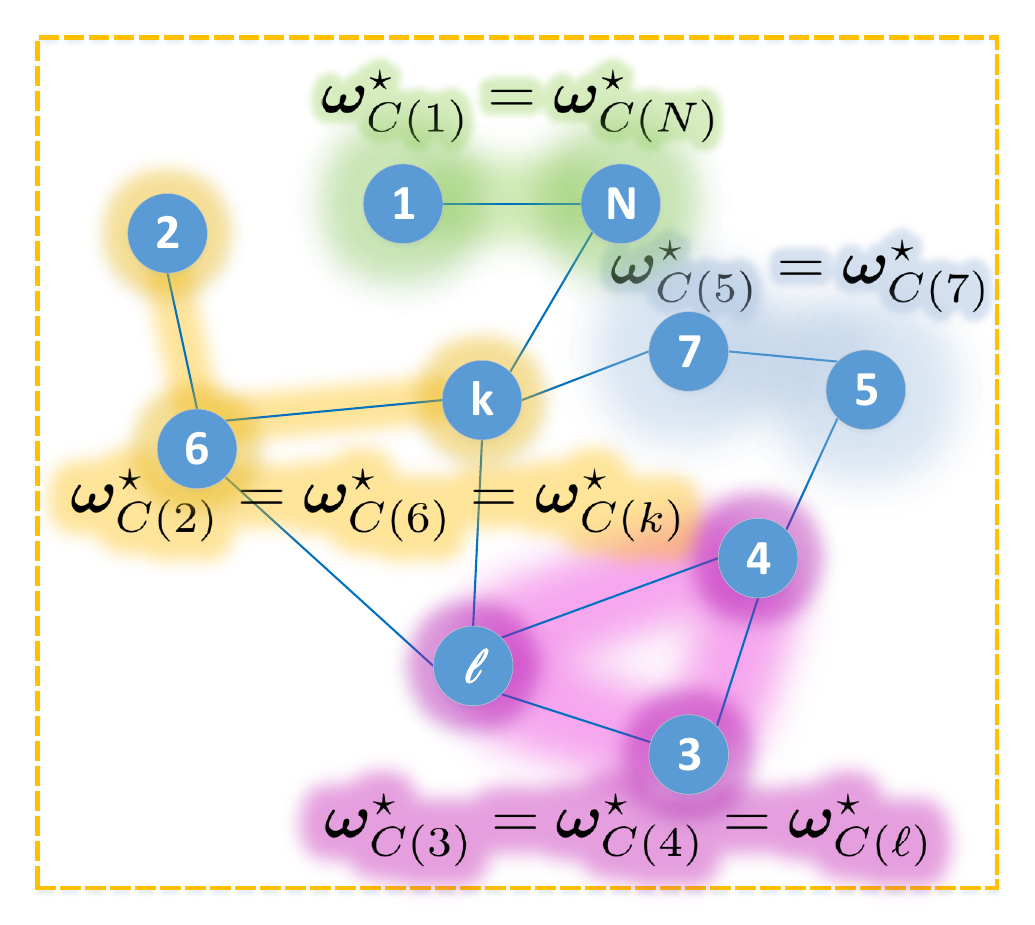}
	\caption{Illustration of diffusion strategies in distributed networks. (Left) Single-task diffusion, where all agents collaborate to estimate a common global parameter $ \bm{\omega}^\star $. (Center) Multi-task diffusion, where each agent estimates its own local parameter $ \bm{\omega}^\star_k $, possibly differing from neighbours. (Right) Clustered multitask diffusion, where agents within the same cluster share a common objective $ \bm{\omega}^\star_{C(k)} $, allowing for both collaboration within clusters and task differentiation across them.}
	\label{fig_MT_DLMS}
\end{figure}

\begin{itemize}
	\item \textbf{Single-task networks:} All nodes collaborate to estimate a common global parameter vector, representing the conventional diffusion approach discussed earlier;
	
	\item \textbf{Multi-task networks:} Each node estimates its own distinct parameter vector, with information exchange leveraging inter-task relationships to improve estimation performance; and
	
	\item \textbf{Clustered multitask networks:} Nodes are organized into clusters, with members of each cluster pursuing a common parameter while different clusters maintain distinct objectives. This hybrid architecture accommodates varying degrees of parameter sharing across the network.
\end{itemize}

The clustered multitask framework represents a generalized approach that encompasses both single-task and multitask models as special cases, offering a flexible framework for modeling complex distributed estimation problems with varying degrees of parameter relatedness.

\subsubsection{Multitask Diffusion Learning}
Building upon the multitask framework introduced previously, this section examines specific algorithms and techniques for implementing diffusion-based learning in multitask environments. We focus particularly on clustered multitask networks, which provide a flexible architecture for balancing localized specialization with collaborative learning across related tasks.

\paragraph{Clustered Optimization Formulation}

In clustered multitask networks, nodes are organized into distinct clusters, with all nodes within the same cluster $C(k)$ collaboratively estimating a shared parameter vector. This structure creates natural boundaries for parameter sharing while still enabling broader collaboration across cluster boundaries through appropriate regularization.
Each node $k$ is associated with a strongly convex, twice-differentiable local cost function $J_k(\bm{\omega}_{\mathcal{C}(k)})$, such as the MSE given by:
\begin{equation}
	\label{eq:JkMSE}
	J_k(\bm{\omega}_{\mathcal{C}(k)}) = \mathbb{E}\left\{ \left| d_k(i) - \bm{u}_{k,i}^\top\, \bm{\omega}_{\mathcal{C}(k)} \right|^2 \right\},
\end{equation}
where $d_k(i)$ and $\bm{u}_{k,i}$ are the measurement and input vector at time $i$.

\paragraph{Inter-Cluster Regularization}

To encourage inter-cluster similarity, regularization terms such as the squared Euclidean distance are added as follows:
\begin{equation}
	\label{eq:l2dist}
	\Delta(\bm{\omega}_{\mathcal{C}(k)}, \bm{\omega}_{\mathcal{C}(\ell)}) = \| \bm{\omega}_{\mathcal{C}(k)} - \bm{\omega}_{\mathcal{C}(\ell)} \|^2,
\end{equation}
for neighboring nodes $k$ and $\ell$. This regularization is applied between parameter vectors of neighboring nodes $k$ and $\ell$ that belong to different clusters. The quadratic penalty encourages neighboring clusters to maintain similar parameter values when supported by the underlying data patterns, while still allowing for necessary differentiation when required by the local objectives.

\paragraph{Global Network Optimization}

Combining the local MSE cost \eqref{eq:JkMSE} with the regularization term \eqref{eq:l2dist}, the global network cost function becomes:
\begin{align}
	\label{eq:MSEl2}
	\overline{J^{\text{glob}}}(\bm{\omega}_{\mathcal{C}_1}, \dots, \bm{\omega}_{\mathcal{C}_Q}) &=
	\sum_{k=1}^{N} \mathbb{E}\left\{ \left| d_k(i) - \bm{u}_{k,i}^\top\, \bm{\omega}_{\mathcal{C}(k)} \right|^2 \right\} \notag\\
	&+ \eta \sum_{k=1}^{N} \sum_{\ell\in\mathcal{N}_k \setminus \mathcal{C}(k)} \rho_{k\ell} \, \| \bm{\omega}_{\mathcal{C}(k)} - \bm{\omega}_{\mathcal{C}(\ell)} \|^2,
\end{align}
where $\bm{\omega}_{\mathcal{C}_q}$ denote the parameter vector for cluster $\mathcal{C}q$ and a hyper-parameter $\eta$ promotes similarity across neighboring clusters, weighted by coefficients $\rho_{k\ell}$.

This formulation balances two essential objectives:

\begin{itemize}
	\item The first term represents the aggregate mean squared error across all nodes, ensuring accurate fitting of each cluster's parameter vector to its local measurements.
	
	\item The second term promotes appropriate similarity between parameter vectors of neighboring clusters, with the strength of this regularization controlled by the following:
	\begin{itemize}
		\item The global regularization parameter $\eta$, which scales the overall importance of inter-cluster similarity; and
		\item The node-specific weights $\rho_{k\ell}$, which adjust the coupling strength between specific pairs of nodes based on their expected task relatedness.
	\end{itemize}
\end{itemize}

The notation $\mathcal{N}_k \setminus C(k)$ refers to the set of neighbors of node $k$ that belong to different clusters, ensuring that regularization is only applied across cluster boundaries rather than within clusters, where nodes already estimate the same parameter vector.

\paragraph{Distributed Implementation}

The global optimization problem can be solved distributively using carefully designed update rules that combine aspects of intra-cluster consensus with inter-cluster regularization. These update equations enable each node to refine its local estimate through a combination of the following:

\begin{itemize}
	\item Adaptation based on local measurement data;
	\item Combination with estimates from other nodes in the same cluster;
	\item Regularized knowledge transfer from neighboring nodes in different clusters.
\end{itemize}

Specifically, the iterative updates at node~$k$ in the clustered multitask setting are given by:

\begin{align}
	\bm{\psi}_{k,i} &= \bm{\omega}_{k,i-1} + \mu \bigg( \sum_{\ell\in\mathcal{N}_k \cap \mathcal{C}(k)}\hspace{-5mm} c_{\ell k} \left( d_\ell(i) - \bm{u}_{\ell,i}^\top\, \bm{\omega}_{k,i-1} \right) \bm{u}_{\ell,i}
	+ \eta \hspace{-5mm}\sum_{\ell\in\mathcal{N}_k \setminus \mathcal{C}(k)} \hspace{-5mm}\rho_{k\ell} \left( \bm{\omega}_{\ell,i} - \bm{\omega}_{k,i} \right) \bigg),\notag \\
	\bm{\omega}_{k,i} &= \sum_{\ell\in\mathcal{N}_k \cap \mathcal{C}(k)} \alpha_{\ell k}\, \bm{\psi}_{\ell,i},
	\label{eq:ATC_MSEl2}
\end{align}
where $\{c_{\ell k}\}$ and $\{\alpha_{\ell k}\}$ are combination weights satisfying suitable stochasticity conditions.
In the standard multitask diffusion case (without clustering), the update rule simplifies to:
\begin{equation}
	\label{eq:ATC_MSE_multi}
	\bm{\omega}_{k,i} = \bm{\omega}_{k,i-1 }
	+ \mu \left( d_k(i) - \bm{u}_{k,i}^\top\, \bm{\omega}_{k,i-1} \right) \bm{u}_{k,i}
	+ \eta \mu \sum_{\ell\in\mathcal{N}_k^{-}} \rho_{k\ell} \left( \bm{\omega}_{\ell,i-1} - \bm{\omega}_{k,i-1} \right),
\end{equation}
where $\mathcal{N}_k^{-}$ denotes the set of neighbouring nodes that are not part of the same cluster as node $k$.
\subsubsection{Effect of Combination Weights and Network Topology}
The performance characteristics of diffusion-based estimation algorithms are strongly influenced by two critical design factors: the combination matrix that governs information exchange among neighboring nodes and the underlying network topology that defines the communication infrastructure. This section examines how these factors impact estimation performance and presents strategies for optimizing them in practical implementations.

\paragraph{Performance-Topology Relationships}

The network topology -- the pattern of connections among nodes -- fundamentally shapes how information propagates through the network during the diffusion process. This relationship creates important trade-offs that system designers must carefully consider:

\begin{itemize}
	\item \textbf{Densely connected networks} typically achieve superior estimation performance, characterized by the following:
	\begin{itemize}
		\item Faster convergence rates due to more efficient information propagation;
		\item Lower mean-square deviation (MSD) in steady-state as nodes benefit from more diverse information sources;
		\item Greater resilience against individual node or link failures; and
		\item Higher implementation costs in terms of communication overhead, power consumption, and infrastructure requirements.
	\end{itemize}
	
	\item \textbf{Sparse network topologies} offer alternative advantages as follows:
	\begin{itemize}
		\item Reduced communication overhead, making them suitable for bandwidth-constrained applications;
		\item Lower power consumption, critical for energy-limited wireless sensor networks;
		\item Simpler implementation with fewer communication links to establish and maintain; and
		\item Potential performance degradation if not carefully designed to preserve essential information flow paths. 
	\end{itemize}
\end{itemize}

This fundamental trade-off between estimation performance and resource efficiency drives much of the research on optimizing network topology for specific application requirements.

\paragraph{Combination Weight Design Strategies}

Equally important to network topology is the design of the combination weights $\{\alpha_{\ell k}\}$ that determine how each node combines information received from its neighbors. These weights directly influence both convergence behavior and steady-state accuracy, with different weighting schemes offering various performance characteristics.

Common rules for setting combination weights include uniform, Laplacian, and Metropolis schemes \cite{sayed2014diffusion, hua2022adaptive, hua2024resilient}:
\begin{align}
	&\text{Uniform weights: }	\alpha_{\ell k} = 
	\begin{cases}
		\frac{1}{|N_k|}, & \text{if } \ell \in N_k, \\
		0, & \text{otherwise},
	\end{cases}\\
	&\text{Metropolis Rule: }\alpha_{\ell k} =
	\begin{cases}
		\frac{1}{\max\{|\mathcal{N}_k|, |\mathcal{N}_\ell|\}}, & \text{if } \ell \in \mathcal{N}_k, \ell \neq k, \\
		1 - \sum_{m \in \mathcal{N}_k \setminus \{k\}} \alpha_{mk}, & \text{if } \ell = k, \\
		0, & \text{otherwise},
	\end{cases}\\
	&\text{Laplacian Rule: }\alpha_{\ell k} =
	\begin{cases}
		\frac{1}{\max\{|\mathcal{N}_\ell|\}}, & \text{if } \ell \in \mathcal{N}_k, \ell \neq k, \\
		1 -\frac{1 - |\mathcal{N}_k|}{\max\{|\mathcal{N}_\ell|\}}, & \text{if } \ell = k, \\
		0, & \text{otherwise},
	\end{cases}
\end{align}
where $|\mathcal{N}_k|$ represents the number of nodes in the neighborhood of node $k$ (including $k$ itself). While computationally efficient and requiring no global network information, this approach may not be optimal when nodes have varying reliability or relevance.

Effective diffusion adaptation thus requires thoughtful design of both the combination policy and network connectivity. In \cite{zhao2012clustering}, the authors approximate the minimization of the instantaneous MSD $\|\bm{\omega}^\star - \bm{\omega}_{k,i}\|^2$ using:
\begin{align}
	\alpha_{\ell k}(i) = \frac{\gamma_{\ell,k}^{-2}(i)}{\sum\limits_{r \in \mathcal{N}_k} \gamma_{r,k}^{-2}(i)},~\text{and}~ \gamma_{\ell,k} = ||\bm{\psi}_{k,i} - \bm{\psi}_{\ell,i} ||^2,
\end{align}
where the dependence on $i$ indicates the time-varying of this weighting independent of the fact that the connection graph is fixed or not.

In \cite{shamsi2021flexible}, the authors analyzed error propagation through the network during the diffusion process. They proposed a flexible weighting strategy based on the similarity between the estimates of neighbouring nodes:
\begin{align}
	\alpha_{\ell k}(i) = \frac{e^{\zeta_{\ell k}(i)}}{\sum\limits_{r \in \mathcal{N}_k} e^{\zeta_{r k}(i)}},~\text{and}~ \zeta_{r k} \triangleq \left(\frac{\text{dist}(\bm{\psi}_{k,i},  \bm{\psi}_{\ell,i})}{a} \right)^b,
\end{align}
where $\text{dist}(\bm{\psi}_{k,i},  \bm{\psi}_{\ell,i})$ denotes a distance measure between two estimated vectors, such as the Euclidean distance $|| \bm{\psi}_{k,i} - \bm{\psi}_{\ell,i} ||^2$ and $a$ and $b$ determine the allowed combination range and its decaying, respectively. It also has been shown effective to isolate the malfunctioning nodes in the network.

In \cite{zayyani2020robust2}, a minimum distance criterion was introduced to prevent the propagation of impulsive noise through the network by optimizing the following problem:
\begin{align}
	\min_{\alpha_{\ell,k},\,\ell \in \mathcal{N}_k} \left\| \sum_{\ell \in \mathcal{N}_k} \alpha_{\ell,k} \bm{\psi}_{\ell,i} - \bm{\omega}_{k,i-1} \right\|_2^2, \quad
	\text{s.t.} \quad \sum_{\ell \in \mathcal{N}_k} \alpha_{\ell,k} = 1.
\end{align}
Using the method of Lagrange multipliers with parameter $\lambda$, the optimal solution can be derived as:
\begin{align}
	{\lambda_{\text{opt}} = \frac{ \bm{w}^\top \bm{v} - 1 }{\bm{w}^\top \boldsymbol{1}}, \quad \bm{a}_{\text{opt}} = (\Psi^\top\Psi)^{-1} \bm{v} + \lambda_{\text{opt}} \bm{w}},
	\label{eq:MD_weight}
\end{align}
where $\Psi \triangleq [\bm\psi_{l_1}, \bm\psi_{l_2}, \cdots, \bm\psi_{l_n}]$ and $\bm{a} \triangleq [\alpha_{l_1}, \alpha_{l_2}, \cdots, \alpha_{l_n}]^\top$, with $l_j \in \mathcal{N}_k$ and $n = |\mathcal{N}_k|$. Moreover, $\bm{v} \triangleq \Psi^\top \bm{\omega}_{k,i}$, and $\bm{w} \triangleq (\Psi^\top\Psi)^{-1}\boldsymbol{1}$. For better readability, the indices $i$ and $k$ have been omitted where appropriate.

\subsubsection{Extensions of Diffusion-based Techniques}
The core diffusion adaptation strategies discussed previously have inspired numerous extensions and enhancements to address specific challenges in distributed estimation. This section surveys recent advances that extend diffusion approaches to handle various practical constraints, improve performance in challenging environments, and enhance resilience against both natural and adversarial disturbances.

\paragraph{Probabilistic Extensions}

Recent probabilistic extensions incorporate Bayesian learning principles into diffusion frameworks, enhancing the ability to handle uncertainty and adapt to changing conditions:

\begin{itemize}
	\item \textbf{Bayesian-learning-based DLMS} \cite{huang2023bayesian} integrates probabilistic modeling into the diffusion framework, enabling more robust adaptation under non-stationary conditions and heavily noise-corrupted environments. By maintaining probabilistic representations of uncertainty, these approaches can better discriminate between measurement noise and actual signal variations.
\end{itemize}

\paragraph{Communication-Efficient Implementations}

Communication overhead represents a critical constraint in many distributed networks, particularly in wireless and energy-limited settings. Several innovative approaches have been developed to reduce communication requirements while preserving estimation performance:

\begin{itemize}
	\item \textbf{Partial diffusion strategies} transmit only subsets of parameter vector entries during each communication round, significantly reducing bandwidth requirements. Notable implementations include the following:
	\begin{itemize}
		\item PDLMS (Partial Diffusion LMS) \cite{arablouei2013distributed}, which selectively communicates the most significant parameter components;
		\item PDRLS (Partial Diffusion Recursive Least Squares) \cite{arablouei2014adaptive}, which extends the partial communication concept to RLS algorithms; and
		\item CR-DLMS (Communication-Reducing DLMS) \cite{arablouei2015analysis}, which dynamically selects communication partners to minimize unnecessary information exchange.
	\end{itemize}
	
	\item \textbf{Complementary approaches} for communication reduction include the following:
	\begin{itemize}
		\item Estimate sparsification \cite{8424904, vahidpour2017analysis}, which exploits parameter sparsity to reduce transmission volume;
		\item Reliability-based node selection \cite{zayyani2022communication}, which prioritizes information from nodes with higher estimation confidence;
		\item Compressive diffusion \cite{sayin2014compressive}, which applies compressed sensing principles to the diffusion framework; and
		\item Frequency-domain updates \cite{peng2023frequency}, which communicate only in selected frequency bands to reduce overall data volume.
	\end{itemize}
\end{itemize}

\paragraph{Handling Missing and Censored Data}

Practical sensing environments frequently encounter measurement limitations such as censoring (where values outside certain ranges cannot be measured) and missing data. Specialized diffusion extensions address these challenges:

\begin{itemize}
	\item \textbf{Censored measurements}, often modeled via Tobit models, are addressed by
	\begin{itemize}
		\item Diffusion-based censored estimation \cite{liu2015distributed}, which incorporates censoring models directly into the adaptation process, and
		\item Bias-compensated algorithms \cite{liu2015censored}, which correct the statistical bias introduced by censoring effects.
	\end{itemize}
	
	\item \textbf{Missing data strategies} depend on the missing data mechanism, i.e.,
	\begin{itemize}
		\item For Missing At Random (MAR) scenarios, multiple imputation techniques are preferred, and
		\item For Missing Completely At Random (MCAR) cases, specialized distributed solutions have been developed \cite{gholami2016diffusion, peng2024frequency}.
	\end{itemize}
	
	\item \textbf{Adaptive participation schemes} intelligently manage node involvement:
	\begin{itemize}
		\item Adaptive censoring approaches \cite{tiglea2023reducing} selectively activate sensors based on information content, and
		\item Low-cost sampling methods \cite{tiglea2020low} dynamically adjust participation rates based on estimation error, optimizing the trade-off between energy consumption and accuracy.
	\end{itemize}
\end{itemize}

\paragraph{Other Relevant Literature}

To improve robustness against impulsive noise, several approaches have been proposed, including error nonlinearities \cite{giv2020robust}, disturbance-based updates \cite{zayyani2020robust, zayyani2020robust2}, mean-$p$ power objectives \cite{korki2019weighted}, and Huber loss approximations \cite{barani2024distributed, ashkezari2019robust}. Despite these, compromised nodes may still degrade performance, motivating node-weighting schemes for enhanced resilience \cite{shamsi2021flexible}.

Recent developments incorporate Maximum Correntropy Criterion (MCC)-based filters \cite{wang2019robust} and generalized MCC frameworks \cite{zhao2023variable} to handle impulsive noise and censored observations. A diffusion framework with partial node visibility was introduced in \cite{shamsi2025joint, shamsi2025distributed, shamsi2025sparse, shamsi2025multi}, leveraging signal flow analysis \cite{shamsi2021flexible} and thresholding-based support identification \cite{shamsi2020nonlinear, shamsi2024acceleration}.

Beyond impulsive noise, the security of distributed networks has received growing attention in signal processing \cite{yang2020adversary} and IoT contexts \cite{chen2018internet}. Malicious agents can disrupt estimation by introducing bias or delay. To mitigate such threats, resilient strategies have emerged \cite{chen2018resilient, liu2018secure, ntemos2017secure, zayyani2022adversary, wan2022secure, meng2021distributed, shi2020secure, li2019resilient}. For example, \cite{zayyani2025secure} proposes the Average Diffusion LMS (ADLMS) with ALRT-based detectors to counteract sensor and link attacks, offering resilience with low complexity. Similarly, \cite{hua2019distributed} presents a DLMS algorithm with adaptive credibility weights to reject unreliable data under channel attacks, enhancing both robustness and accuracy.

\subsection{Nonlinear Models} \label{sec_nonlin}

While the majority of distributed filtering techniques focus on linear dynamical systems, some real-world cyber-physical systems exhibit inherent nonlinearities that cannot be adequately captured by linear models. This section explores key extensions of distributed filtering approaches to nonlinear systems, reviewing recent developments that enable collaborative state estimation in complex nonlinear environments\footnote{This paper primarily focuses on distributed estimation and filtering for ``linear" dynamical systems, where the theoretical foundations, background, and algorithmic structures are presented in details. The brief discussion of nonlinear cooperative state estimation is included to provide context and to highlight extensions of the linear framework, and to enhance the completeness of the survey. This section aims to outline the existing modeling approaches and key ideas, by providing references and relevant literature for interested readers.}.

As noted in comprehensive surveys \cite{BATTILOTTI2019562,Hu01012020}, distributed nonlinear filtering represents an active research area with both theoretical challenges and practical applications. Unlike linear filtering, where optimal solutions often exist in closed form, nonlinear filtering typically requires approximation techniques to make the estimation problem tractable in distributed settings. Below, we examine four principal approaches to distributed nonlinear filtering, each offering distinct advantages for different application scenarios.

\subsubsection*{Distributed Moving Horizon Estimation for Nonlinear Systems}
Moving horizon estimation (MHE) is an optimization-based state estimation framework that explicitly incorporates system constraints, nonlinear dynamics, and bounded disturbances by solving a finite-horizon estimation problem at each sampling time \cite{battistelli2018distributed}. In distributed settings, MHE provides a flexible alternative to Kalman-filter-based approaches, particularly for  constrained nonlinear networked systems.
Consider the discrete-time nonlinear system as,
\begin{align}
	x(t+1) &= f(x(t),u(t)) + \nu(t),
\\
	y(t) &= h(x(t)) + \mu(t).
\end{align}
In centralized MHE, the state estimate at time $t$ is obtained by solving an optimization problem over a sliding horizon of length $L$ as given below \cite{rao2003constrained}:
\begin{equation}
	\begin{aligned}
		\min_{\{x(k)\}} \quad &
		\|x(t-L)-\widehat{x}(t-L)\|_{P}^{2}
		+ \sum_{k=t-L}^{t-1} \|y(k) - h(x(k))\|_{R}^{2} \\
		& + \sum_{k=t-L}^{t-1} \|x(k+1)-f(x(k),u(k))\|_{Q}^{2} \\
		\text{s.t.} \quad &
		x(k+1) = f(x(k),u(k)), \quad k = t-L,\dots,t-1, \\
		& x(k) \in \mathcal{X}, \quad v(k):=y(k) - h(x(k)) \in \mathcal{V}
	\end{aligned}
\end{equation}
where $P$, $Q$, and $R$ are positive definite weighting matrices, $\widehat{x}_{k-L}$ is a-priori estimate, and $\mathcal{X}$ and $\mathcal{V}$ denote the admissible constraint sets. The estimate $\widehat{x}(t)$ is given by the optimizer's terminal state.
\\
A common distributed MHE approach decomposes the centralized cost into local objective functions \cite{farina2012,philipp2011moving,wang2023distributed,zou2019moving2}. At agent $i$, the local MHE problem over horizon $L$ is given by:
\begin{equation}
	\begin{aligned}
		\min_{\{x_i(k)\}} \quad &
		\|x_i(t-L)-\widehat{x}_i(t-L)\|_{P_i}^{2}
		+ \sum_{k=t-L}^{t-1} \|y_i(k) - h_i(x_i(k))\|_{R_i}^{2} \\
		& + \sum_{k=t-L}^{t-1} \|x_i(k+1)-f(x_i(k),u(k))\|_{Q_i}^{2} \\
		\text{s.t.} \quad &
		x_i(k+1) = f(x_i(k),u(k)), \\
		& x_i(k) \in \mathcal{X}, \quad v_i(k):=y_i(k) - h_i(x(k)) \in \mathcal{V}.
	\end{aligned}
\end{equation}
In consensus-based distributed MHE, consensus constraints are imposed and the coupling constraints are relaxed by adding disagreement penalties to the local cost functions \cite{borelle2025robust}.
Alternatively, ADMM-based distributed MHE introduces local copies of the state and corresponding Lagrange multipliers to enforce agreement among agents \cite{kim2021distributed,distrib_estim_kekatos}. This results in iterative local MHE updates combined with neighbor communication of primal and dual variables until convergence (or until a predefined number of iterations) is reached.

\subsubsection*{Consensus + Innovation Filtering for Nonlinear Systems}

Consensus + Innovation filtering extends the linear filtering framework to nonlinear systems by combining two complementary mechanisms:

\begin{itemize}
	\item A \textbf{consensus term} that fuses estimates from neighboring nodes, promoting agreement across the network, and
	\item An \textbf{innovation term} derived from nonlinear measurement models and local observations.
\end{itemize}

In this approach, each agent updates its state estimate based on a weighted combination of its neighbors' estimates (the consensus component) and the innovation from its local nonlinear measurements. The consensus step ensures that the global estimate of the nonlinear system state is maintained with reasonable consistency across all agents, while the innovation term incorporates new measurement information to refine accuracy.

This methodology has been successfully applied to various nonlinear estimation problems, with notable implementations and extensions described in \cite{mohammadi2014distributed,kar6934985,karnonlin}. The framework's flexibility makes it particularly suitable for systems with moderately nonlinear dynamics or measurement models, offering a natural extension of linear consensus-based techniques.

\subsubsection*{Distributed Extended Kalman Filtering}

The Extended Kalman Filter (EKF) represents one of the most widely used approaches for nonlinear state estimation in centralized settings. It operates by linearizing the nonlinear system around the current state estimate, applying standard Kalman filter equations to this linearized model, and then updating the state estimate accordingly.

Although traditionally implemented in centralized architectures, the EKF has been successfully adapted for distributed scenarios through several key innovations:

\begin{itemize}
	\item Each node maintains its own local EKF to process nonlinear measurements and update its state estimate;
	\item Nodes periodically share their state estimates and associated uncertainty information with neighbors; and
	\item Various fusion rules combine local and neighboring estimates to refine the overall estimation accuracy.
\end{itemize}

This distributed EKF approach has been implemented with numerous variations, including consensus-based fusion \cite{BATTISTELLI2016169}, information-form implementations \cite{LI20177983}, covariance intersection methods \cite{8740882}, and adaptive architectures \cite{liIET,REZAEI2021102957}. These approaches balance estimation accuracy with communication efficiency, enabling effective nonlinear state estimation across networks with diverse topologies and resource constraints.

\subsubsection*{Distributed Unscented Kalman Filtering}

The Unscented Kalman Filter (UKF) represents an alternative approach to nonlinear filtering that avoids explicit linearization. Instead, the UKF employs a deterministic sampling technique to capture the statistical properties of the state distribution:

\begin{itemize}
	\item A set of carefully selected sample points (sigma points) is chosen to represent the state distribution;
	\item These points are propagated through the exact nonlinear system dynamics; and
	\item The transformed points are used to reconstruct the posterior mean and covariance.
\end{itemize}

This approach typically achieves higher accuracy than the EKF for systems with significant nonlinearities, as it better captures the effect of nonlinear transformations on probability distributions.

Distributed implementations of the UKF \cite{TNUNAY2020270,7118687,Liu2024,Lv2021,9797037} follow similar principles to distributed EKF approaches, with nodes maintaining local UKFs and exchanging information with neighbors. The primary difference lies in the local filtering algorithm, with UKF-based methods generally offering improved performance for highly nonlinear systems at the cost of somewhat increased computational complexity.

\subsubsection*{Distributed Particle Filtering}

For systems with severe nonlinearities or non-Gaussian noise characteristics, particle filters provide a powerful estimation framework. These methods represent probability distributions using sets of weighted samples (particles) rather than parametric distributions \cite{978374}, enabling them to capture multi-modal and heavily skewed distributions that arise in many complex nonlinear systems.

In distributed implementations \cite{6375933,6450113,6804018,ije,9311256,10132554}, particle filtering operates through several coordinated mechanisms:

\begin{itemize}
	\item Each node maintains its own local set of particles representing possible system states;
	\item Nodes exchange particle information or statistical summaries with neighbors according to various communication protocols;
	\item Fusion algorithms combine local and neighboring particle representations to refine the state estimation; and
	\item Resampling strategies prevent particle degeneracy while maintaining estimation accuracy.
\end{itemize}

Distributed particle filtering approaches offer unparalleled flexibility for handling complex nonlinear dynamics and non-Gaussian uncertainties. However, this flexibility comes at the cost of increased computational and communication requirements compared to EKF and UKF-based methods. Recent research has focused on developing communication-efficient variants that preserve estimation quality while reducing resource demands, making distributed particle filtering increasingly practical for resource-constrained cyber-physical systems.

These four complementary approaches to distributed nonlinear filtering provide system designers with a rich toolkit for addressing nonlinear estimation challenges across diverse application domains. The choice among these methods typically depends on the specific characteristics of the nonlinear system, the available computational and communication resources, and the required estimation accuracy.

\section{Distributed Fault Detection}
Modern cyber-physical systems often involve highly interconnected networks of components that must operate reliably despite potential faults and failures. This section examines distributed approaches to fault detection and isolation (FDI) that enable robust system monitoring without relying on centralized processing architectures. We explore how the distributed estimation techniques developed in previous sections can be extended to detect, isolate, and mitigate faults across networked systems.

\subsection*{Motivation and Challenges}

Complex interconnected systems such as smart grids, autonomous robotic networks, and industrial automation infrastructures face unique fault detection challenges:

\begin{itemize}
	\item \textbf{Diverse fault sources:} Faults can emerge from numerous sources, including environmental disturbances, component degradation, sensor malfunctions, communication failures, or unexpected interactions among subsystems;
	
	\item \textbf{Cascading failure risks:} Undetected localized faults can propagate through interconnected systems, potentially triggering cascading failures that affect large portions of the network; and
	
	\item \textbf{System-wide consequences:} Faults often lead to degraded performance, unplanned downtime, safety hazards, or even catastrophic system failure if not promptly detected and isolated.
\end{itemize}

Traditional centralized fault detection approaches \cite{SAMY2011658,5282515,survey_sandberg}, while well-established in theory and practice, face significant limitations in large-scale networked settings:

\begin{itemize}
	\item \textbf{Scalability constraints:} Centralized methods struggle to process the volume and diversity of measurements generated by large-scale systems;
	
	\item \textbf{Single-point vulnerability:} Reliance on centralized processing creates a critical vulnerability -- failure of the central node can disable the entire monitoring system;
	
	\item \textbf{Communication bottlenecks:} Routing all measurements to a central processor creates bandwidth congestion and often introduces unacceptable latency; and
	
	\item \textbf{Limited resilience:} Centralized approaches typically lack graceful degradation capabilities when parts of the system become inaccessible or compromised.
\end{itemize}

\subsection*{Distributed Fault Detection Paradigm}
In contrast to centralized approaches, distributed fault detection leverages the distributed estimation and filtering techniques explored in previous sections to implement fault detection capabilities across networked nodes. This distributed paradigm offers several key advantages:

\begin{itemize}
	\item \textbf{Local processing:} Fault detection algorithms operate primarily on local measurements, with strategic information exchange among neighboring nodes;
	
	\item \textbf{Scalable architecture:} The computational burden scales with system size as processing is distributed across the network;
	
	\item \textbf{Enhanced robustness:} No single point of failure exists, as detection capabilities are distributed throughout the system; and
	
	\item \textbf{Communication efficiency:} Only relevant diagnostic information needs to be shared, rather than raw measurements.
\end{itemize}

In this section, we focus primarily on distributed observer-based fault detection techniques that build upon the distributed estimation methods described earlier. While we emphasize observer-based approaches due to their strong theoretical foundations and practical effectiveness, we also survey alternative distributed fault detection methodologies that offer complementary capabilities for specific application scenarios.

The distributed fault detection approaches we examine operate under a common framework where individual nodes satisfy the following specifications:

\begin{enumerate}
	\item Maintain local observers/estimators based on system models and available measurements;
	\item Generate residual signals by comparing predicted and actual system behavior;
	\item Apply detection logic to identify faults based on residual patterns;
	\item Share relevant diagnostic information with neighboring nodes; and
	\item Fuse local and neighbor diagnostic assessments to enhance detection reliability.
\end{enumerate}

Through these mechanisms, distributed fault detection enables cyber-physical systems to maintain operational integrity despite the occurrence of faults, supporting critical objectives including system reliability, availability, maintainability, and safety.

\subsection*{Mathematical Formulation of Distributed Fault Detection}

To develop a rigorous framework for distributed fault detection, we consider noise-corrupted physical systems in the form of equation \eqref{eq_A} with the addition of potential sensor faults. The measurement model with fault terms is given by
\begin{equation} \label{eq_sysf}
	\mathbf{y}(t) = \mathbf{C}\mathbf{x}(t) + \mathbf{f}(t) + \boldsymbol{\mu}(t),
\end{equation}
where $\boldsymbol{\mu}(t)$ represents the general measurement noise typically modelled as Gaussian random variable and $\mathbf{f}(t) \in \mathbb{R}^N$ represents the fault vector affecting sensor outputs at time $t$. Each component of this vector follows a binary pattern:
\begin{equation} \label{eq_ff}
	f_i(t) = 
	\begin{cases}
		0, & \text{no fault at node } i , \\
		\neq 0, & \text{faulty node } i.
	\end{cases}
\end{equation}
In general, fault variable $f(t)$ is considered a random-valued variable and no specific constraint is considered for $f(t)$. In general, fault term and noise term differ in the following aspects:
	\begin{itemize}
		\item Fault term is not necessarily zero-mean variable while noise term is considered Gaussian zero-mean random variable;
		\item Faults are of larger magnitude relative to normal noise and assumed to persist over longer time-scales (sudden jumps or slow drifts). Noise, on the other hand, typically is smaller, bounded (or with known variance) and with high-frequency.
	\end{itemize}	  
	
Some literature consider fast time-varying and unbounded faults. The work \cite{hosseini2025integrated} fuses a distributed fault identification observer with a fault-tolerant control law while considering time-varying delays. The work \cite{tahoun2020fault} studies fault diagnosis and resilient control for networked multi-agent nonlinear systems subject to time-varying sensor faults. The paper \cite{pourasghar2025guaranteed} develops a robust estimator-based fault-detection scheme for nonlinear systems using a combined $H_2$ performance and $L_\infty$ robustness design. The work \cite{han2026disturbance} develops an integrated observer and fault-tolerant control framework for switched nonlinear systems subject to stochastic noise and time-varying disturbances.

The formulation given by \eqref{eq_sysf}-\eqref{eq_ff} captures the practical scenario where individual sensors within the network may experience faults independently, requiring detection mechanisms that can identify which specific sensors are providing corrupted measurements.

\paragraph{Observer-Based Fault Detection Principles}

Observer-based fault detection has emerged as a powerful methodology for real-time monitoring of cyber-physical systems in the presence of faults. This approach leverages the distributed state estimation techniques discussed in previous sections, extending them to enable fault detection capabilities. The core principles of this approach include:

\begin{itemize}
	\item \textbf{Continuous state estimation:} Distributed observers/estimators continuously track the system state based on available measurements and system models;
	
	\item \textbf{Residual generation:} By comparing actual measurements with those predicted by the observers, residual signals are generated that quantify the discrepancy between expected and observed system behavior;
	
	\item \textbf{Threshold-based detection:} When residuals exceed certain thresholds (determined either probabilistically or deterministically), fault conditions are flagged; and
	
	\item \textbf{Distributed implementation:} Each node maintains its own local observer/estimator, enabling localized fault detection while strategic information sharing enhances system-wide diagnostic capabilities.
\end{itemize}

\paragraph{Implementation Using Distributed Estimation}

The consensus-based distributed estimator presented in equations \eqref{eq_p}-\eqref{eq_m} provides a natural framework for implementing distributed fault detection. This estimator can be applied directly to fault-corrupted measurements, serving the dual purpose of state tracking and fault detection. When faults occur, they manifest as anomalies in the estimation process that can be detected through careful analysis of estimation errors.

The resulting error dynamics follow equation \eqref{eq_err1}, with fault terms incorporated into the $\boldsymbol{\zeta}(t)$ term. Under fault-free conditions, the steady-state estimation error has been shown to satisfy $\lim_{t\rightarrow\infty} \mathbb{E}(\mathbf{e}_i(t)) = 0$ \cite{jstsp,usman_cdc:11}, providing a baseline expectation for normal system operation.

The error covariance in this case is defined as $\Sigma_e := \lim_{t\rightarrow\infty} \mathbb{E}(\mathbf{e}_i(t)^{\top}\mathbf{e}_i(t))$ and satisfies the relationship \cite{usman_cdc:11,ecc_attack,tnse_attack}:
\begin{equation} \label{eq_sigma_e}
	\|\Sigma_e\|_2 = \frac{a_1\frac{1}{N}\|\Sigma_{\nu}\|_2 + a_2\|\Sigma_{\mu}\|_2}{1 - b^2},
\end{equation}
where
\begin{itemize}
	\item $b := \|(W \otimes A)-K D_C(W \otimes A)\|_2 < 1$,
	\item $a_1 := \|I_{Nn}-K D_C\|^2_2$,
	\item $a_2 := \|K\|^2_2$, 
	\item $\Sigma_{\mu} := \text{diag}[\sum_{j\in N_i} \mathbf{C}^{\top}_j \Sigma^j_{\mu}\mathbf{C}_j]$ where $\Sigma^j_{\mu}$ denotes the $j$th diagonal entry of $\Sigma_{\mu}$ as the variance of the measurement noise term $\boldsymbol{\mu}$, and
	\item $\Sigma_{\nu}$ represents the variance of the process noise term $\boldsymbol{\nu}$.
\end{itemize}

This analytical characterization of estimation error behavior provides the foundation for designing effective fault detection mechanisms that can discriminate between normal system variations due to noise and actual fault conditions.

\paragraph{Residual-Based Detection Approaches}

In the context of residual-based fault detection, two main paradigms have emerged in the literature, each representing a different philosophy regarding how detection mechanisms process temporal information:

\begin{itemize}
	\item \textbf{Stateless fault detection:} This approach makes detection decisions based solely on the current residual values, without incorporating historical information. Detection thresholds are typically designed based on statistical properties of the residuals under fault-free conditions; and
	
	\item \textbf{Stateful fault detection:} This approach maintains historical information about residual behavior over time, often using sliding windows or exponentially weighted averages. By incorporating temporal information, stateful methods can detect subtle faults that develop gradually or manifest intermittently.
\end{itemize}

These complementary approaches offer different trade-offs between detection speed, sensitivity, and false alarm rates, providing system designers with flexible options to match specific application requirements. The following sections examine each of these detection paradigms in greater detail, exploring their mathematical formulations, implementation considerations, and performance characteristics.

\subsection{Stateless Detection} \label{sec_stateless}
Stateless fault detection represents a fundamental approach to distributed fault detection that focuses on instantaneous analysis of system behavior. In this paradigm, the detection process operates without maintaining historical information about previous residuals or system states -- there is no monitoring time-window. Instead, each detection decision is made based solely on the current residual error at a specific time, providing computational simplicity and rapid response to abrupt fault conditions.

\paragraph{Residual Generation}

The core mechanism of stateless detection begins with residual generation. At each local node $i$, the residual is defined as the difference between the actual measurement and the estimated output:
\begin{equation}
	r_i(t) = y_i(t) - C_i\hat{x}_i(t|t).
\end{equation}

Under fault-free conditions, this residual is primarily influenced by system and measurement noise. However, when faults occur, they introduce additional components that shift the residual statistics in detectable ways.

\paragraph{Residual Analysis}

Given the error dynamics established previously, the residual at sensor $i$ (which may include a possible fault $f_i$) at time $t$ can be expressed as
\begin{equation} \label{eq_residual}
	r_i(t) = C_i\hat{A}e_i(t-1) + C_i\zeta_i(t) + \mu_i(t) + f_i(t),
\end{equation}
where $\hat{A} := (W \otimes A - KD_C(W \otimes A))$ represents the effective system matrix for the error dynamics, and $\zeta_i(t)$ encompasses various noise and interaction terms.

\paragraph{Observer Gain Design for Fault Isolation}

For effective detection and isolation of faults at specific nodes, the gain matrix $K$ must be carefully designed to ensure that faults at one node have minimal impact on the residuals at other nodes. This is achieved by satisfying the constraint:
\begin{equation} \label{eq_const_K}
	|C_i^{\top}K_iC_j| \leq \varepsilon|1 - C_j^{\top}K_jC_j|, \quad \forall j \in \mathcal{N}_i, j \neq i,
\end{equation}
for sufficiently small $\varepsilon < 1$ as a design parameter. This constraint ensures that the fault-related term $C_i^{\top}K_iC_jf_j$ in the residual $r_i(t)$ due to a fault $f_j$ at node $j$ is scaled down by a factor of $\varepsilon$ compared to the fault-related term in the residual $r_j(t)$ of sensor $j$ itself. This isolation property is crucial for determining which specific node has experienced a fault. The gain design can be formulated as a Linear Matrix Inequality (LMI) problem \cite{ecc_attack,tnse_attack}.

\paragraph{Statistical Hypothesis Testing}

The threshold-based detection operates on statistical hypothesis testing principles. In the absence of faults, the residual follows a zero-mean Gaussian distribution whose variance is determined by system and measurement noise characteristics. When a fault occurs, the residual becomes biased, following a Gaussian distribution with non-zero mean.

As an example consider the observer in \cite{khan2014collaborative,tcns20,tnse_attack}. Given the noise variance $\Sigma_\nu$ and $\Sigma_\mu$ and the residuals $r_i(t)$ from Eq.~\eqref{eq_residual}, the detection threshold for a detection-level ${m \in \mathbb{R}_{>0}}$ is ,
\begin{align} \label{eq_thresold}
	\theta_p \coloneqq m\Sigma_r^i,~\Sigma_r^i \coloneqq |C_i|\|\Sigma_e\|_2 + \Sigma_\mu^i,
\end{align}
where ${\kappa = \mbox{erf}(\frac{m}{\sqrt{2}})}$
is detection probability (with $\mbox{erf}(\cdot)$ as the \textit{Gauss error function}) and $\|\Sigma_e\|_2$ from \eqref{eq_sigma_e}.

\begin{figure}[t]
	\centering
	\includegraphics[width=3.5in]{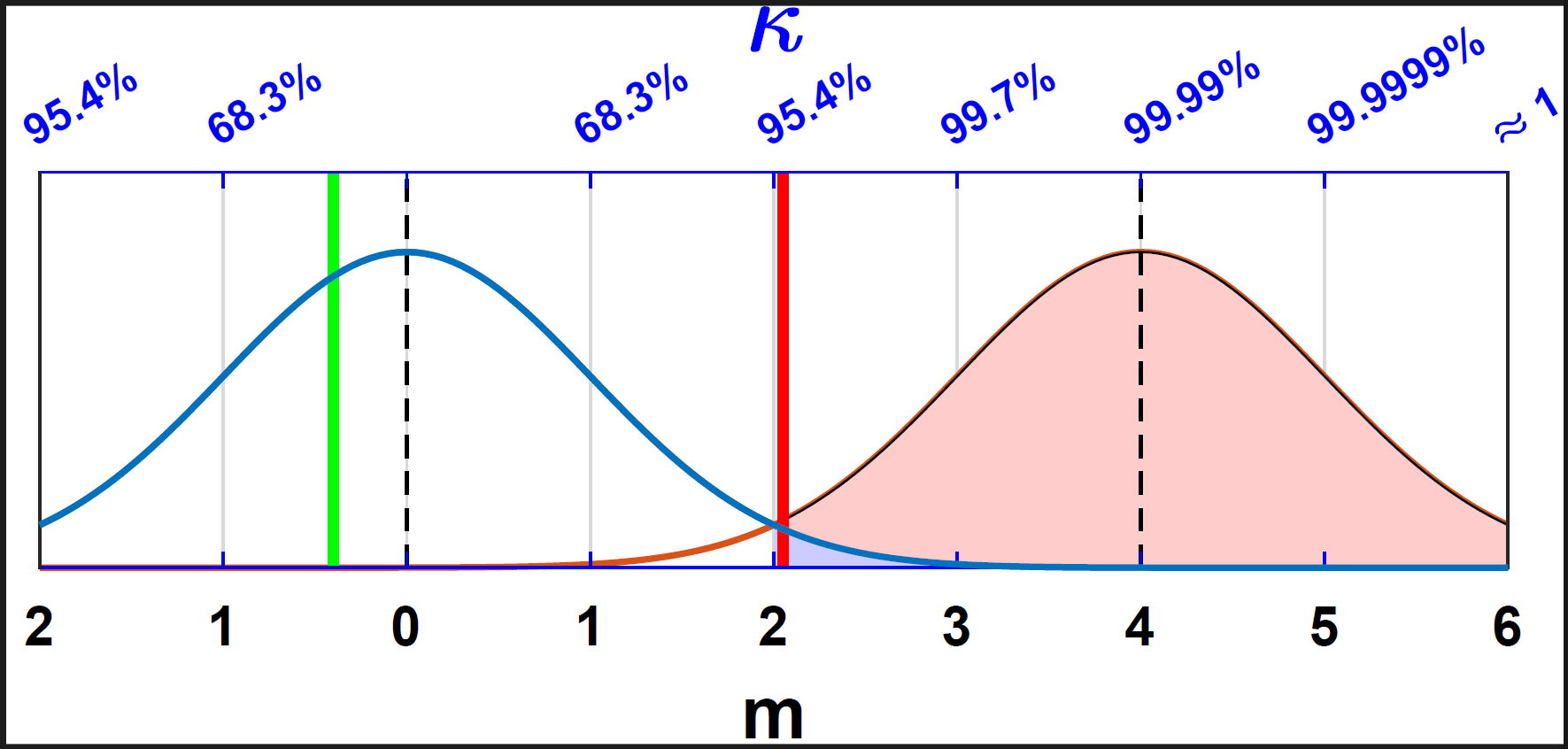} \vspace{-0.3cm}
	\caption{This figure shows the confidence intervals for the normalized residual in the fault-free case (blue curve) and the faulty case (red curve). Each $m$ value in Eq. \eqref{eq_thresold} is associated with a confidence interval and implies a probability threshold $\kappa$ associated with the Gaussian distribution for the residual.  Two example red and green lines are associated with two values of the normalized residual $\frac{r_i(t)}{\Sigma_r^i}$ via Eq. \eqref{eq_thresold}. Following the maximum-likelihood testing, the threshold is defined as the intersection (midpoint) of the two probability distributions, where the residual is associated with the faulty distribution (red curve).
	}
	\label{fig_normal_dist}
\end{figure}

By applying a statistical hypothesis test with these two distributions (fault-free and faulty), we can determine the probability that an observed residual indicates a fault. Specifically, if the magnitude of the residual $|r_i(t)|$ exceeds the threshold $m\Sigma_r^i$, then a fault is detected with probability $\kappa$, while the probability of a false alarm is $1-\kappa$. This is better illustrated in Fig.~\ref{fig_normal_dist}.

\paragraph{Threshold Selection and Performance Trade-offs}

The parameter $m$ in the threshold definition directly controls the sensitivity of the detection system, creating an explicit trade-off between detection sensitivity and false alarm rates. Larger values of $m$ result in higher thresholds, reducing false alarms but potentially delaying fault detection or missing smaller faults. Conversely, smaller values of $m$ increase detection sensitivity but at the cost of more frequent false alarms.

For practical implementations, $m$ is typically chosen to achieve a desired false alarm rate. Common threshold configurations use integer values of $m$, with each value corresponding to a specific detection probability $\kappa$ and false alarm rate as shown in Table~\ref{tab_kappa}. For instance, $m=2$ corresponds to a 95.4\% detection probability with a 4.6\% false alarm rate, while $m=3$ provides a 99.7\% detection probability with only a 0.3\% false alarm rate.

\begin{table} [h]
	\centering
	\caption{Different threshold probabilities $\kappa$ for integer $m$ in Eq. \eqref{eq_thresold}. }
	\label{tab_kappa}
	\begin{tabular}{|c|c|c|c|c|}
		\hline
		$\frac{m}{2}$& $1$ & $2$ & $3$ & $4$  \\
		\hline
		Threshold probability $\kappa$ &  $68.3\%$ & $95.4\%$ & $99.7\%$  & $99.99\%$ \\
		\hline
		False alarm rate $\varkappa = 1-\kappa$ & $31.7\%$  &$4.6\%$& $0.3\%$ &$0.01\%$  \\
		\hline
		\hline
	\end{tabular}
\end{table}

These statistically designed thresholds provide a rigorous foundation for fault detection decisions, allowing system designers to explicitly balance detection sensitivity against false alarm resilience based on specific application requirements and operating contexts.

\subsection{Statefull Detection}
Unlike stateless approaches that consider only instantaneous measurements, stateful fault detection maintains historical information about system behavior over time. This temporal integration allows detection algorithms to identify subtle or developing faults that might not be apparent from single-point measurements. By monitoring residual patterns across a defined time window, stateful detection can achieve higher sensitivity and reduced false alarm rates, particularly for gradually evolving fault conditions.

\paragraph{Time-Window Distance Measures}

The foundation of stateful detection lies in quantifying deviations between expected and actual system behavior over an extended time period. This is accomplished through distance measures that aggregate residual information across multiple time steps. The basic distance measure for a sliding time window of length $\mc{T}$ is defined as
\begin{equation} \label{eq_z}
	\iota_i^\mc{T}(t) = \sum_{m=t-\mc{T}+1}^t \frac{r_i(m)^2}{\Sigma_r^i}.
\end{equation}

This measure effectively normalizes each squared residual by its expected variance and sums these values over the time window. The normalization ensures that the measure appropriately accounts for the anticipated noise characteristics of each measurement, while the summation captures persistent deviations that may indicate fault conditions.

\paragraph{Statistical Properties for Detection}

The statistical properties of this distance measure provide a rigorous foundation for fault detection decisions. Under fault-free conditions, the summation of squared normalized random variables from a normal distribution follows a Chi-squared distribution with $\mc{T}$ degrees of freedom (denoted by $\chi^2_\mc{T}$), with an expected value of $\mathbb{E}(\iota_i^\mc{T}) = \mc{T}$ \cite{survey_sandberg,Bausch2013OnTE}.

This property enables the application of well-established statistical tests to evaluate how well the observed residuals conform to their expected distribution. When system behavior deviates from nominal conditions due to faults, the distance measure tends to increase beyond what would be expected from noise alone, providing a statistical basis for fault detection.

\paragraph{Threshold Design for Chi-Squared Detectors}

Similar to stateless detection, stateful approaches require carefully designed detection thresholds that balance sensitivity against false alarm rates. For a Chi-squared detector with a desired false alarm rate of $1-\kappa$, the appropriate threshold is given by
\begin{equation} \label{eq_theta_T}
	\theta_\kappa^\mc{T} = 2\Gamma^{-1}(\kappa,\frac{\mc{T}}{2}),
\end{equation}
where $\Gamma^{-1}(\cdot,\cdot)$ denotes the inverse regularized lower incomplete gamma function. If the computed $\chi^2_\mc{T}$ detector value exceeds this preset threshold $\theta_\kappa^\mc{T}$, a fault is detected with probability $\kappa$.

The relationship between the false alarm rate and threshold can be expressed as
\begin{equation}
	1-\kappa = 1-\frac{\gamma(\frac{\iota_i^T}{2},\frac{\mc{T}}{2})}{\Gamma(\frac{\mc{T}}{2})},
\end{equation}
where $\gamma(\cdot,\cdot)$ denotes the lower incomplete gamma function.

\paragraph{Weighted Distance Measures}

A significant enhancement to the basic time-window approach is the incorporation of temporal weighting that places greater emphasis on recent measurements \cite{umsonst2019tuning}. This weighted distance measure is defined as
\begin{equation} \label{eq_z2}
	\bar{\iota}_i(t) = \sum_{m=t-\mc{T}+1}^t \varrho^{t-m} \frac{r_i(m)^2}{\Sigma_r^i},
\end{equation}
where $0 < \varrho \leq 1$ serves as a weighting factor that determines how rapidly the influence of past measurements diminishes. This detector, known as the weighted sum of Chi-squared distributions \cite{Bausch2013OnTE}, offers improved sensitivity to recent changes in system behavior while still considering historical context.

The corresponding detection threshold for this weighted measure is
\begin{equation}	\label{eq_theta_T2}
	\theta_\kappa^\varrho = 2\Gamma^{-1}\left(\kappa,\frac{1-\varrho^{T}}{2-2\varrho}\right)
\end{equation}
with the false alarm rate relationship given by
\begin{equation} \label{eq_p_Tmu}
	1-\kappa = 1-\frac{\gamma (\frac{\iota_i^\mc{T}}{2},\frac{\mc{T}}{2})}{\Gamma(\frac{\mc{T}}{2})},~1-\kappa = 1-\frac{\gamma (\frac{\overline{\iota}_i}{2},\frac{1-\varrho^{\mc{T}}}{2-2\varrho})}{\Gamma(\frac{1-\varrho^{\mc{T}}}{2-2\varrho})}. 
\end{equation}

\paragraph{Design Considerations and Trade-offs}

Statefull detection approaches introduce additional design parameters beyond those in stateless methods, creating expanded opportunities for performance optimization:

\begin{itemize}
	\item \textbf{Window length $\mc{T}$:} Longer windows provide greater noise suppression and sensitivity to subtle faults, but increase detection latency and computational requirements. The window length should be chosen based on the time scale of expected faults and the desired balance between sensitivity and responsiveness;
	
	\item \textbf{Weighting factor $\varrho$:} Values closer to $1$ result in more uniform weighting across the window, while smaller values emphasize recent measurements more strongly. This parameter allows the detector to be tuned for different fault development rates, from abrupt changes to gradual drift; and
	
	\item \textbf{Detection probability $\kappa$:} As with stateless detection, this parameter directly influences the trade-off between false alarms and missed detections, though stateful methods generally achieve better performance at the same false alarm rate compared to stateless approaches.
\end{itemize}

The integration of temporal information in stateful detection provides a powerful complement to instantaneous stateless methods. By leveraging both approaches in combination, distributed fault detection systems can achieve robust performance across a wide range of fault scenarios, from sudden, large-magnitude faults to subtle, slowly developing anomalies.

\subsection{Relevant Literature}
The decentralized nature of modern cyber-physical systems has stimulated extensive research into distributed FDI algorithms. These approaches leverage collaborative processing and consensus mechanisms across multiple agents or sensors to ensure system reliability without relying on centralized architectures. This section provides an overview of key research directions and innovations in distributed FDI, categorized by their primary focus areas and methodological approaches.

\paragraph{Resource-Efficient Detection Mechanisms}

Communication and computational resources represent critical constraints in many CPS deployments. Several research directions have emerged to address these limitations:

\begin{itemize}
	\item \textbf{Event-triggered FDI mechanisms} \cite{davoodi2019,10398421,wang2024distributed} optimize communication efficiency by transmitting information only when significant changes occur or when predefined conditions are met. These approaches substantially reduce bandwidth requirements while maintaining detection performance, making them particularly valuable for wireless sensor networks with energy constraints.
	
	\item \textbf{Low-complexity detection algorithms} \cite{khan2022design} focus on minimizing computational demands through streamlined processing techniques, enabling implementation on resource-constrained embedded platforms common in distributed sensing applications.
\end{itemize}

\paragraph{Resilience Against System Uncertainties}

Practical CPS deployments invariably encounter various forms of uncertainty that can compromise detection reliability. Several research streams have developed robust approaches to maintain detection performance despite these challenges:

\begin{itemize}
	\item \textbf{Model uncertainty resilience} \cite{6898006} enables detection algorithms to function effectively despite imperfect system models, parameter variations, or unmodeled dynamics. These approaches typically incorporate adaptive mechanisms or robust design techniques that account for bounded uncertainty; and
	
	\item \textbf{Communication imperfection handling} \cite{davoodi2018} addresses practical networking challenges such as packet drops, transmission delays, and intermittent connectivity. These methods maintain detection functionality despite unreliable information exchange among distributed nodes.
\end{itemize}

\paragraph{Architectural and System-Level Approaches}

Beyond algorithmic innovations, several research directions have explored broader architectural approaches to distributed fault detection:

\begin{itemize}
	\item \textbf{Heterogeneous system designs} \cite{davoodi2013distributed,davoodi2017} accommodate networks composed of diverse sensor types, varying processing capabilities, and mixed measurement modalities. These approaches often employ unknown input observer design techniques to handle the complexity introduced by system heterogeneity; and
	
	\item \textbf{Integrated control and detection frameworks} \cite{davoodi2016simultaneous} simultaneously address the challenges of distributed control and fault detection, creating unified approaches that maintain both control performance and fault detection capabilities in a coordinated manner.
\end{itemize}

\paragraph{Advanced Detection Paradigms}

Several innovative detection paradigms have emerged that extend distributed fault detection beyond traditional approaches:

\begin{itemize}
	\item \textbf{Probabilistic threshold-based methodologies} have been developed for various system classes, including the following:
	\begin{itemize}
		\item Full-rank dynamical systems \cite{tcns20}, where complete state information can be reconstructed from available measurements;
		\item Rank-deficient systems \cite{ijc2023}, where structural limitations prevent complete state reconstruction from any single measurement set; and
		\item Nonlinear dynamical systems \cite{DONG2024115243,nazifi2024hybrid}, which incorporate detection mechanisms that account for nonlinear system behaviors.
	\end{itemize}
	
	\item \textbf{Data-driven approaches} leverage advances in machine learning and statistical analysis:
	\begin{itemize}
		\item Clustering-based methods \cite{7402814} identify and group similar fault patterns, enabling more nuanced detection and classification;
		\item Machine-learning techniques \cite{jan2021distributed} employ supervised and unsupervised learning to identify complex fault signatures without requiring explicit system models; and
		\item Data-driven frameworks \cite{10189367,DING2023105718,yazdanpanah2024data} develop detection algorithms directly from system data, particularly valuable when accurate analytical models are unavailable or prohibitively complex.
	\end{itemize}
\end{itemize}

These diverse research directions collectively advance the state-of-the-art in distributed fault detection, providing system designers with a rich toolkit of methodologies that can be selected and combined based on specific application requirements.

\section{Applications} \label{sec_app}
The theoretical foundations and algorithmic approaches detailed in previous sections find concrete expression in diverse cyber-physical system applications. This section examines how distributed estimation, filtering, and fault detection techniques are implemented across four key domains, each presenting unique challenges and requirements that highlight different aspects of distributed algorithm design.

We begin by exploring applications in smart grids and power networks (Section~\ref{sec_grid}), where distributed algorithms enable monitoring and control of geographically dispersed energy infrastructure. Here, we examine how these techniques support renewable energy integration, economic dispatch optimization, and fault management in microgrids -- applications where system scale and reliability requirements make centralized approaches impractical.

Next, we investigate social systems (Section~\ref{sec_social}), where distributed algorithms monitor and analyze human interaction dynamics across social networks. This domain showcases how techniques developed for physical systems can be adapted to track opinion dynamics, information propagation, and collective behaviors, revealing the versatility of distributed estimation approaches beyond traditional engineering applications.

The third application domain focuses on target tracking and localization (Section~\ref{sec_target}), where distributed algorithms enable multiple sensors to collaboratively track moving objects. This application area highlights the importance of measurement fusion, dynamic model selection, and real-time processing in scenarios ranging from autonomous vehicle navigation to aerial surveillance systems.

Finally, we examine intelligent transportation systems (Section~\ref{sec_ITS}), where vehicle-to-vehicle and vehicle-to-infrastructure communications create new opportunities for distributed coordination. Applications including collaborative localization, traffic state estimation, and secure vehicle platooning demonstrate how distributed algorithms enhance safety, efficiency, and reliability in increasingly connected transportation networks.

Through these application examples, we illustrate how theoretical concepts and algorithmic innovations translate into practical solutions for complex monitoring and control challenges across diverse cyber-physical domains.

\subsection{Smart Grid and Power Network} \label{sec_grid}
Modern power networks and smart grids represent quintessential examples of large-scale cyber-physical systems that benefit substantially from distributed monitoring and control approaches \cite{usman_smc:08}. As these infrastructures evolve to incorporate renewable energy sources, distributed generation, and advanced demand-response capabilities, the need for decentralized estimation and fault detection becomes increasingly critical. The dynamical system representations of power grid systems commonly follows the state-space representation in Section \ref{sec_lindyn} given by Eqs.~\eqref{eq_A}-\eqref{eq_C}, see details in \cite{zhao2019power}.

\paragraph{Challenges in Modern Power Infrastructure}

The integration of renewable energy sources and electric vehicles \cite{manousakis2023integration,hasan2023smart,yu2022electric,dsp} has fundamentally transformed traditional power networks, introducing several key challenges that distributed algorithms are uniquely positioned to address:

\begin{itemize}
	\item \textbf{Intermittent generation:} Renewable sources like wind and solar exhibit inherent variability and limited predictability, requiring sophisticated estimation techniques to forecast generation patterns and balance loads;
	
	\item \textbf{Geographic dispersion:} Generation assets are increasingly distributed across wide geographic areas, making centralized monitoring impractical;
	
	\item \textbf{Heterogeneous components:} Modern grids incorporate diverse technologies with varying characteristics, communication capabilities, and control requirements; and
	
	\item \textbf{Critical reliability requirements:} Power infrastructure must maintain high availability while preventing cascading failures that could affect large service areas.
\end{itemize}

\paragraph{Distributed Monitoring Architecture}

\begin{figure} 
	\centering
	\includegraphics[width=3in]{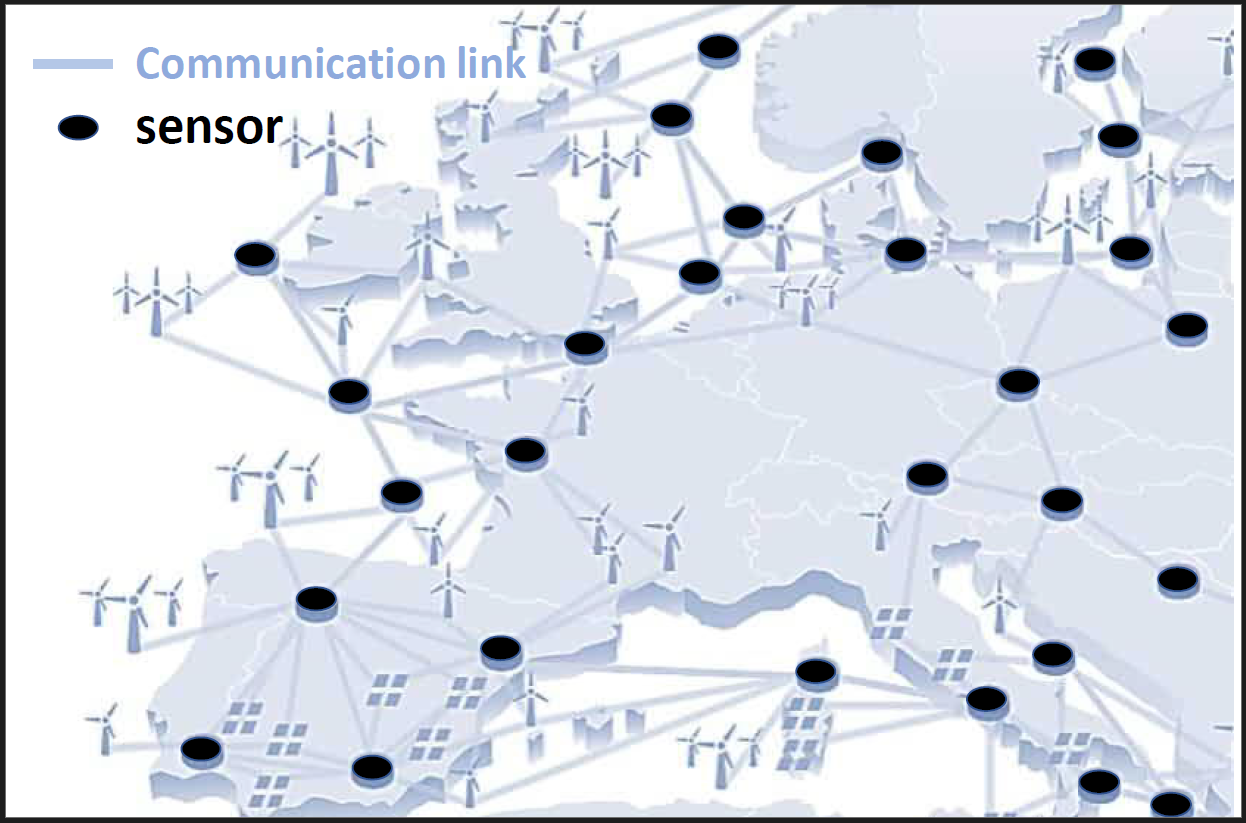}
	\caption{This figure shows a sample sensor network monitoring the geographically distributed power grid. Distributed estimation techniques play a key role in monitoring such large-scale systems using distributed fusion algorithms with no need for a centralized data-processing unit.
	} \label{fig_grid}
\end{figure}

Fig.~\ref{fig_grid} illustrates a representative distributed monitoring architecture for renewable-energy grids incorporating wind farms, solar installations, and conventional generation. In this configuration, a geographically distributed sensor network enables localized processing and collaborative state estimation through distributed algorithms.

This approach delivers several key advantages compared to traditional centralized architectures:

\begin{itemize}
	\item \textbf{Enhanced resilience:} The distributed processing architecture eliminates single points of failure, preventing global system shutdown when individual components malfunction;
	
	\item \textbf{Localized fault management:} Distributed fault detection enables precise isolation of faulty assets \cite{ijc2023}, containing problems before they propagate through the network;
	
	\item \textbf{Scalable deployment:} The distributed architecture naturally scales with system expansion, accommodating new generation sources without requiring fundamental redesign; and
	
	\item \textbf{Efficient communication:} By processing data near its source, the system minimizes communication bandwidth requirements and reduces latency.
\end{itemize}

\paragraph{Distributed Estimation Applications in Power Systems}

The power industry has adopted various distributed estimation approaches to address specific operational challenges:

\begin{itemize}
	\item \textbf{Renewable generation forecasting:} Distributed filtering techniques help predict generation patterns from variable renewable sources by combining local measurements from wind turbines and solar panels with meteorological data \cite{ranjan2022literature,tshenyego2021wide};
	
	\item \textbf{Power quality monitoring:} Distributed processing enhances the accuracy of voltage and frequency estimation across the grid, improving power quality assessment capabilities \cite{camsap11,9359483};
	
	\item \textbf{Network state estimation:} Various distributed approaches have been developed for monitoring the operating condition of power networks, including the following:
	\begin{itemize}
		\item Distributed estimation methods for control centers \cite{PASQUALETTI2012747};
		\item Distributed maximum a posteriori estimation \cite{SUN201627};
		\item Distributed Kalman filtering in batch-mode regression form \cite{NGUYEN2021107510};
		\item Distributed unscented information filtering \cite{Yang2019power}; and
		\item Reliable estimation over unreliable communication networks \cite{Rana2017power}.
	\end{itemize}
\end{itemize}

\paragraph{Economic Dispatch and Energy Management}

Beyond physical state monitoring, distributed algorithms play a crucial role in optimizing power system economics through distributed energy management and economic dispatch. These applications ensure that electricity generation and consumption are balanced economically and optimally \cite{kar6934985,lcss_energy}.

The economic dispatch problem is mathematically formulated as an optimization of generation costs:
\begin{equation}
	\begin{aligned}
		\min_{\mathbf{x}} ~~ & F(\mathbf{x}) = \sum_{i=1}^{n} f_i(x_i) \\
		\text{s.t.} ~~ & \sum_{i=1}^{n} x_i = P_{mis},
	\end{aligned}
\end{equation}
where state $x_i$ represents the generated power at node $i$ and $P_{mis}$ is the power mismatch between the generated power and demand. The cost function typically takes a quadratic form:
\begin{equation}
	f_i(x_i) = \gamma_i x_i^2 + \beta_i x_i + \alpha_i,
\end{equation}
with parameters $\gamma_i$, $\beta_i$, and $\alpha_i$ defined based on the generator type (fueled by oil, gas, coal, just to mention a few) as documented in \cite{wood2013power,yang2013consensus}.

Consensus-based distributed estimation protocols, such as the consensus + innovation approach \cite{kar6934985}, can iteratively solve this optimization problem in a fully distributed manner. This decentralized approach allows individual generators to optimize their output based on local costs while accounting for neighboring units' output levels \cite{7913707,ZHOU2022119641}. Furthermore, integrating interval observers for fault diagnosis significantly enhances the robustness of distributed economic dispatch against potential system anomalies \cite{10843114}.

\paragraph{Microgrid Fault Management}

In microgrids -- localized energy systems composed of multiple interconnected distributed generation units -- distributed fault detection and estimation techniques are especially valuable. These approaches enable quick localization and isolation of faults, preventing cascading failures that could compromise the entire microgrid \cite{9265410,Sistani,LIU2025106216}. The distributed nature of these techniques aligns perfectly with the inherently decentralized architecture of microgrids, providing natural fault containment boundaries and facilitating autonomous operation during main grid disconnection.

\subsection{Social Systems} \label{sec_social}
Beyond physical and engineered systems, distributed estimation and detection techniques find powerful applications in social systems -- complex networks of interacting individuals whose collective behavior generates emergent phenomena of significant societal importance. This section explores how the distributed algorithms examined earlier can be adapted to monitor, analyze, and understand social dynamics across diverse contexts.

\paragraph{Diversity of Social Networks and Dynamics}

Social networks manifest across an extraordinarily wide spectrum of contexts, spanning human interactions, animal communities, economic systems, market behaviors, online platforms, citation networks, and numerous other domains \cite{jstsp14}. These networks serve as fundamental infrastructures for understanding how information, influence, and behaviors propagate through interconnected social entities.

The underlying social phenomena of interest exhibit similar diversity, including the following:
\begin{itemize}
	\item \textbf{Collective decision processes:} Voting behaviors, consensus formation, and group judgment;
	\item \textbf{Coordinated movements:} Flocking, herding, and synchronized behaviors;
	\item \textbf{Information diffusion:} Rumor propagation, viral content spread, and information cascades;
	\item \textbf{Economic patterns:} Stock price movements, market trends, and consumer behaviors; and
	\item \textbf{Community evolution:} Formation, growth, and dissolution of social groups.
\end{itemize}

Each state within these social dynamics typically represents an opinion, belief, preference, or behavioral characteristic of an individual actor or community member. The interconnected nature of these states -- with each potentially influencing and being influenced by others -- creates complex dynamical systems that present unique monitoring and analysis challenges. 

\paragraph{Cyber-Social Monitoring Architecture}

\begin{figure} 
	\centering
	\includegraphics[width=3.5in]{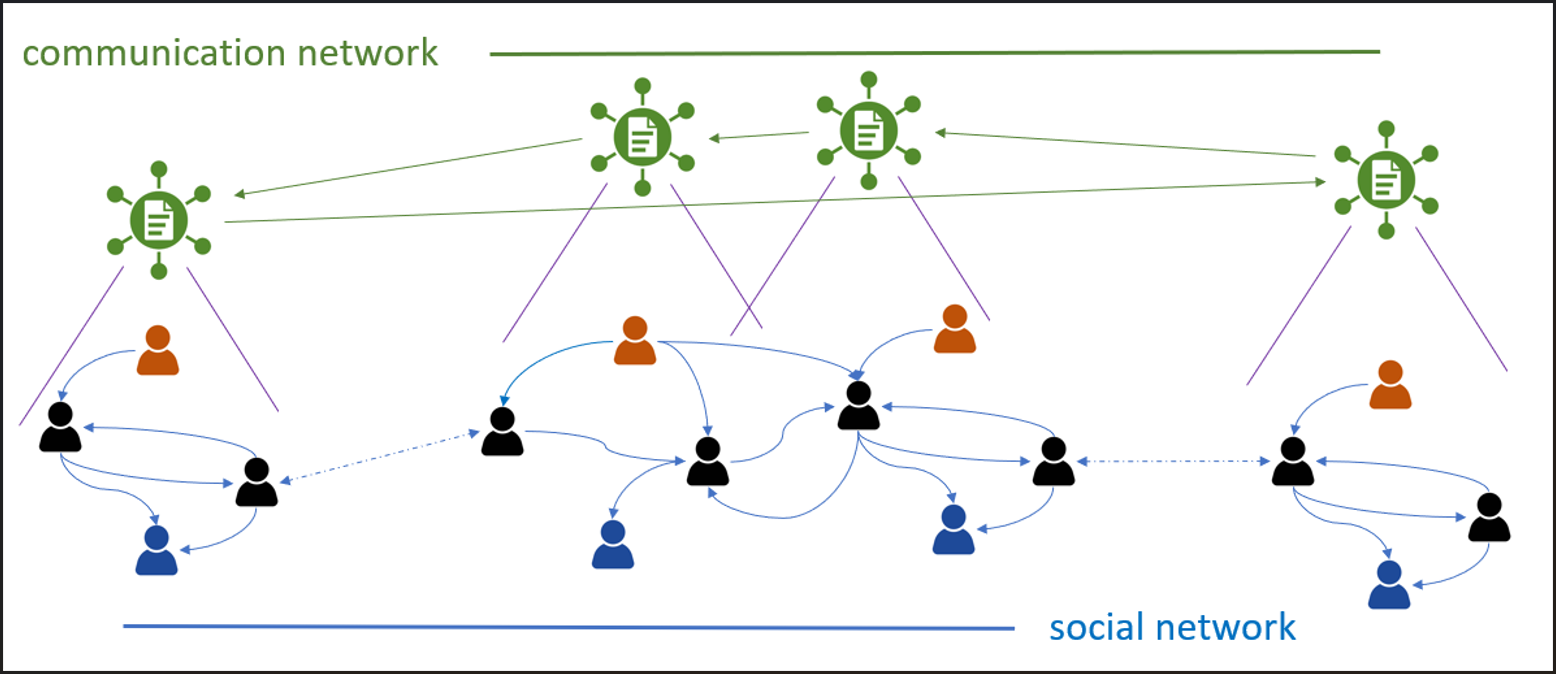}
	\caption{This figure illustrates a social system monitored by a multi-agent network. The social system has the interactions between individuals depicted by blue links, where some individuals might be just influencers (in red) and others just followers (in blue). Additionally, agents depicted in green, gather data from the selected individuals. These collectors then exchange their estimated data through a dedicated communication network, illustrated by green lines and make localized decisions via proper distributed algorithms.
	} \label{fig_social}
\end{figure}

Fig.~\ref{fig_social} illustrates an innovative cyber-social system architecture that leverages distributed estimation for monitoring social dynamics. In this framework, a network of computational agents -- whether autonomous monitoring entities or specialized algorithms implemented within digital platforms \cite{isj_cyber} -- observes the states of individuals across a social network and collaboratively processes this information. 

The cyber layer (green nodes and connections) forms a processing infrastructure that monitors the social layer (blue and red nodes with blue connections), where
\begin{itemize}
	\item Red nodes represent influential actors who primarily impact others' states;
	\item Blue nodes represent more passive actors whose states are predominantly influenced by others;
	\item Green monitoring agents collect data from selected individuals; and
	\item Green connections enable information exchange among monitoring agents.
\end{itemize}

This architecture enables distributed analysis of social phenomena without requiring centralized data collection, preserving privacy while providing meaningful insights into collective behaviors and opinion dynamics \cite{khan2014collaborative,pequito_gsip}.

\paragraph{Social Opinion Dynamics Models}

The evolution of states (opinions, beliefs, preferences) across social networks follows various models that capture how individuals influence each other over time. Several well-established linear models include the following:
\begin{itemize}
	\item \textbf{Friedkin-Johnson social influence networks} \cite{Friedkin:1999vd}, which balance individual stubbornness with social influence;
	\item \textbf{The French model} \cite{French}, which focuses on interpersonal pressure toward uniformity; and
	\item \textbf{The DeGroot model} \cite{dong2023social}, which implements weighted averaging of neighboring opinions.
\end{itemize}

These models are typically linear as in Section~\ref{sec_lindyn} and share structural similarities with the consensus dynamics described in Section~\ref{sec_consensus}, though they typically incorporate additional parameters to capture the complex nature of social influence. A representative state update equation in these models takes the form \cite{isj_cyber,jstsp14}:
\begin{equation}
	x_i(t+1) = \sum_{j \in \mathcal{N}_i \cup \{i\}} a_{ij} x_j(t),
\end{equation}
where $x_i$ represents the state (opinion) of individual $i$ at time step $t$, and $0 \leq a_{ij} \leq 1$ denotes the influence weight that neighboring individual $j$ exerts on individual $i$'s state.

\paragraph{Distributed Estimation and Detection Approaches}

The complexity and diversity of social systems have inspired numerous specialized adaptations of distributed estimation and detection techniques:
\begin{itemize}
	\item \textbf{Trust-weighted distributed filtering} \cite{matei2009composite} adjusts the influence of different information sources based on their assessed reliability, similar to how individuals in social networks assign varying credibility to different peers;
	
	\item \textbf{Discrete distribution estimation} \cite{sarwate2014distributed} enables monitoring of categorical beliefs or preferences that don't follow continuous state models;
	
	\item \textbf{Interaction graph-based learning} \cite{8550320,jstsp14,khan2014collaborative} leverages the structure of social connections to improve estimation accuracy;
	
	\item \textbf{Observability-based tracking approaches} examine opinion dynamics through various interaction models:
	\begin{itemize}
		\item Ergodic interaction models \cite{ravazzi2021ergodic} that capture time-varying influence patterns;
		\item Cooperative/antagonistic interaction frameworks [296] that differentiate between positive and negative social influences; and
		\item Structural analysis methods \cite{pequito_gsip,isj_cyber} that identify critical monitoring points in complex social networks.
	\end{itemize}
	
	\item \textbf{Gossip-based learning algorithms} \cite{jiang2013understanding,wai2016active,xing2023community} mimic the natural information diffusion processes in social networks, where information spreads through pairwise interactions; and
	
	\item \textbf{Bias detection algorithms} \cite{icas_social,tnse_attack} identify potentially manipulated or misleading information flows within social monitoring systems.
\end{itemize}

These approaches demonstrate how distributed estimation and detection techniques can be effectively adapted to the unique challenges of social system monitoring, enabling insights into collective behaviors while respecting the privacy and autonomy of individual actors. As social interactions increasingly occur across digital platforms that generate vast quantities of behavioral data, distributed approaches to social system monitoring will continue to gain importance for both research and practical applications.

\subsection{Target Tracking and Localization} \label{sec_target}
Accurate determination of object positions and trajectories represents one of the most fundamental and widely applicable tasks in cyber-physical systems. This section examines how distributed estimation and detection techniques enable robust tracking and localization across diverse application domains, from autonomous vehicles and robotics to smart infrastructure and sensor networks.

\paragraph{Fundamental Concepts and Challenges}

Localization and tracking encompass related but distinct objectives in cyber-physical systems:

\begin{itemize}
	\item \textbf{Localization} determines the position of an object or agent within a defined environment, serving as the foundational capability for navigation, interaction, and coordination; and
	
	\item \textbf{Target tracking} extends localization by continuously monitoring an object's state (position, velocity, and potentially acceleration) over time, enabling trajectory prediction and informed decision-making based on anticipated movements.
\end{itemize}

Both capabilities represent quintessential cyber-physical tasks that integrate physical sensing with computational algorithms to achieve robust performance under uncertainty. The distributed nature of modern sensing systems, where multiple sensors observe a target from different locations, creates both opportunities and challenges for estimation accuracy, reliability, and scalability.

\begin{figure} [t]
	\centering
	\includegraphics[width=3.5in]{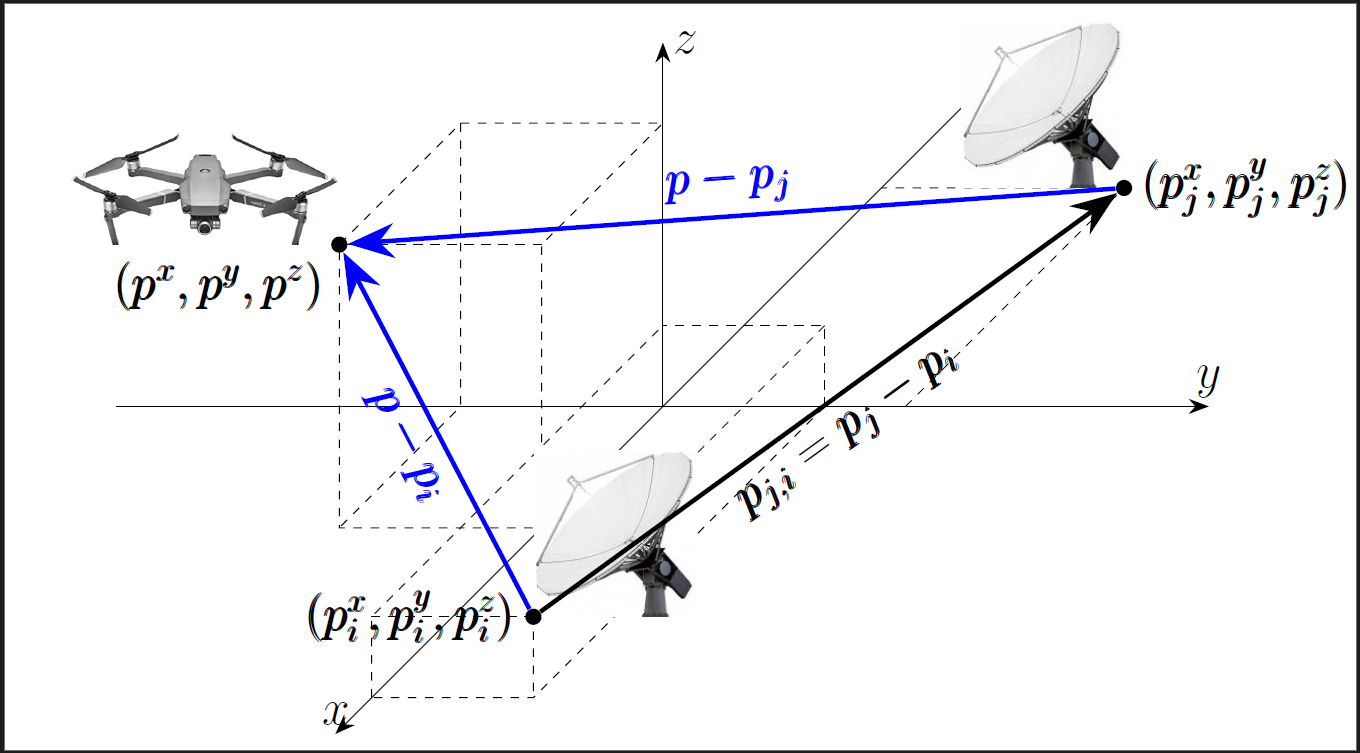}
	\caption{A group of distributed static sensors (radars) receive a beacon signal from a target (drone) and locally track this target via distributed estimation and localized fault detection techniques.   }
	\label{fig_target}
\end{figure}

In this section, we focus specifically on relative localization, which determines an entity's position relative to reference points or other entities within the network. This approach leverages geometric principles such as triangulation \cite{5352246} and multilateration \cite{scl_target} based on distance measurements or angle calculations \cite{mohammadi2015distributed} from known locations, as illustrated in Fig.~\ref{fig_target}.

\paragraph{Time-Difference-of-Arrival Measurement Framework}

As a representative example of distributed localization, we examine time-difference-of-arrival (TDOA) measurements. This approach determines the target position by comparing the arrival times of signals at different sensors \cite{vtc22}, eliminating the need for precise time synchronization between the target and sensing network.

Consider a network of $n$ sensors with positions $\mathbf{p}_i = (p_{x,i}; p_{y,i}; p_{z,i})$ performing localization of a mobile target. The measurement process follows these steps:

\begin{enumerate}
	\item Each sensor receives a beacon signal (with known propagation speed $c$) from the mobile target;
	
	\item Each sensor records the time $t_i = \frac{1}{c}\|\mathbf{p}(t) - \mathbf{p}_i(t)\|$ corresponding to the distance between itself and the target; 
	
	\item Sensors share their time measurements and positions with neighboring sensors in set $\mathcal{N}_i$;
	
	\item By subtracting measurements, each sensor computes TDOA values, which convert to range-difference information:
\end{enumerate}

\begin{align}
	\frac{1}{2}&\Big(\|\mathbf{p}(t)-\mathbf{p}_i(t)\|^2-\|\mathbf{p}(t)-\mathbf{p}_{j}(t)\|^2\Big) =\nonumber\\
	&\frac{1}{2} \Big((p_{x,j}-p_{x,i})(2p_x-p_{x,j}-p_{x,i}) \nonumber\\
	&~ +(p_{y,j}-p_{y,i})(2p_y-p_{y,j}-p_{y,i}) \nonumber\\
	&~ + (p_{z,j}-p_{z,i})(2p_z-p_{z,j}-p_{z,i}) \Big) \nonumber\\
	&= C_i \mathbf{x}(t) - \frac{1}{2}(\|\mathbf{p}_j\|^2 - \|\mathbf{p}_i\|^2),
\end{align}
where the observation matrix $C_i$ is defined as
\begin{align}
	C_i=\left(
	\begin{array}{cccccc}
		p_{x,j,i} & p_{y,j,i} &  p_{z,j,i} & 0 & 0 & 0\\
		\vdots & \vdots & \vdots & \vdots & \vdots & \vdots\\
		p_{x,j_{|\mathcal{N}_i|},i} & p_{y,j_{|\mathcal{N}_i|},i} &  p_{z,j_{|\mathcal{N}_i|},i} & 0 & 0 & 0\\
	\end{array} \right),
\end{align}
with $\mathbf{p}_{j,i} := \mathbf{p}_{j}-\mathbf{p}_{i} := (p_{x,j,i}; p_{y,j,i}; p_{z,j,i})$ defined as the relative position vector between sensors $j$ and $i$.

For fixed sensor positions, the bias term $\|\mathbf{p}_j\|^2 - \|\mathbf{p}_i\|^2$ remains constant and known. By incorporating this known bias, the TDOA measurement can be expressed in standard form:
\begin{align}
	\mathbf{y}_i(t) = C_i\mathbf{x}(t) + \boldsymbol{\mu}_i(t),
\end{align}
where $\boldsymbol{\mu}_i(t)$ represents the additive measurement noise.

\paragraph{Target Dynamic Modeling}

In real-world tracking scenarios, target dynamics are generally unknown and must be approximated using mathematical models. The most widely used target motion models follow the general linear form:
\begin{equation}
	\mathbf{x}(t+1) = A\mathbf{x}(t) + G\boldsymbol{\nu}(t),
\end{equation}
where $\mathbf{x}(t)$ represents the target state vector, matrices $A$ and $G$ describe the transition and input dynamics, and $\boldsymbol{\nu}(t)$ represents process noise or random inputs.

Three primary models dominate the literature \cite{roy2006target,gustafsson2002particle,ennasr2020time,ennasr2016distributed,bar2004estimation,li2003survey,scl_target}:

\begin{itemize}
	\item \textbf{Nearly-Constant-Velocity (NCV) Model:} This approach models position and velocity in 3D space with state vector
	\begin{align} \label{eq_pxyz}
		\mathbf{x} = \begin{pmatrix} p_x \\ p_y \\ p_z \\ \dot{p}_x \\ \dot{p}_y \\ \dot{p}_z \end{pmatrix}.
	\end{align}
	
	The transition and input matrices are defined as
	\begin{align} \label{eq_ncv}
		A = \begin{pmatrix} \mathbf{I}_3 & T\mathbf{I}_3 \\ \mathbf{0}_3 & \mathbf{I}_3 \end{pmatrix}, \quad
		G = \begin{pmatrix} \frac{T^2}{2}\mathbf{I}_3 \\ T\mathbf{I}_3 \end{pmatrix},
	\end{align}
	where $\mathbf{I}_3$ and $\mathbf{0}_3$ represent 3×3 identity and zero matrices, and $T$ denotes the sampling interval.
	
	\item \textbf{Nearly-Constant-Acceleration (NCA) Model:} This model incorporates acceleration terms, expanding the state vector to
	\begin{align} \label{eq_paxyz}
		\mathbf{x} = \begin{pmatrix} p_x \\ p_y \\ p_z \\ \dot{p}_x \\ \dot{p}_y \\ \dot{p}_z \\ \ddot{p}_x \\ \ddot{p}_y \\ \ddot{p}_z \end{pmatrix},
	\end{align}
	with corresponding matrices
	\begin{align} \label{eq_nca}
		A = \begin{pmatrix} \mathbf{I}_3 & T\mathbf{I}_3 & \frac{T^2}{2}\mathbf{I}_3 \\ \mathbf{0}_3 & \mathbf{I}_3 & T\mathbf{I}_3 \\ \mathbf{0}_3 & \mathbf{0}_3 & \mathbf{I}_3 \end{pmatrix}, \quad \text{ and } \quad
		G = \begin{pmatrix} \frac{T^2}{2}\mathbf{I}_3 \\ T\mathbf{I}_3 \\ \mathbf{I}_3 \end{pmatrix}.
	\end{align}
	
	\item \textbf{Singer Model:} This probabilistic approach enhances the NCA model by incorporating a maneuvering parameter $\alpha = \frac{1}{\theta}$, where $\theta$ represents the maneuver time constant:
	\begin{align}
		A = \begin{pmatrix} \mathbf{I}_3 & T\mathbf{I}_3 & \frac{\alpha T-1+e^{-\alpha T}}{\alpha^2}\mathbf{I}_3 \\ \mathbf{0}_3 & \mathbf{I}_3 & \frac{1-e^{-\alpha T}}{\alpha}\mathbf{I}_3 \\ \mathbf{0}_3 & \mathbf{0}_3 & e^{-\alpha T}\mathbf{I}_3 \end{pmatrix}, \quad
		G = \begin{pmatrix} \mathbf{0}_3 \\ \mathbf{0}_3 \\ \mathbf{I}_3 \end{pmatrix}.
	\end{align}
\end{itemize}

These models are interrelated and can transform into one another under specific conditions. For instance, the Singer model approaches the NCV model as the maneuver time constant $\theta$ decreases, and converges toward the NCA model as $\theta$ increases \cite{singer1970estimating}. Larger $\theta$ values represent gradual, predictable maneuvers, while smaller values indicate abrupt, evasive movements.

\paragraph{Distributed Tracking Approaches}

The literature on distributed algorithms for localization and target tracking has expanded significantly in recent years, addressing various aspects of the tracking problem:

\begin{itemize}
	\item \textbf{Consensus-based localization and tracking} \cite{5352246,7737071,10321726,ghods2025resilient} leverages the consensus techniques discussed in Section~\ref{sec_cons_est} to enable collaborative position estimation across sensor networks;
	
	\item \textbf{TDOA-based distributed tracking} has been developed for both delay-free networks \cite{ennasr2020time,ennasr2016distributed,TASE_target,panetta2020distributed}  and time-delayed setups \cite{icrom_target,8972461,scl_target}, addressing practical implementation challenges in real-world sensing systems;
	
	\item \textbf{Advanced filtering approaches} for distributed tracking include:
	\begin{itemize}
		\item Distributed maximum likelihood Kalman filters \cite{9599711} that optimize estimation under probabilistic uncertainty models;
		\item Consensus Kalman filters \cite{Safarinejadian} that combine consensus mechanisms with optimal filtering;
		\item Adaptive filters \cite{10584424} that dynamically adjust to changing target dynamics; and
		\item Event-triggered filters \cite{li2022event,gao2019event} that reduce communication overhead by transmitting only significant updates.
	\end{itemize}
	
	\item \textbf{Specialized tracking scenarios} have been addressed through the following approaches:
	\begin{itemize}
		\item Nonholonomic target models \cite{8713859} for targets with constrained motion capabilities;
		\item Asynchronous communication networks \cite{li2020distributed2} that accommodate realistic timing variations;
		\item Tracking under cyber attacks \cite{LIANG201944} to maintain performance despite malicious interference;
		\item Fault-tolerant approaches for leader-follower structures \cite{haeri2023}; and
		\item Belief propagation techniques \cite{xue2025distributed} that leverage probabilistic graphical models for efficient information fusion.
	\end{itemize}
\end{itemize}

These diverse approaches demonstrate how distributed estimation and detection techniques can be effectively applied to the fundamental challenge of target tracking, enabling robust performance across a wide range of sensing conditions, target behaviors, and network configurations.

\subsection{Intelligent Transportation Systems} \label{sec_ITS}
Transportation networks are rapidly evolving from passive infrastructure into dynamic, interconnected cyber-physical systems that actively monitor, manage, and optimize traffic flow. This section examines how distributed estimation and fault detection techniques enable the emerging paradigm of cooperative intelligent transportation systems (ITS), creating safer, more efficient, and more reliable mobility solutions.

\begin{figure} 
	\centering
	\includegraphics[width=4in]{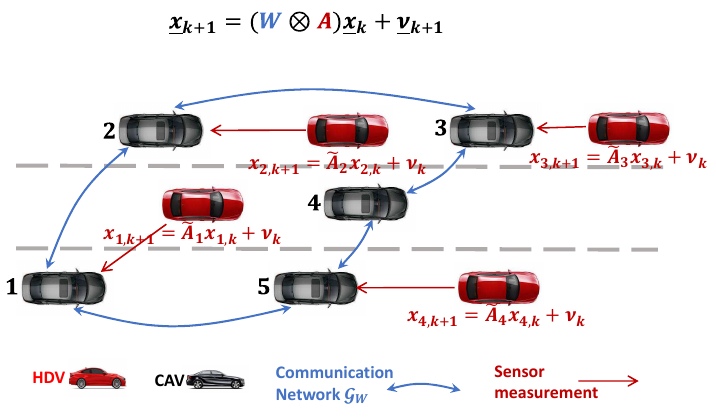}
	\caption{This figure presents a mixed traffic transportation setup including CAVs and HDVs. The goal is to enable each CAV to track the state of other HDVs. Some nearby CAVs take measurements of the state of the HDVs and share it over a communication network with nearby CAVs. Then, a distributed filter is used to estimate the state of HDVs (e.g., position, velocity) via local sensor data and shared information.} \label{fig_mixedtraffic}
\end{figure}

\paragraph{Mixed Traffic Transportation Systems}
Mixed traffic transportation systems consist of connected autonomous vehicles (CAVs) and human-driven vehicles (HDVs) sharing the same roadway, see Fig.~\ref{fig_mixedtraffic}. In these settings, accurate knowledge of other vehicles' states (e.g., positions, velocities) is crucial for safe driving. Traditional centralized estimation methods collect all sensor data at a central unit to infer traffic states, but they face scalability issues and vulnerability to single-point failures. To address these challenges, distributed estimation has emerged as a promising alternative, where individual CAVs estimate the state of HDVs via local sensing and information exchange with neighboring CAVs, see \cite{eurasip,ghanavati2025autonomous,zhang2023privacy,fu2025motion} for details.
To formulate the most general scenario, consider the system-measurement model~\eqref{eq_A}-\eqref{eq_C} as a general model for group of $N$ HDVs. The dynamics of every HDV is modelled by the NCV and NCA dynamics described by Eqs.~\eqref{eq_pxyz}-\eqref{eq_nca}, where the HDV's state is a variable in $\mathbb{R}^m$. The state of HDVs is then tracked by every CAV $i$ denoted by ~$\widehat{\mb{x}}^i_{k|k}\in\mathbb{R}^{N m}$ as the estimate of $\mb{x}_k\in\mathbb{R}^{N m}$. Every CAV uses all the available  measurements over the communication network. Then, the \emph{global} estimate of the states of HDVs is defined as, 
\begin{eqnarray}
	\widehat{\mb{x}}_{k|k}:=
	\left(
	\begin{array}{c}
		\widehat{\mb{x}}^1_{k|k}\\
		\widehat{\mb{x}}^2_{k|k}\\
		\vdots \\
		\widehat{\mb{x}}^n_{k|k}
	\end{array}
	\right)\in\mathbb{R}^{nN m},
\end{eqnarray}
which represents the estimate of the global state of HDVs as a networked system defined as, 
\begin{eqnarray}\label{eq_xx}
	\underline{\mb{x}}_{k}:=
	\left(
	\begin{array}{c}
		\mb{x}_k\\
		\mb{x}_k\\
		\vdots \\
		\mb{x}_k
	\end{array}
	\right) = \mb{1}_n \otimes \mb{x}_k.
\end{eqnarray}
Then, the mixed traffic ITS dynamics associated with $\underline{\mb{x}}_{k}$ is \cite{eurasip}:
\begin{eqnarray}
	\underline{\mb{x}}_{k+1} &=& \mb{1}_n\otimes \mb{x}_{k+1}\nonumber \\
	&=& \mb{1}_n\otimes (A\mb{x}_{k}+\nu_k)\nonumber \\ \label{eq_sys5}
	&=& {(W\otimes A)}\underline{\mb{x}}_{k}+{\mb{1}_n\otimes\nu_k}, 
\end{eqnarray}
which follows the sochasticity of $W$ matrix defined by \eqref{eq_stochastic}. Then, the distributed estimation of the mixed traffic ITS modelled by a network of $n$ CAVs tracking the state of $N$ HDVs follows the observability of the pair
\begin{equation} \label{eq_dist_obsrv}
	(W \otimes A, D_C),
\end{equation}
with $D_C$ representing the shared measurements as defined in~\eqref{eq_D_C}. This is
referred to as \textit{distributed observability} in Section~\ref{sec_single}. Then, a distributed estimator, e.g., \eqref{eq_p}-\eqref{eq_m}, can be adopted to address this tracking problem.

\paragraph{Cooperative Intelligent Transportation Framework}

Cooperative ITS represents a transformative approach to transportation management that integrates advanced communication technologies with distributed computational capabilities. By enabling vehicle-to-vehicle (V2V) and vehicle-to-infrastructure (V2I) communications \cite{ELASSY2024100252}, these systems create rich information ecosystems that support real-time decision-making across multiple scales -- from individual vehicles to entire transportation networks.

In this cooperative framework, distributed algorithms play a critical role in the following:
\begin{itemize}
	\item Enhancing scalability to accommodate growing numbers of connected vehicles and infrastructure elements;
	\item Improving robustness against individual component failures or communication disruptions;
	\item Enabling localized decision-making while maintaining global coordination;
	\item Reducing latency for safety-critical applications through edge processing; and
	\item Preserving privacy by minimizing the centralized collection of sensitive movement data.
\end{itemize}

\paragraph{Distributed Traffic State Estimation}

A fundamental challenge in managing transportation networks involves accurate estimation of critical traffic states, including vehicle speeds, traffic density, and travel times. Distributed algorithms address this challenge by aggregating data from heterogeneous sources -- including roadside cameras, in-vehicle GPS units, infrastructure sensors, and mobile devices -- to create comprehensive yet efficient traffic models \cite{Yuan2020its,Gao2018its}.

Various distributed fusion techniques have been developed for traffic state estimation, e.g. 
\begin{itemize}
	\item \textbf{Model-based approaches} that leverage dynamic traffic flow models:
	\begin{itemize}
		\item Distributed unscented Kalman filtering  \cite{Manogaran2020its} for nonlinear traffic dynamics;
		\item Consensus filter-based distributed variational Bayesian algorithms \cite{Safarinejadian2015its} that incorporate uncertainty quantification; and
		\item Parallelized particle filtering \cite{Mihaylova2012its} for handling complex, non-Gaussian traffic patterns.
	\end{itemize}
	
	\item \textbf{Constraint-based methods} that incorporate physical limitations and boundary conditions:
	\begin{itemize}
		\item Probabilistic-constrained distributed fusion filters \cite{Qu26042022} that maintain estimation consistency; and
		\item Set-membership filtering with attack detection capabilities \cite{Mousavinejad2019its} to ensure reliable operation despite malicious interference.
	\end{itemize}
	
	\item \textbf{Resource-aware frameworks} that balance computational loads:
	\begin{itemize}
		\item Distributed dynamic computation offloading \cite{Xia2022its} that optimizes processing across available resources.
	\end{itemize}
\end{itemize}

\paragraph{Collaborative Localization and Navigation}

Precise vehicle positioning represents a critical capability for advanced transportation systems, particularly for autonomous driving applications. Distributed approaches enable vehicles to share positioning data with nearby units, significantly improving localization accuracy beyond what individual vehicles could achieve in isolation \cite{Lu2021its,Fayyad2020its}.

These collaborative systems employ sophisticated sensor fusion techniques that integrate data from multiple sources, including the following:
\begin{itemize}
	\item Global Navigation Satellite Systems (GNSS);
	\item Inertial Measurement Units (IMUs);
	\item Light Detection and Ranging (LiDAR) sensors;
	\item Computer vision systems; and
	\item Infrastructure-based reference points.
\end{itemize}

By combining these diverse information sources through distributed algorithms, vehicles can achieve robust localization even in challenging urban environments with GPS signal obstruction, adverse weather conditions, or ambiguous visual features \cite{GAO2023110862}.

\paragraph{Coordinated Traffic Management}

Beyond individual vehicle capabilities, distributed algorithms enable system-wide coordination that improves overall traffic efficiency:
\begin{itemize}
	\item \textbf{Adaptive signal control} systems allow vehicles to communicate expected arrival times at intersections, enabling dynamic adjustments to signal timing that reduce unnecessary stops and congestion \cite{ISLAM2017272,Jaleel2020its}; and
	
	\item \textbf{Real-time routing optimization} leverages distributed fusion of traffic data to identify congestion patterns and suggest alternative pathways, reducing travel times across the network \cite{Kumar2023its,ZHANG2024103656}.
\end{itemize}

These distributed approaches offer significant advantages over traditional centralized traffic management by
\begin{itemize}
	\item Responding more rapidly to changing conditions;
	\item Scaling efficiently as network complexity grows;
	\item Continuing to function even when some components fail; and
	\item Adapting to localized traffic patterns while maintaining global coordination.
\end{itemize}

\paragraph{Vehicle Platooning Systems}

\begin{figure}
	\centering
	\includegraphics[width=4in]{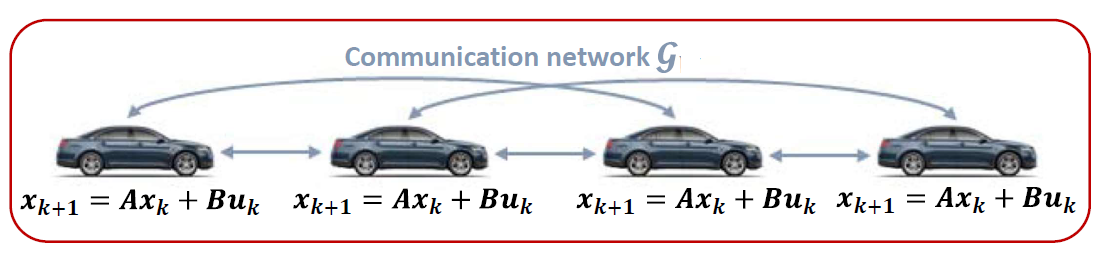}
	\caption{A group of connected and autonomous vehicles as a platoon communicating over an information-sharing network. Distributed algorithms are used to facilitate localized decision-making at vehicles while cooperating with the rest of platoon.
	} \label{fig_platoon}
\end{figure}

Vehicle platooning represents one of the most promising applications of distributed algorithms in transportation systems. This approach coordinates multiple automated vehicles travelling in close proximity, as illustrated in Fig.~\ref{fig_platoon}, offering benefits including reduced fuel consumption through improved aerodynamics, increased road capacity, and enhanced safety through coordinated maneuvers.

Platooning is typically facilitated through vehicular ad-hoc networks \cite{Abolfazli2023its} that enable real-time communication among vehicles and with surrounding infrastructure. Within these networks, distributed algorithms support several critical functions:
\begin{itemize}
	\item \textbf{Distributed observer-based tracking} enables vehicles to maintain precise inter-vehicle spacing while accounting for dynamic environmental conditions \cite{icrom24,Huang2023its,eurasip};
	
	\item \textbf{Security-enhanced platooning algorithms} address the unique vulnerabilities introduced by vehicle connectivity:
	\begin{itemize}
		\item Resilient distributed event-triggered approaches that maintain functionality despite denial-of-service attacks \cite{Zhao2023its,9440748};
		\item Secure platooning protocols designed to withstand various cyber threats \cite{MOUSAVINEJAD2022229};
		\item Robust estimation techniques that maintain reliability despite compromised measurements \cite{HE2021109953};
		\item Specialized detection algorithms for identifying false-data injection attacks \cite{10669813}; and
		\item Finite-time attack detection and estimation methods that provide rapid response to potential threats \cite{Guo2024its}.
	\end{itemize}
\end{itemize}

The integration of distributed estimation and fault detection techniques into vehicle platooning systems creates resilient, adaptive formations that can maintain safe operation even under challenging conditions, including communication disruptions, sensor failures, or malicious interference.
As transportation infrastructure continues to evolve toward greater connectivity and automation, the role of these distributed techniques will become increasingly central to ensuring reliable, efficient mobility.

\section{Conclusion and Future Directions} \label{sec_conclusion}
This comprehensive survey has examined the theoretical foundations, algorithmic developments, and practical applications of distributed estimation, filtering, and fault detection techniques within cyber-physical systems. By synthesizing insights across these interconnected domains, we have provided researchers and practitioners with a unified perspective on how distributed algorithms enable robust monitoring and control of complex, large-scale systems.

\subsection*{Key Contributions and Insights}

Our survey has yielded several important insights that collectively advance the understanding of distributed algorithms for CPS:

\begin{itemize}
	\item \textbf{Theoretical Foundations:} We have established a rigorous mathematical framework that integrates concepts from linear dynamical systems, graph theory, and observability analysis. This interdisciplinary foundation provides the necessary theoretical tools for understanding when and how distributed estimation can succeed in complex networked environments.
	
	\item \textbf{Algorithm Taxonomy:} Our systematic comparison of consensus-based approaches -- distinguishing between single-time-scale and double-time-scale algorithms -- clarifies the fundamental trade-offs between communication efficiency and estimation performance. This taxonomy enables system designers to select appropriate algorithms based on specific application constraints and requirements.
	
	\item \textbf{Resilience Mechanisms:} Through our analysis of diffusion-based filtering algorithms and observationally redundant designs, we have highlighted how distributed approaches can enhance system resilience against failures, disturbances, and malicious attacks. These mechanisms are particularly valuable in critical infrastructure applications where reliability is paramount.
	
	\item \textbf{Fault Management:} Our examination of distributed fault detection and isolation techniques demonstrates how localized monitoring can prevent cascading failures in interconnected systems. By detecting and isolating faults at their source, these approaches maintain overall system integrity even when individual components malfunction.
	
	\item \textbf{Resource Optimization:} The communication and computation complexity analysis provided throughout this survey offers valuable insights into the resource requirements of various distributed algorithms, informing implementation decisions for resource-constrained environments.
	
	\item \textbf{Real-World Applications:} Through detailed case studies spanning smart grids, social systems, target tracking, and intelligent transportation, we have illustrated how theoretical advances translate into practical solutions across diverse application domains.
\end{itemize}

\subsection*{Future Research Directions}

While significant progress has been made in distributed algorithms for cyber-physical systems, numerous challenges and opportunities remain for future research:

\paragraph{Communication Efficiency}

As distributed systems continue to scale, communication efficiency becomes increasingly critical. Future research could focus on the following:

\begin{itemize}
	\item \textbf{Event-triggered protocols} that dynamically adjust communication rates based on system conditions, reducing unnecessary data transmission while preserving estimation performance;
	
	\item \textbf{Compressed sensing techniques} that enable accurate reconstruction of system states from sparse measurements, minimizing the volume of data exchanged between nodes; and
	
	\item \textbf{Strategic information sharing} that optimizes which data to exchange based on information content rather than predetermined schedules, enhancing efficiency in bandwidth-constrained environments.
\end{itemize}

\paragraph{Data-Driven Approaches}

The increasing availability of operational data creates opportunities to enhance distributed algorithms through machine learning:

\begin{itemize}
	\item \textbf{Hybrid model-based and data-driven techniques} that combine the interpretability of first-principles models with the adaptability of learning-based approaches;
	
	\item \textbf{Distributed reinforcement learning} for adaptive estimation and filtering that improves performance through experience without requiring centralized training; and
	
	\item \textbf{Transfer learning methods} that enable knowledge sharing between related but distinct monitoring tasks, reducing the data requirements for new applications.
\end{itemize}

\paragraph{Privacy and Security}

As distributed algorithms process increasingly sensitive information, privacy and security considerations become paramount. Specifically, we have the following:

\begin{itemize}
	\item \textbf{Privacy-preserving distributed estimation} techniques that enable collaboration without exposing raw measurements, using approaches such as differential privacy, secure multi-party computation, or homomorphic encryption;
	
	\item \textbf{Resilient distributed algorithms} designed specifically to withstand sophisticated cyber attacks, including data manipulation, topology poisoning, and inference attacks; and
	
	\item \textbf{Trust mechanisms} that dynamically assess the reliability of information sources and adjust fusion weights accordingly.
\end{itemize}

\paragraph{Computational Architecture}

Emerging computational paradigms offer new possibilities for implementing distributed algorithms:

\begin{itemize}
	\item \textbf{Edge-cloud collaborative architectures} that strategically partition computation between local devices and cloud resources based on latency requirements, computational intensity, and available bandwidth;
	
	\item \textbf{IoT integration frameworks} that accommodate the heterogeneity, resource constraints, and intermittent connectivity characteristic of IoT deployments; and
	
	\item \textbf{Specialized hardware accelerators} for distributed estimation that enable more efficient implementation of key computational kernels.
\end{itemize}

\paragraph{Quality of Service Guarantees}

As distributed algorithms support increasingly critical applications, performance guarantees become essential to ensure the following:

\begin{itemize}
	\item \textbf{QoS-aware distributed designs} that explicitly consider application-specific performance metrics during algorithm selection and parameter tuning;
	
	\item \textbf{Resource allocation frameworks} that dynamically balance computational and communication resources across competing objectives based on current system priorities; and
	
	\item \textbf{Formal verification methods} for distributed algorithms that provide provable guarantees about estimation performance, fault detection reliability, and system stability.
\end{itemize}

These research directions collectively represent pathways toward more efficient, reliable, secure, and capable distributed algorithms for cyber-physical systems. As these systems continue to grow in scale, complexity, and importance, advances in distributed estimation, filtering, and fault detection will play an increasingly critical role in ensuring their safe and effective operation.

\bibliographystyle{elsarticle-num}
\bibliography{bibliography}

\end{document}